\documentclass[12pt]{article}

\usepackage{graphicx}
\usepackage{fancybox}
\usepackage{amsmath}
\usepackage{amssymb}
\usepackage{latexsym}
\usepackage{epsfig}

\newcommand{\Om}{\Omega}

\newcommand{\al}{\alpha}
\newcommand{\ep}{\epsilon}
\newcommand{\vep}{\varepsilon}

\newcommand{\la}{\lambda}

\newcommand{\R}{\mbox{R}}
\newcommand{\tR}{\widetilde{\mbox{R}}}

\newcommand{\lb}{\lbrack}
\newcommand{\rb}{\rbrack}

\newcommand{\msc}[1]{\mbox{\scriptsize #1}}
\newcommand{\dsp}{\displaystyle}

\newcommand{\bc}{\Bbb C}
\newcommand{\br}{\Bbb R}
\newcommand{\bz}{\Bbb Z}

\newcommand{\bsz}{\Bbb Z}
\newcommand{\bsr}{\Bbb R}

\newcommand{\s}{\mbox{{\bf s}}}
\newcommand{\sms}{\msc{{\bf s}}}

\newcommand{\cO}{{\cal O}}
\newcommand{\cN}{{\cal N}}

\newcommand{\cC}{{\cal C}}

\newcommand{\cD}{{\cal D}}

\newcommand{\cK}{{\cal K}}
\newcommand{\cZ}{{\cal Z}}

\newcommand{\tJ}{\tilde{J}}

\newcommand{\tj}{\tilde{j}}

\newcommand{\tell}{\tilde{\ell}}

\newcommand{\hc}{\hat{c}}

\newcommand{\hcK}{\widehat{\cal K}}

\newcommand{\Th}[2]{\Theta_{#1,#2}}
\renewcommand{\th}{{\theta}}

\newcommand{\chic}{{\chi_{\msc{con}}}}
\newcommand{\chid}{{\chi_{\msc{dis}}}}

\newcommand{\hchid}{\widehat{\chi}_{\msc{dis}}}

\newcommand{\tr}{\mbox{Tr}}

\newcommand{\tpsi}{\tilde{\psi}}

\newcommand{\erf}{\mbox{Erf}}
\newcommand{\erfc}{\mbox{Erfc}}
\newcommand{\sgn}{\mbox{sgn}}

\newcommand{\nn}{\nonumber\\}

\newcommand{\dis}{\rm dis}
\newcommand{\con}{\rm con}
\newcommand{\rem}{\rm rem}
\newcommand{\reg}{\rm reg}

\renewcommand{\Re}{{\rm Re}}
\renewcommand{\Im}{{\rm Im}}

\newcommand{\any}{{}^{\forall}}

\renewcommand{\mod}{\, \mbox{mod} ~ }

\newcommand {\eqn}[1]{(\ref{#1})}

\makeatletter
\@addtoreset{equation}{section}
\def\theequation{\thesection.\arabic{equation}}
\makeatother

\begin{document}

%%% Title page %%%%%
\begin{titlepage}
 \
 \renewcommand{\thefootnote}{\fnsymbol{footnote}}
 \font\csc=cmcsc10 scaled\magstep1
 {\baselineskip=14pt
 \rightline{
 \vbox{\hbox{December, 2010}
\hbox{YITP-94}
       }}}

 \baselineskip=20pt
%\vskip 1cm
 
\begin{center}

{\bf \Large  Non-holomorphic Modular Forms 
\\
and 
\\
$SL(2,\br)/U(1)$ Superconformal Field Theory} 

 \vskip 1.2cm
 
\noindent{ \large Tohru Eguchi}\footnote{\sf eguchi@yukawa.kyoto-u.ac.jp}
 \\
%{\sf eguchi@yukawa.kyoto-u.ac.jp}

\medskip

{\it Yukawa Institute for Theoretical Physics, Kyoto University, Kyoto 606-8502, Japan}

\vskip 8mm
\noindent{ \large Yuji Sugawara}\footnote{\sf ysugawa@se.ritsumei.ac.jp}
\\
%{\sf ysugawa@se.ritsumei.ac.jp}

\medskip

 {\it Department of Physical Science, 
 College of Science and Engineering, \\ 
Ritsumeikan University,  
%\\
%1-1-1 Noji-Higashi, 
%Kusatsu
Shiga 525-8577, Japan}
 
%\vskip 2mm

\end{center}

\bigskip

\begin{abstract}

We study the torus partition function of the $SL(2,\br)/U(1)$ SUSY gauged WZW model
coupled to $\cN=2$ $U(1)$ current. 
Starting from the path-integral formulation of the theory, 
we introduce an infra-red  regularization 
which preserves good modular properties  
and discuss the decomposition of the partition function in terms of the $\cN=2$ characters of discrete (BPS) and continuous (non-BPS) representations. 
Contrary to our naive expectation, we find a non-holomorphic dependence 
(dependence on $\bar{\tau}$) in the 
expansion coefficients of continuous representations. This non-holomorphicity appears in such a way that the anomalous modular behaviors of the discrete (BPS) characters are compensated by  the transformation law of the non-holomorphic coefficients of the continuous (non-BPS) characters.  Discrete characters together with the non-holomorphic continuous characters combine into   
real analytic Jacobi forms and these combinations exactly agree with  the ``modular completion" of discrete characters known in the theory of Mock theta functions \cite{Zwegers}. 
 
We consider this to be a general phenomenon: we expect to  encounter 
``holomorphic anomaly" ($\bar{\tau}$-dependence) in string partition function on non-compact target manifolds. The anomaly occurs due to the incompatibility of holomorphy and modular invariance of the theory. Appearance of non-holomorphicity in $SL(2,\br)/U(1)$ elliptic genus has recently been observed by Troost \cite{Troost}.

\end{abstract}

%\vfill

\setcounter{footnote}{0}
\renewcommand{\thefootnote}{\arabic{footnote}}

\end{titlepage}

\baselineskip 18pt

\vskip2cm 
%\newpage

%%%%%%%%%%%%%%%%%%%%%%%%%%%%%%%%%%%%%%%%%%%%%%%%%%%%%%%%%%%%%%%%%%%%%%%%%%%%%%%
%%%%%%%%%%%%%%%%%%%%%%%%%%%%%%%%%%%%%%%%%%%%%%%%%%%%%%%%%%%%%%%%%%%%%%%%%%%%%%%
%%%%%%%%%%%%%%%%%%%%%%%%%%%%%%%%%%%%%%%%%%%%%%%%%%%%%%%%%%%%%%%%%%%%%%%%%%%%%%%

\section{Introduction}

The torus partition function of 2D conformal field theory plays a basic role 
in unraveling the physical spectrum of the theory, 
and the modular invariance plays a key role in this analysis.

Roughly speaking, we have the next three possibilities for the Hilbert space of
a sensible conformal theory; 
\begin{itemize}
\item The space of normalizable states only includes a discrete set of primary states.

\item The space of normalizable states only includes a continuous family of primary states.

\item The space of normalizable states includes both discrete and continuous spectra. 

\end{itemize}
The first case is most familiar, and corresponds to compact backgrounds of string 
theory. 
The partition function of a closed string theory is schematically written in a holomorphically factorized form,  
$$
Z= \sum_{j, \tj} \, N_{j,\tj}\, \chi_j(\tau) \, \chi_{\tj}(\tau)^*, 
$$
where the conformal blocks $\chi_j(\tau)$ are identified with characters of 
some chiral algebra.
%\footnote
 %  {Here we do not necessarily assume `rational' conformal field theories. For instance, 
%the conformal theory of a single compact boson is well-known to be rational 
%iff the compactification radius is equal $R=\sqrt{\al' \frac{k}{k'}}$, 
%($k,k' \in \bz_{>0}$).  However, even in case  
%$R\neq \sqrt{\al' \frac{k}{k'}}$, the partition function is written 
%in a simple discrete sum, though it includes infinite chiral blocks. 
%}.
 $\chi_j(\tau)$ usually possesses 
simple modular transformation law and 
the modular invariance of the partition function is achieved in a relatively easy way.

The second case describes non-compact backgrounds. 
This class of models still has a simple structure of modular invariant partition functions, which includes integrals over parameters 
of continuous representations. 
The bosonic or $\cN=1$ Liouville theories are non-trivial examples of this class.  
These theories also contain a discrete set of {\em non-normalizable\/} primary fields, which correspond to degenerate representations of (super-) Virasoro algebra. These states, however, do not appear in the torus partition function\footnote
  {This feature is often phrased as the lack of operator-state correspondence.}.

The third one is the most intriguing class. The partition function is expected 
to be decomposed into a sum of the `discrete' and `continuous' parts; 
\begin{equation}
Z(\tau)= Z_{\dis}(\tau) + Z_{\con}(\tau). 
\label{Z decomp}
\end{equation}
These models are expected to describe non-compact backgrounds 
with the continuous part $Z_{\con}(\tau)$ representing the propagating degrees of freedom. However, in contrast to the second case, we also have the discrete part $Z_{\dis}(\tau)$  
which represents normalizable bound states localized in some curved region of space-time.  
A non-trivial issue in this class of conformal field theories 
is how to achieve the modular invariance of the theory.  
In fact, in these models, non-trivial mixing of discrete and continuous representations often take place 
under the modular S-transformation, 
which makes it difficult to assure modular invariance in a simple manner. 
Such an anomalous modular behavior has been first observed in 
the massless (BPS) representation of $\cN=4$ superconformal algebra \cite{ET}.  
Similar behavior appears in the discrete (BPS) characters 
in the $\cN=2$ Liouville or the $SL(2,\br)/U(1)$ SUSY gauged WZW model \cite{ES-L,ES-BH}
(see also \cite{Odake,Miki}).
In our previous attempts \cite{ES-BH,ES-C} (see also \cite{HPT,IPT}), 
these models were assumed to have the partition function 
of the form \eqn{Z decomp}, and we have identified the discrete spectra $Z_{\dis}(\tau)$ of the theory.
However, the issue of modular invariance remained unsolved  
due to the complexity of modular property of discrete (BPS) representations mentioned above.

In this paper, we study the $SL(2,\br)/U(1)$ SUSY gauged WZW model, 
focusing mainly on this issue. 
We shall analyze the torus partition function with a  modulus 
coupled to $\cN=2$ $U(1)$ current. 
We begin our analysis based on the approach of path-integration, 
which automatically ensures the modular invariance.  
%%%%%%%%%%%%%%%%%%%%%%%%%%%%%%%%%%%%%%%%%%%%%%%%%%
%%%%%%%%%%%%%%%%%%%%%%%%%%%%%%%%%%%%%%%%%%%%%%%%%%
%%%%%%%%%%%%%%%%%%%%%%%%%%%%%%%%%%%%%%%%%%%%%%%%%%
We then introduce a suitable regularization preserving good modular
behaviors\footnote
 {This in fact means that the regularized partition function 
slightly breaks the modular S-invariance. However, the violation is 
under a good control by a small parameter characterizing the regularization.  
T-invariance is kept intact by the regularization. 
See subsection 2.3 for the detail.
},
and discuss the decomposition of regularized partition function into ${\cal N}=2$ characters. 
%%%%%%%%%%%%%%%%%%%%%%%%%%%%%%%%%%%%%%%%%%%%%%%%%%
%%%%%%%%%%%%%%%%%%%%%%%%%%%%%%%%%%%%%%%%%%%%%%%%%%
%%%%%%%%%%%%%%%%%%%%%%%%%%%%%%%%%%%%%%%%%%%%%%%%%%
It turns out that characters of the discrete representations in the
 partition function 
 are always accompanied by a series of continuous representations with 
{\em non-holomorphic} 
($\bar{\tau}$-dependent) coefficients in such a way that together they transform like Jacobi forms. This is in fact the structure known as the  ``modular completion" of discrete representations  in the theory of  Mock theta functions \cite{Zwegers}.
Thus the correct partition function is given by (\ref{Z decomp}) with discrete part replaced by its modular completion.

We consider that this is a general phenomenon and string theory amplitudes on non-compact manifolds possess an anomaly coming from the incompatibility between  holomorphy and modular invariance. When we insist on strict modular invariance, we may loose holomorphy, while if we relax modular invariance, for instance, to subgroups of $SL(2,\bz)$ we may possibly keep holomorphy intact. These alternatives will correspond to the choice of boundary conditions at infinity of non-compact manifolds.

In this paper we further study the elliptic genus of $SL(2,\bz)/U(1)$ theory based on 
the character decomposition of partition function mentioned above
and also the direct evaluation of path-integral representation of elliptic genus.
The latter approach has an advantage of no need of regularization and can be compared closely with the  
paper \cite{Troost} which inspired the present work. It turns out that both analyses will lead us to an 
identical result given again in terms of real analytic Jacobi forms having 
$\bar{\tau}$ dependence. 

Unexpectedly, we find the use of the mathematical theory of Mock theta functions in the analysis of non-compact geometry in string theory\footnote{Other applications of mock theta functions to studies of superconformal 
field theories are given in recent papers \cite{EH}.}.

~

%%%%%%%%%%%%%%%%%%%%%%%%%%%%%%%%%%%%%%%%%%%%%%%%%%%%%%%%%%%%%%%%%%%%%%%%%%%%%%%%%%%%%%%%%%%%%%%%%%%%%%%%%%%%%%%%%%%%%%%%%%%%%%%%%%%%%%%%%%%%%%%%%%%%%%%%%%%%%%%%%%%%%%%%%%%%%%%%%%%%%%%%%%%%%%%%%%%%%%%%%%%%%%%%%%%%%%%%%%%%%%%%%%%%%%%%%%%%%%%%%%%%%%%%

\section{Partition Function of $SL(2,\br)/U(1)$ SUSY Gauged WZW Model}

\subsection{$SL(2,\br)/U(1)$ SUSY Gauged WZW Model}

We shall first introduce the model which we study in this paper, summarizing relevant 
notations.
We consider the $SL(2,\br)/U(1)$ SUSY gauged WZW model with level $k$ \footnote
   {$k$ is the level of the total $SL(2,\br)$-current, whose bosonic part has the level $\kappa \equiv k+2$.}, 
%$k=\frac{N}{K}$, ($N, K $ are positive co-prime integers), in other words, with 
%$\hc(\equiv \frac{c}{3})= 1+ \frac{2K}{N}$.
which is well-known  \cite{KS} to have 
$\cN=2$ superconformal symmetry with central 
charge;
%
%The Kazama-Suzuki supercoset model \cite{KS} for $SL(2;\br)_k/U(1)$ 
%is defined as the coset CFT
%\begin{eqnarray}
%\frac{SL(2;\br)_{\kappa}\times SO(2)_1}{U(1)_{-(\kappa-2)}}~,
%\end{eqnarray}
%which is an $\cN=2$ SCFT with the central charge and level \footnote
% {Throughout this paper we denote the level of super $SL(2;\br)$ as $k$, 
%   and the level of bosonic $SL(2;\br)$ as $\kappa=k+2$.}
%
\begin{equation}
\hat{c} \equiv \frac{c}{3} =1+\frac{2}{k}, ~~~ k\equiv \kappa-2.
\label{def c k}
\end{equation}
The world-sheet action in the present convention is written as
\begin{eqnarray}
 S(g,A,\psi^{\pm}, \tpsi^{\pm}) &=& \kappa S_{\msc{gWZW}}(g,A)
 + S_{\psi}(\psi^{\pm}, \tpsi^{\pm}, A), 
\label{total action}\\
%%%
\kappa S_{\msc{gWZW}} (g,A) &=& \kappa S^{SL(2,\bsr)}_{\msc{WZW}} (g)
+ \frac{\kappa}{\pi}\int_{\Sigma}d^2v\, 
\left\{\tr \left(\frac{\sigma_2}{2}g^{-1}\partial_{\bar{v}}g\right)A_v
+ \tr \left(\frac{\sigma_2}{2} \partial_v g g^{-1}\right)A_{\bar{v}}    
\right. \nn
&& \hspace{3cm} \left. 
+ \tr\left(\frac{\sigma_2}{2}g\frac{\sigma_2}{2} g^{-1} \right)
   A_{\bar{v}}A_v + \frac{1}{2} A_{\bar{v}}A_v  \right\} , 
\label{gWZW action} \\
S^{SL(2,\bsr)}_{\msc{WZW}} (g) &=& -\frac{1}{8\pi} \int_{\Sigma} d^2v\,
\tr \left(\partial_{\al}g^{-1}\partial_{\al}g\right) +
\frac{i}{12\pi} \int_B \,\tr\left((g^{-1}dg)^3\right) ,
\label{SL(2) WZW action} \\
%%%
 S_{\psi}(\psi^{\pm}, \tpsi^{\pm}, A)& =& 
\frac{1}{2\pi}\int d^2v\, \left\{
\psi^+(\partial_{\bar{v}}+A_{\bar{v}}) \psi^-
+\psi^-(\partial_{\bar{v}}-A_{\bar{v}}) \psi^+ \right. \nn 
&& \hspace{3cm} \left. +\tpsi^+(\partial_{v}+A_{v}) \tpsi^-
+\tpsi^-(\partial_{v}-A_{v}) \tpsi^+
\right\} ,
\label{fermion action}
\end{eqnarray}
where the complex fermions $\psi^{\pm}$ (and $\tpsi^{\pm}$)
have charge $\pm 1$ with respect to  the $U(1)$-gauge group. 
The bosonic part $\kappa S_{\msc{gWZW}}(g,A)$ is 
the gauged WZW action for the coset $SL(2, \br)_{\kappa}/U(1)_A$ 
\cite{KarS,GawK}, 
where $U(1)_A$ indicates the gauging of 
axial $U(1)$-symmetry; $g\,\rightarrow\, \Omega\, g \,  \Omega$,
$\Omega(v,\bar{v}) = e^{iu(v,\bar{v}) \sigma_2}$ ($u(v,\bar{v})\in
\br$, $\sigma_2$ is the Pauli matrix\.)\footnote
    {We take $\{\frac{i\sigma_2}{2}, ~ \frac{\sigma_3}{2}, ~ \frac{\sigma_1}{2}
\}$ for the basis of $SL(2, \br) \cong SU(1,1)$.}. 
It is well-known that this model describes the string theory on 2D Euclidean 
black-hole with the cigar geometry \cite{2DBH}. 
The WZW action $\kappa S^{SL(2,\bsr)}_{\msc{WZW}} (g)$ is 
formally equal to  $-\kappa S^{SU(2)}_{\msc{WZW}}(g)$, 
and has a negative signature in $i\sigma_2$-direction.
Since we have $H^3(SL(2, \br))=0$,   
the action $\kappa S^{SL(2, \bsr)}_{\msc{WZW}} (g)$
can be rewritten in a purely two dimensional form and the level $\kappa$  
need not be an integer.

%%%%%%%%%%%%%%%%%%%%%%%%%%%%%%%%%%%%%%%%%%%%%%%%%%%%%%%%%%%%%%%%%%%%%%%%%%%%%%%%%%%%%%
%%%%%%%%%%%%%%%%%%%%%%%%%%%%%%%%%%%%%%%%%%%%%%%%%%%%%%%%%%%%%%%%%%%%%%%%%%%%%%%%%%%%%%
According to the familiar treatment of gauged WZW models, 
one can easily separate the anomalous degrees of freedom originating from chiral gauge 
transformations (the compact scalar field $Y$  given in the next subsection), 
which makes the relevant field contents to be free and chiral \cite{KarS,GawK,Schnitzer}.
%%%%%%%%%%%%%%%%%%%%%%%%%%%%%%%%%%%%%%%%%%%%%%%%%%%%%%%%%%%%%%%%%%%%%%%%%%%%%%%%%%%%%%
%%%%%%%%%%%%%%%%%%%%%%%%%%%%%%%%%%%%%%%%%%%%%%%%%%%%%%%%%%%%%%%%%%%%%%%%%%%%%%%%%%%%%%
After this procedure, the chiral currents defined by  
\begin{eqnarray}
&& j^A(v) = \kappa \tr \left(T^A\partial_v g g^{-1}\right), 
~~~ \tilde{j}^A(\bar{v}) = -\kappa \tr 
\left(\overline{T^A} g^{-1}\partial_{\bar{v}} g\right) , 
\label{WZW currents} \\
&& T^3 = \frac{1}{2}\sigma_2~,~~ T^{\pm} = \pm
\frac{1}{2}\left(\sigma_3\pm i\sigma_1\right) 
\end{eqnarray}
satisfy the affine $\widehat{SL}(2, \br)_{\kappa}$ current algebra
(we write the left-mover only);
\begin{equation}
\left\{
\begin{array}{lll}
j^3(v)j^3(0) &\sim& \dsp -\frac{\kappa/2}{v^2} \\
j^3(v)j^{\pm}(0)&\sim& \dsp \frac{\pm 1}{v}j^{\pm}(0) \\
j^+(v)j^-(0) &\sim& \dsp \frac{\kappa}{v^2}-\frac{2}{v}j^3(0)
\end{array}
\right.
\label{SL 2}
\end{equation}
and the pair of free fermions $\psi^+$, $\psi^-$ satisfy 
the OPE's $\dsp \psi^+(v)\psi^-(0) \sim 1/v$,
$\psi^{\pm}(v)\psi^{\pm}(0)\sim 0$. 
The explicit realization of $\cN=2$ SCA is given by  
\begin{eqnarray}
&&T(v)= \frac{1}{k} \eta_{AB}j^Aj^B + \frac{1}{k} J^3 J^3 
- \frac{1}{2}(\psi^+\partial_v \psi^- - \partial_v \psi^+ \psi^-),~~~
(\eta_{AB}= \mbox{diag}(1,1,-1)), \nn
&& J = \psi^+\psi^- + \frac{2}{k}J^3, ~~~
 G^{\pm} = \frac{1}{\sqrt{k}} \psi^{\pm}j^{\mp} , 
\label{N=2 SCA}
\end{eqnarray}
where we introduced the total $U(1)$-current  $J^3 \equiv j^3 + \psi^+\psi^-$, 
which couples with the gauge field $A$ and commuetes with all the generators of 
$\cN=2$ SCA \eqn{N=2 SCA}.

~

%%%%%%%%%%%%%%%%%%%%%%%%%%%%%%%%%%%%%%%%%%%%%%%%%%%%%%%%%%%%%%%%%%%
%%%%%%%%%%%%%%%%%%%%%%%%%%%%%%%%%%%%%%%%%%%%%%%%%%%%%%%%%%%%%%%%%%%

\subsection{Path Integral Evaluation of Torus Partition Function}

The main purpose of this subsection is to evaluate the torus partition 
function of the SUSY gauged WZW model \eqn{total action} by 
path-integration. 
We define the world-sheet torus $\Sigma$ 
by the identifications  
$(w,\bar{w}) \sim (w+2\pi,\bar{w}+2\pi)
\sim(w+2\pi \tau, \bar{w}+2\pi \bar{\tau})$ ($\tau\equiv
\tau_1+i\tau_2$, $\tau_2>0$, 
and use the convention $v=e^{iw}$, $\bar{v}=e^{-i\bar{w}}$).
We call the cycles defined by these two identifications
as the $\al$ and $\beta$-cycles as usual.

The relevant calculation is  carried out in a way parallel to that given in
\cite{ES-BH}, although we have the following differences;
%%%
\begin{itemize}
\item We shall focus on the $\tR$-sector ($\R$-sector with $(-1)^F$ insertion) of the theory.  
Character formulas in the following sections are 
those in the $\tR$ sector. Formulas in other sectors are obtained by spectral flow.
%%%
\item We introduce the complex moduli  $z$, $\bar{z}$
which couple with the zero-modes of  $\cN=2$ $U(1)$-currents 
$J(v)$, $\tJ(\bar{v})$ \eqn{N=2 SCA}.
In other words, we would like to evaluate 
the partition sum weighted by $e^{2\pi i \left(z J_0 - \bar{z} \tJ_0\right)}$.
\end{itemize}
%%%

We shall start with the Wick rotated model in order to make 
the gauged WZW action \eqn{gWZW action} positive definite. 
This means replacing 
$
g(v,\bar{v}) \in SL(2, \br) 
$
with 
$
g(v,\bar{v}) \in H^+_3 \cong SL(2, \bc)/SU(2), 
$
and the gauge field 
$ \dsp
A \equiv \left(A_{\bar{v}}d\bar{v}+ A_v dv\right)\frac{\sigma_2}{2}
$
should be regarded as a hermitian 1-form. 
%%%%%%%%%%%%%%%%%%%%%%%%%%%%%%%%%%%%%%%%%%%%%%%%%%%%%%%%%%%%%%%%%%%%%%%%
It is convenient to reexpress our (axial-like) gauged WZW action 
$S^{SL(2,\br)}_{\msc{gWZW}}(g,A)$ \eqn{gWZW action} 
in terms of $S^{(A)}_{\msc{gWZW}}(g,h,h^{\dag})$
given in  \eqn{gWZW A} 
with the identification 
$
\dsp 
A_{\bar{v}}\frac{\sigma_2}{2} = \partial_{\bar{v}} h h^{-1},
$
$
\dsp 
A_{v}\frac{\sigma_2}{2} = \partial_{v} h^{\dag} h^{\dag\, -1}.
$
We also parameterize $h \in \exp \left(\bc \sigma_2\right)$ as 
\begin{eqnarray}
h=  \Omega h[u] , ~~~ \Omega \equiv e^{(X+iY)\frac{\sigma_2}{2}}, ~~~
h[u] \equiv e^{i\Phi[u] \frac{\sigma_2}{2}},
\label{parameterization h}
\end{eqnarray}
where real scalar fields $X$, $Y$ correspond 
to the axial ($\br_A$) and vector ($U(1)_V$) 
gauge transformations respectively.
$\Phi[u] (w,\bar{w})$ is associated with the modulus of a holomorphic
line bundle; 
$u\equiv s_1\tau + s_2 \in \mbox{Jac}(\Sigma) \cong \Sigma$, 
$(0\leq s_1,s_2 <1)$, conventionally defined as 
\begin{eqnarray}
\Phi[u] (w,\bar{w})= \frac{i}{2\tau_2}\left\{
(\bar{w}\tau-w\bar{\tau})s_1+(\bar{w}-w)s_2\right\}
\equiv \frac{1}{\tau_2} \Im (w \bar{u}).~
%(u\equiv s_1\tau-s_2,~0\leq s_1,s_2 <1)~.
\label{Phi u}
\end{eqnarray} 
It is a real harmonic function 
satisfying the twisted boundary conditions;
\begin{eqnarray}
\hspace{-1cm}
\Phi[u](w+2\pi,\bar{w}+2\pi)=\Phi[u](w,\bar{w}) - 2\pi s_1,~~~
\Phi[u](w+2\pi\tau,\bar{w}+2\pi\bar{\tau})=\Phi[u](w,\bar{w}) + 2\pi s_2.
\label{bc Phi u}
\end{eqnarray}
Emphasizing the modulus dependence, 
we shall denote the corresponding gauge field as $A[u]$, namely,
\begin{eqnarray}
&& A[u]_{\bar{w}}= \partial_{\bar{w}} X  + i \partial_{\bar{w}} Y + a[u]_{\bar{w}}, 
~~~  A[u]_{w}=  \partial_{w} X  - i \partial_{w} Y + a[u]_{w},
\label{Au} \\
%%%
&& a[u]_{\bar{w}} \equiv i \partial_{\bar{w}} \Phi[u] \equiv - \frac{u}{2\tau_2},
~~~ a[u]_{w} \equiv - i \partial_{w} \Phi[u] \equiv - \frac{\bar{u}}{2\tau_2}.
\label{au}
\end{eqnarray}
%%%%%%%%%%%%%%%%%%%%%%%%%%%%%%%%%%%%%%%%%%%%%%%%%%%%%%%
%%%%%%%%%%%%%%%%%%%%%%%%%%%%%%%%%%%%%%%%%%%%%%%%%%%%%%%
%It will be useful to point out that our modulus parameter $u$ yields 
%the insertion of $e^{2\pi i \left(u J^3_0 -\bar{u} \tJ^3_0 \right)}$
%in the operator formalism. 
%%%%%%%%%%%%%%%%%%%%%%%%%%%%%%%%%%%%%%%%%%%%%%%%%%%%%%%
%%%%%%%%%%%%%%%%%%%%%%%%%%%%%%%%%%%%%%%%%%%%%%%%%%%%%%%
It will be useful to point out that our modulus parameter $u$
is normalized so that the partial derivative of classical action 
\eqn{total action} with respect to it yields 
$$
\left. 
- \frac{\partial}{\partial u} S(g,a[u], \psi^{\pm}, \tpsi^{\pm}) \right|_{u=0} 
= 2\pi i J^3_0, ~~~ 
\left. 
- \frac{\partial}{\partial \bar{u}} S(g,a[u], \psi^{\pm}, \tpsi^{\pm}) \right|_{u=0} 
= - 2\pi i \tJ^3_0.
$$

%%%%%%%%%%%%%%%%%%%%%%%%%%%%%%%%%%%%%%%%%%%%%%%%%%%%%%%
%%%%%%%%%%%%%%%%%%%%%%%%%%%%%%%%%%%%%%%%%%%%%%%%%%%%%%%

~

Now, the desired  partition function is schematically written as 
\begin{eqnarray}
%&& Z^{(\stR)}(\tau, \bar{\tau}; \z, \bar{\z}) 
&& Z(\tau,  z) 
= \int_{\Sigma}\frac{d^2u}{\tau_2} \, 
\int \cD\lb g, A[u], \psi^{\pm}, \tpsi^{\pm}\rb\, 
\nn
&& \hspace{2cm} \times 
\exp \left[-\kappa S_{\msc{gWZW}}\left(g,A[u+\frac{2}{k}z]\right) - 
S_{\psi}\left(\psi^{\pm},\tpsi^{\pm}, A[u+\frac{k+2}{k} z ]\right)\right],
\label{part fn 0}
\end{eqnarray}
where $ \frac{d^2u}{\tau_2} \equiv ds_1 ds_2$ is the modular invariant measure 
of modulus parameter $u$, and we work in  the $\tR$-sector for world-sheet fermions. 
%%%%%%%%%%%%%%%%%%%%%%%%%%%%%%%%%%%%%%%%%%%%%%%%%%%%%%%%%%%%%%%%%%%%%%%%%%
%Note that the inclusion of complex parameter $z$ corresponds to  
%the insertion of an operator 
%$e^{2\pi i \left(z J_0 - \bar{z} \tJ_0\right)}$ since 
%the $\cN=2$ $U(1)$-current is given by
%$$
%J(v) = \psi^+\psi^- + \frac{2}{k} J^3 \equiv \frac{2}{k}j^3+ %\frac{k+2}{k}\psi^+\psi^-.
%$$
%%%%%%%%%%%%%%%%%%%%%%%%%%%%%%%%%%%%%%%%%%%%%%%%%%%%%%%%%%%%%%%%%%%%%%%%%%%

%%%%%%%%%%%%%%%%%%%%%%%%%%%%%%%%%%%%%%%%%%%%%%%%%%%%%%%%%%%%%%%%%%%%%%%
%%%%%%%%%%%%%%%%%%%%%%%%%%%%%%%%%%%%%%%%%%%%%%%%%%%%%%%%%%%%%%%%%%%%%%%
%%%%%%%%%%%%%%%%%%%%%%%%%%%%%%%%%%%%%%%%%%%%%%%%%%%%%%%%%%%%%%%%%%%%%%%
As expected, the inclusion of complex parameter $z$ corresponds to  
the marginal deformation\footnote
  {Here we use  the word ``marginal'' in the sense that 
   it preserves the exact conformal symmetry (while violating supersymmetry).} described by 
the insertion of $e^{2\pi i \left(z J_0 - \bar{z} \tJ_0\right)}$. 
In fact, one can directly confirm from \eqn{part fn 0} that 
$
\left.
\frac{\partial}{\partial z} Z(\tau,z)\right|_{z=0}
$
yields the insertion of zero-mode $2\pi i J_0$,
after making a suitable chiral gauge transformation\footnote
   {A non-trivial point is the contribution from 
the anomaly for chiral gauge transformations
(coupling to the scalar field $Y$). 
As we will observe later, the linear couplings of modulus $z$ with 
the currents $i \partial Y$ cancel out, and the $z$-dependence
 appears at a quadratic order, which does not spoil 
the interpretation of $\left. \frac{\partial}{\partial z} Z \right|_{z=0}$ as 
the insertion of $2\pi i J_0$.}.
Moreover, this deformation is strictly marginal 
{\em even} under the $u$-integration, because the modulus $u$   
couples with the zero-mode of current $ j^3 + \psi^+\psi^- + (\mbox{anomaly 
contribution})$ associated to the $U(1)$-coseting,  
which has no singular OPE with the $\cN=2$ $U(1)$-current $J$. 
These facts imply that the complex parameter $z$ appearing 
in \eqn{part fn 0} should be captured by the insertion 
of $e^{2\pi i z J_0}$ in the operator formalism.
(The same is true for $\bar{z}$.)
%%%%%%%%%%%%%%%%%%%%%%%%%%%%%%%%%%%%%%%%%%%%%%%%%%%%%%%%%%%%%%%%%%%%%%%%%%%%%
%%%%%%%%%%%%%%%%%%%%%%%%%%%%%%%%%%%%%%%%%%%%%%%%%%%%%%%%%%%%%%%%%%%%%%%%%%%%%
%%%%%%%%%%%%%%%%%%%%%%%%%%%%%%%%%%%%%%%%%%%%%%%%%%%%%%%%%%%%%%%%%%%%%%%%%%%%%

A simple way to evaluate \eqn{part fn 0} is to
convert the functional integral of gauge field 
into those of real scalar fields
$X$, $Y$ and modulus $u$,  as is manipulated in \cite{GawK}.
In doing so, we have to be a bit careful, because 
different values of moduli couple with bosonic 
and fermionic parts. 

%%%%%%%%%%%%%%%%%%%%
%\newpage
%%%%%%%%%%%%%%%%%%%%

Namely, 
\begin{itemize}
\item For the bosonic sector, 
the modulus parameter takes a value 
$u+\frac{2}{k}z$. The scalar field of anomaly free 
direction $X$ 
decouples, while the contribution from 
anomalous $U(1)_V$ ($Y$ direction) is 
extracted by utilizing the identity;
\begin{eqnarray}
%&& S_{\msc{gWZW}} ({}^{\Om}g,{}^{\Om}A) \equiv  
&& 
\hspace{-1cm}
-\kappa S^{(A)}_{\msc{gWZW}} (\Om^{-1} g\Om^{\dag\, -1},\, 
\Om h[u+\frac{2}{k}z],\, h[u+\frac{2}{k}z]^{\dag} \Om^{\dag})  
\nn
&& 
%\hspace{1cm}
= -\kappa S^{(V)}_{\msc{gWZW}}(g,h[u+\frac{2}{k}z],h^{\dag}[u+\frac{2}{k}z]) 
+\kappa S^{(A)}_{\msc{gWZW}}(\Om \Om^{\dag\, -1},\, h[u+\frac{2}{k}z], 
\,h[u+\frac{2}{k}z]^{\dag\,-1}),
\nn
&&
\label{PW2}
\end{eqnarray} 
which is readily derived from the definitions 
\eqn{gWZW V}, \eqn{gWZW A}.
We here assume the gauge invariance of path-integral measure as usual;
$\cD (\Om^{-1} g \Om^{\dag\,-1})=\cD g$. 
%%%%

\item For the fermionic sector,
the modulus parameter takes a value $u+\frac{k+2}{k}z$.
We should regularize the fermion determinant so as to be gauge invariant 
along the $\br_A$-direction, and the anomalous 
$U(1)_V$-direction is again described by 
the gauged WZW action \\
$ -2 S^{(A)}_{\msc{gWZW}}(\Om \Om^{\dag\, -1},\, h[u+\frac{k+2}{k}z], \,
h[u+\frac{k+2}{k}z]^{\dag\,-1})$.
%with a different coefficient. 
 
%%%%

\item We have to also introduce the standard $bc$-ghosts with 
spin (1,0) to rewrite the Jacobian for the transformation 
of path-integral measure $\cD A\, \rightarrow \, \cD X \cD Y$.

\end{itemize}

Explicitly, the partition function of each sector is  evaluated as follows;
\begin{description}
\item
[\underline{$H_3^+$-sector :} ] 

For the $H^3_+$-sector, 
we obtain by using the formulas \eqn{Gaw formula 2} (up to a normalization
factor);
\begin{eqnarray}
Z_g\left(\tau, u+\frac{2}{k}z\right) &\equiv & 
\int \cD g\, \exp\left[-\kappa S^{(V)}_{\msc{gWZW}}\left(g, \, h[u+\frac{2}{k}z],\, 
h[u+\frac{2}{k}z]^{\dag}\right)\right] 
\nn
&=& \frac{e^{2\pi \frac{(u_2+\frac{2}{k}z_2)^2}{\tau_2}}}{\sqrt{\tau_2}
|\th_1(\tau,u+\frac{2}{k}z)|^2}.
\label{Z g}
\end{eqnarray}

%%%%%%%%%%%%%%%%%%%%%%%%%%%%%%%%%%%%%%%%%%%%%%%%%%%%%%%%%%%%%
\item
[\underline{fermion and ghost sectors :} ]

The path-integration of fermionic and ghost sectors yields  
the standard fermion determinants with 
periodic boundary conditions, since we are working in the $\tR$-sector for
$\psi^{\pm}$, $\tpsi^{\pm}$.  
\begin{eqnarray}
%Z^{(\stR)}_{\psi}\left(\tau,u+\frac{k+2}{k}\z\right)
Z_{\psi}\left(\tau,u+\frac{k+2}{k}z\right)
&\equiv & 
\int \cD\lb\psi^{\pm},\tpsi^{\pm}\rb\,
\exp\left[ -S_{\psi}\left(\psi^{\pm},\tpsi^{\pm},a[u+\frac{k+2}{k}z]\right)\right] 
\nn
&=& e^{-2\pi \frac{(u_2+ \frac{k+2}{k}z_2)^2}{\tau_2}}
\frac{\left|\th_1(\tau,u+\frac{k+2}{k}z)\right|^2}
{\left|\eta(\tau)\right|^2},
\label{Z psi} \\
%%%
Z_{\msc{gh}}(\tau)
&\equiv& 
\int \cD\lb b, \tilde{b}, c, \tilde{c}\rb\,
\exp \left[-S_{\msc{gh}}(b,\tilde{b},c,\tilde{c})\right] = \tau_2\left|\eta(\tau)\right|^4~.
\label{Z gh}
\end{eqnarray}

%%%%%%%%%%%%%%%%%%%%
\item
[ \underline{$U(1)_V$-sector :} ]

This sector is described by a single compact boson $Y$~ ($Y\sim Y+2\pi$).
Recalling \eqn{def c k},
%$k= \kappa-2$ and $\dsp \hc = \frac{k+2}{k}$, 
we can explicitly compute the relevant world-sheet action as
%%%%%%%%%%%%%%%%%%%%%%%%%%%%%%%%%%%%%%%%
%%%%%%%%%%%%%%%%%%%%%%%%%%%%%%%%%%%%%%%%
\begin{eqnarray}
S_Y(Y;u, z)&\equiv& 
- \kappa S^{(A)}_{\msc{gWZW}}\left(e^{iY\sigma_2}, \, 
h[u+\frac{2}{k}z], \, h[u+\frac{2}{k}z]^{\dag \, -1}\right)
\nn
&& \hspace{1.5cm}
+2 S^{(A)}_{\msc{gWZW}}\left(e^{iY\sigma_2}, \,
h[u+\frac{k+2}{k}z], \, h[u+\frac{k+2}{k}z]^{\dag \, -1}\right)
\nn
&=& \frac{k+2}{\pi}\int d^2w\, 
\left|\partial_{\bar{w}}\left(Y+\Phi[u+\frac{2}{k}z]\right)\right|^2 
\nn
&& \hspace{1.5cm} 
-  \frac{2}{\pi}\int d^2w\, 
\left|\partial_{\bar{w}}\left(Y+\Phi[u+\frac{k+2}{k}z]\right)\right|^2 
\nn
&=&  \frac{k}{\pi} \int d^2w\, \left|\partial_{\bar{w}} Y^u\right|^2 
+ \frac{k+2}{\pi} \int d^2w\, 
\left|\partial_{\bar{w}} \Phi[\frac{2}{k}z]\right|^2
-  \frac{2}{\pi} \int d^2w\, 
\left|\partial_{\bar{w}} \Phi[\frac{k+2}{k}z]\right|^2
\nn
&=& \frac{k}{\pi} \int d^2w\, \left|\partial_{\bar{w}} Y^u\right|^2 
-\frac{2\pi}{\tau_2} \hc |z|^2,
\label{SYuz}
\end{eqnarray}
%%%%%%%%%%%%%%%%%%%%%%%%%%%%%%%%%%%%%%%%%%%%%
%%%%%%%%%%%%%%%%%%%%%%%%%%%%%%%%%%%%%%%%%%%%%
where $Y^u\equiv Y +\Phi[u]$ and satisfies the twisted boundary conditions;
\begin{eqnarray}
&&Y^u(w+2\pi,\bar{w}+2\pi)=Y^u(w,\bar{w})- 2\pi (m_1+s_1), \nn
&&Y^u(w+2\pi\tau,\bar{w}+2\pi\bar{\tau})=Y^u(w,\bar{w})+ 2\pi (m_2+s_2)
, ~~~(m_1, m_2 \in \bz).
\end{eqnarray}
%where we set $u\equiv s_1\tau-s_2 \in \Sigma$ ($0\leq s_1,s_2 <1$). 
Rescaling canonically the twisted boson $Y^u$ as $Y^u\, \rightarrow\, Y^u/\sqrt{\al' k}$,
we arrive at the theory of a twisted compact boson
with radius $R=\sqrt{\al' k}$. 
Therefore, the relevant path-integration yields
\begin{eqnarray}
\hspace{-1cm}
Z_Y(\tau,u, z)&\equiv & 
\int \cD Y^u\, \exp\left[-\frac{1}{\pi \al'} \int d^2w\, 
\left|\partial_{\bar{w}} Y^u\right|^2
+ \frac{2\pi}{\tau_2} \hc \left|z \right|^2 \right] \nn
&=&  e^{\frac{2\pi}{\tau_2} \hc \left|z \right|^2}\,
\frac{\sqrt{k}}{\sqrt{\tau_2}\left|\eta(\tau)\right|^2}\,
\sum_{m_1,m_2\in \bsz}\, \exp\left(-\frac{\pi k}{\tau_2}
\left|(m_1+s_1)\tau+(m_2+s_2)\right|^2\right).
\label{Z Y} 
\end{eqnarray}
%%%%%%%%%%%%%%%%%%%%%%%%%%%%%%%%%%%%%%%%%%%%%%%%%%%%%%%%%%%%%
%%%%%%%%%%%%%%%%%%%%%%%%%%%%%%%%%%%%%%%%%%%%%%%%%%%%%%%%%%%%%
%%%%%%%%%%%%%%%%%%%%%%%%%%%%%%%%%%%%%%%%%%%%%%%%%%%%%%%%%%%%%
Note that the linear couplings with the modulus parameters $z$, $\bar{z}$ with 
the currents $i \partial_w Y^u$, $i \partial_{\bar{w}} Y^u$  
cancel out in \eqn{SYuz}. This fact justifies the identification 
of parameters $z$, $\bar{z}$ with the marginal deformation 
by $e^{2\pi i \left(z J_0-\bar{z} \tJ_0\right)}$ mentioned before. 
%%%%%%%%%%%%%%%%%%%%%%%%%%%%%%%%%%%%%%%%%%%%%%%%%%%%%%%%%%%%%
%%%%%%%%%%%%%%%%%%%%%%%%%%%%%%%%%%%%%%%%%%%%%%%%%%%%%%%%%%%%%
%%%%%%%%%%%%%%%%%%%%%%%%%%%%%%%%%%%%%%%%%%%%%%%%%%%%%%%%%%%%%

\end{description}

~

%%%%%%%%%%%%%%%%%%%%%%%%%%%%%%%%%%%%%%%%%%%%%%%%%%%%%%%%%%%%%%%%%%%%%%%

Each sector \eqn{Z g}, 
\eqn{Z psi}, \eqn{Z gh}, and \eqn{Z Y} of the partition function is separately
invariant under modular transformation 
\begin{equation}
\tau \, \mapsto \, \frac{a\tau+b}{c\tau+d}, ~~~ 
u\, \mapsto \, \frac{u}{c\tau+d}, ~~~ z\, \mapsto \, \frac{z}{c\tau+d}, 
\hspace{1cm}
\forall 
\left(
\begin{array}{cc}
a & b \\
c & d
\end{array}
\right) \in SL(2,\bz).
\end{equation}

Combining all of them, 
we finally obtain the desired  partition function;
%%%%
\begin{eqnarray}
Z(\tau,z) &=& \int_{\Sigma} \frac{d^2 u}{\tau_2} \, 
Z_g \left(\tau, u+\frac{2}{k}z\right) \, Z_{\psi}\left(\tau, u+ \frac{k+2}{k}z\right) \,
Z_Y (\tau, u, z) \, Z_{\msc{gh}}(\tau)
\nn
&=& \cN \, e^{\frac{2\pi}{\tau_2} \left( \hc \left|z\right|^2 - \frac{k+4}{k} z_2^2\right)}\,
\sum_{m_1,m_2\in \bz}\, \int_{\Sigma} \frac{d^2 u}{\tau_2} \, 
\left|
\frac{\th_1\left(\tau,u+\frac{k+2}{k}z\right)}{\th_1\left(\tau, u+\frac{2}{k}z\right)}
\right|^2 \,
e^{-4\pi \frac{u_2z_2}{\tau_2}} \, e^{-\frac{\pi k}{\tau_2} \left|m_1\tau+m_2+ u\right|^2}
\nn 
&=& \cN \, e^{\frac{2\pi}{\tau_2} \left( \hc \left|z\right|^2 - \frac{k+4}{k} z_2^2\right)}\,
\int_{\bc} \frac{d^2 u}{\tau_2} \, 
\left|
\frac{\th_1\left(\tau,u+\frac{k+2}{k}z\right)}{\th_1\left(\tau, u+\frac{2}{k}z\right)}
\right|^2 \,
e^{-4\pi \frac{u_2z_2}{\tau_2}} \, e^{-\frac{\pi k}{\tau_2} \left|u\right|^2},
\label{part fn 1}
\end{eqnarray}
where $\cN$ is a normalization constant to be fixed later. 
In deriving the last line we made use of the identity; 
$
\th_1(\tau,z+m\tau+n) = (-1)^{m+n}\, q^{- \frac{m^2}{2}} e^{-2\pi i m z}\,
\th_1(\tau,z).
$

This result is a natural generalization of the formula given in \cite{ES-BH}. 
By construction, we expect that the obtained partition function \eqn{part fn 1}
is modular invariant. However, there exists a subtlety here 
since the $u$-integration 
diverges logarithmically. This is due to the $\sim \left|u+\frac{2}{k}z \right|^{-2}$ behavior 
around $u + \frac{2}{k}z =0$, $(\mod \bz\tau + \bz)$
of the integrand, and such an IR divergence is inevitable for non-compact models  
like $SL(2;\br)/U(1)$. 
It is thus important to introduce a suitable regularization not violating 
good modular behaviors, 
and  we will discuss this issue in the next subsection.

~

%%%%%%%%%%%%%%%%%%%%%%%%%%%%%%%%%%%%%%%%%%%%%%%%%%%%%%%%%%%%%%%%%%%%%%%%%%%%%%%%%
%%%%%%%%%%%%%%%%%%%%%%%%%%%%%%%%%%%%%%%%%%%%%%%%%%%%%%%%%%%%%%%%%%%%%%%%%%%%%%%%%

\subsection{Regularized Partition Function}

Our next task is to extract the content of ${\cal N}=2$ representations  
of the partition function \eqn{part fn 1}. 
Actually the partition function is infra-red divergent because 
 the theta function in the denominator acquires zeros under $u$-integration.
This is due to the non-compactness of the target manifold $SL(2,\br)/U(1)$. 
Thus we have to introduce a suitable cut-off which preserves the good modular behaviors  
of the theory. In the following we take $k=\frac{N}{K}$ ($N, K $ are positive co-prime integers) and consider models with the central charge $\hc(\equiv \frac{c}{3})= 1+ \frac{2K}{N}=1+{2\over k}$.

For convenience of analysis, we introduce minor modifications 
of the formula \eqn{part fn 1};
%%%
\begin{itemize}
%\item We rewrite the modulus parameter $\z$ simply by $z$ 
%from now on. 

\item We fix the normalization constant $\cN$ as $\cN=k$ so that
\begin{equation}
\lim_{z\, \rightarrow \,0} Z(\tau,z) = 1,
\end{equation}
which we find reasonable after performing the character decomposition. 

%\item We shall just drop off the modular invariant factor $e^{\frac{2\pi}{\tau_2} \hc |z|^2}$.

\item We redefine the integration variable $u\, \rightarrow \, -u$. 
\end{itemize}
%%%%%%%%%%%%%%%%%%%%%%%%%%%%%%%%%%%%%%%%%%%%%%%%%%%%%%%
Namely, we shall start with the formula;
% ($z\equiv z_1+iz_2$, $u\equiv u_1+iu_2$)
\begin{eqnarray}
%\hspace{-5mm}
%Z^{(\stR)}(\tau,\bar{\tau},z,\bar{z}) &=& 
Z(\tau,z) &=&
k e^{\frac{2\pi}{\tau_2}\left( \hc |z|^2 -\frac{k+4}{k} z_2^2\right)} \,
\int_{\bc} \frac{d^2u}{\tau_2}\, 
\left|\frac{\th_1\left(\tau, -u + \left(1+ \frac{2}{k}\right)z\right)}
{\th_1\left(\tau, -u + \frac{2}{k} z \right)} \right|^2 
\, e^{ 4\pi \frac{u_2 z_2}{\tau_2}}
 e^{-\frac{\pi k}{\tau_2}\left| u \right|^2},
\label{Z0 tR}
\end{eqnarray}
where we use the notations as $\Re\, z= z_1$, $\Im\, z=z_2$ etc. for the 
complex numbers $z, \, \tau, \,u$.

%%%%%%%%%%%%%%%%%%%%%%%%%%%%%%%%%%%%%%%%%%%%%%%%%%%%%%%%%%%%%%%%%%%%%%%%%%%%%%%

We first discuss a suitable regularization of \eqn{Z0 tR} which preserves 
good modular properties. 
By shifting the integration variable $u$ as $u\,\rightarrow \, u+\frac{2}{k}z$ 
and introducing winding numbers
$w$, $m$ by $u=(s_1+w)\tau+(s_2+m)$, we can rewrite \eqn{Z0 tR} as 
%%%
\begin{eqnarray}
\hspace{-5mm}
Z(\tau, z) &=& 
k e^{\frac{2\pi}{\tau_2} \left( \hc |z|^2 -z_2^2\right) }\,
\sum_{w,m\in \bsz}\, \int_0^1 ds_1  \int_0^1 ds_2\, 
\left| \frac{\th_1\left(\tau, -s_1\tau-s_2+ z \right)}
{\th_1\left(\tau, -s_1\tau-s_2 \right)} \right|^2 
\nn
&& \hspace{3cm}
\times  e^{4\pi s_1 z_2}
\, e^{-\frac{\pi k }{\tau_2} \left|\left(s_1+ w \right)\tau 
+ \left(s_2+ m \right) + \frac{2}{k}z\right|^2}.
\label{Z0' tR}
\end{eqnarray}
%%%%%%%%%%%%%%%%%%%%%%%%%%%%%%%%
%%%%%%%%%%%%%%%%%%%%%%%%%%%%%%%%
Now, we define the regularized partition function  
simply by replacing the integration region $D\equiv (0,1) \times (0,1)$ 
by $D(\ep):=(\ep,1-\ep) \times (0,1)$ where $\ep$ is a small positive constant;   
%%%%%%%%%%%%%%%%%%%%%%%%%%%%%%%%%
%%%%%%%%%%%%%%%%%%%%%%%%%%%%%%%%%
\begin{eqnarray}
\hspace{-5mm}
%Z^{(\stR)}_{\reg}(\tau,\bar{\tau}; z,\bar{z}; \ep) &=& 
Z_{\reg}(\tau, z; \ep) &=& 
k e^{\frac{2\pi}{\tau_2} \left( \hc |z|^2 -z_2^2\right) }\,
\sum_{w,m\in \bsz}\, 
%\int_{D(\ep)} ds_1 ds_2\, 
\int_{\ep}^{1-\ep}ds_1 \, \int_{0}^{1} ds_2\, 
\left| \frac{\th_1\left(\tau, -s_1\tau-s_2+ z \right)}
{\th_1\left(\tau, -s_1\tau-s_2 \right)} \right|^2 
\nn
&& \hspace{3cm}
\times  e^{4\pi s_1 z_2}
\, e^{-\frac{\pi k }{\tau_2} \left|\left(s_1+ w \right)\tau 
+ \left(s_2+ m \right) + \frac{2}{k}z\right|^2}.
\label{Zreg}
\end{eqnarray}
%%%%%%%%%%%%%%%%%%%%%%%%%%%%%%%%%%%%%%%%%%%%%%%%%%%%%%%%%%%%%%%%%%%%%%%%%%%%
We expect \eqn{Zreg} to be convergent since all poles 
in the integrand are removed, and to exhibit a 
logarithmic divergence in  $\ep\,\rightarrow \,0$ limit. 
%%%%%%%%%%%%%%%%%%%%%%%%%%%%%%%%%%%%%%%%%%%%%%%%%%%%%%%%%%%%%%%%%%%%%%%%%%%%%
%%%%%%%%%%%%%%%%%%%%%%%%%%%%%%%%%%%%%%%%%%%%%%%%%%%%%%%%%%%%%%%%%%%%%%%%%%%%%
The integrand of \eqn{Zreg} has a $\bz\times \bz$-periodicity with respect to $s_1$, $s_2$,
and behaves like a modular form under the change of integration variables $(s_1,s_2)$ as 
\begin{eqnarray}
&& T~:~ (s_1, s_2) ~\longrightarrow ~ (s_1, s_2+s_1) , \hspace{1cm}
S~:~ ~ (s_1,s_2) ~ \longrightarrow ~ (s_2 , -s_1).
\nonumber
\end{eqnarray}
This regularization does not strictly preserve the modular invariance, however, its violatation is under a good control;
\begin{equation}
Z_{\reg}(\tau+1, z; \ep)= Z_{\reg}(\tau, z; \ep), \hspace{1cm}
Z_{\reg}\left(-\frac{1}{\tau}, \frac{z}{\tau}; \ep \right)
- Z_{\reg}(\tau,z) = \cO(\ep \log \ep, \ep).
\end{equation}
The violation of $S$-invariance comes from the small change  
of the integration region $D(\ep)$ due to the $S$-transformation, 
while the $T$-transformation preserves it (up to the periodicity of $s_1$, $s_2$).

%%%%%%%%%%%%%%%%%%%%%%%%%%%%%%%%%%%%%%%%%%%%%%%%%%%%%%%%%%%%%%%%%%%%%%%%%%%%%%%%%%%%
%%%%%%%%%%%%%%%%%%%%%%%%%%%%%%%%%%%%%%%%%%%%%%%%%%%%%%%%%%%%%%%%%%%%%%%%%%%%%%%%%%%%

We shall now analyze  $Z_{\reg}(\tau,z;\ep)$ in detail. 
By dualizing the temporal winding number $m$ into the KK momentum $n$
by means of the Poisson resummation formula \eqn{PR formula}, 
we can rewrite  $Z_{\reg}(\tau,z;\ep)$ as follows;
\begin{eqnarray}
%\hspace{-5mm}
Z_{\reg}(\tau,z;\ep) &=& 
\sqrt{k\tau_2} 
%%%%%%%%%%%%%%%%%%%%%%%%%%%%%%%%%%%%%%%%%%%%%%%%%%%%%%%%%%%%
e^{\frac{2\pi}{\tau_2} \left( \hc |z|^2 -z_2^2\right) }\,
%%%%%%%%%%%%%%%%%%%%%%%%%%%%%%%%%%%%%%%%%%%%%%%%%%%%%%%%%%%%
\sum_{w,n \in \bsz}\, 
\int_{\ep}^{1-\ep}ds_1 \, \int_{0}^{1} ds_2\, 
\left| \frac{\th_1\left(\tau, -s_1\tau-s_2+ z\right)}
{\th_1\left(\tau, -s_1\tau-s_2 \right)} \right|^2 
\nn
&& \hspace{1.5cm}
\times  e^{4\pi s_1 z_2}
\, e^{-\pi \tau_2 \left\{ \frac{n^2}{k} + k \left(s_1+ w  
+ \frac{2z_2}{k\tau_2}\right)^2\right\}
+ 2\pi i n \left\{\left(s_1+w \right)\tau_1 + s_2 + \frac{2z_1}{k} \right\}}
\nn
&=& 
\sqrt{k\tau_2}
%%%%%%%%%%%%%%%%%%%%%%%%%%%%%%%%%%%%%%%%%%%%
e^{2\pi \frac{\hc}{\tau_2}z_1^2} \,
%%%%%%%%%%%%%%%%%%%%%%%%%%%%%%%%%%%%%%%%%%%
\sum_{w,n \in \bsz}\, 
\int_{\ep}^{1-\ep}ds_1 \, \int_{0}^{1} ds_2\, 
\left| \frac{\th_1\left(\tau, -s_1\tau-s_2+ z\right)}
{\th_1\left(\tau, -s_1\tau-s_2 \right)} \right|^2 
\nn
&& \hspace{1.5cm}
\times  
\, e^{-\pi \tau_2 \left\{ \frac{n^2}{k} + k \left(s_1+ w  \right)^2\right\}
+ 2\pi i n \left\{\left(s_1+w \right)\tau_1 + s_2 + \frac{2z_1}{k} \right\}
- 4\pi w z_2}
\nn
&=& 
\sqrt{k\tau_2} 
%%%%%%%%%%%%%%%%%%%%%%%%%%%%%%%%%%%%%%%%%%%%
e^{2\pi \frac{\hc}{\tau_2}z_1^2} \,
%%%%%%%%%%%%%%%%%%%%%%%%%%%%%%%%%%%%%%%%%%%
\sum_{w,n \in \bsz}\, 
\int_{\ep}^{1-\ep}ds_1 \, \int_{0}^{1} ds_2\, 
\left| \frac{\th_1\left(\tau, -s_1\tau-s_2+ z\right)}
{\th_1\left(\tau, -s_1\tau-s_2 \right)} \right|^2 
\nn
&& \hspace{1.5cm}
\times  
\, e^{-\pi \tau_2 \left\{ \frac{n^2}{k} + k \left(s_1+ w  \right)^2\right\}
+ 2\pi i n \left\{\left(s_1+w \right)\tau_1 + s_2 \right\}} \,
y^{w+\frac{n}{k}} \bar{y}^{w-\frac{n}{k}}.
\label{Zreg 2}
\end{eqnarray}
%%%%%%%%
%%%%%%%%
We next expand the ratio of  $\th_1$ functions by using 
the identity \eqn{th1/th1 formula 2};
%%%%%%%%
%%%%%%%%
\begin{eqnarray}
&& \left| \frac{\th_1\left(\tau, -s_1\tau-s_2+ z\right)}
{\th_1\left(\tau, -s_1\tau-s_2 \right)} \right|^2 = 
\left|\frac{-i \th_1(\tau,z)}{\eta(\tau)^3}\right|^2\,
\sum_{\ell, \tell \in \bz} \, \frac{(yq^{\ell})} 
{(1-yq^{\ell})}\cdot \left[{(y q^{\tell})} \over{(1- y q^{\tell})}\right]^*
\nn
&& \hspace{4cm}
\times 
e^{-2\pi i (s_1\tau_1+s_2) (\ell-\tell) + 2\pi s_1 \tau_2 (\ell+\tell)}.
\label{exp th1/th1}
\end{eqnarray}
After substituting \eqn{exp th1/th1} into \eqn{Zreg 2}, one can easily  
integrate $s_2$ out, which just yields the constraint 
\begin{equation}
n= \ell-\tell.
\label{cond n l}
\end{equation}
%%%%%%%%%%%%%%%%%%%%%%%%%%%%%%%%%%%%%%%%%%%%%
We next evaluate the $s_1$-integral. 
Picking up relevant terms, we obtain
\begin{eqnarray}
&& e^{-\pi \tau_2 \frac{N}{K} s_1^2 
-2\pi s_1 \left\{\tau_2 \frac{N}{K} w - i\tau_1 n 
+ i \tau_1 (\ell-\tell)-\tau_2(\ell+\tell) \right\}
} 
%\nn
%&& ~~~ 
= e^{-\pi \tau_2 \frac{N}{K} s_1^2 - 2\pi s_1 \tau_2 \frac{v}{K}},
\end{eqnarray}
where we set 
\begin{equation}
v:= Nw- K(\ell+\tell),
\label{v def}
\end{equation}
and used the condition \eqn{cond n l}. 
With the help of a simple Gaussian integral, we  further obtain
\begin{eqnarray}
\int_{\ep}^{1-\ep} ds_1 
\, e^{-\pi \tau_2 \frac{N}{K} s_1^2 -2\pi s_1\tau_2 \frac{v}{K}}
&=& \sqrt{\frac{\tau_2}{NK}} \,  \int_{\ep}^{1-\ep} ds_1\, \int_{\br-i0} dp\,
e^{-\frac{\pi}{NK}\tau_2 p^2 - 2\pi i \tau_2 \frac{s_1}{K}(p-iv)}
\nonumber \\
&=& \sqrt{\frac{K}{N\tau_2}} \frac{1}{2\pi i} \,
\int_{\br-i0} dp\, 
\frac{e^{-\frac{\pi}{NK} \tau_2 p^2  }} {p-iv}
\, \left\{ 
e^{-\vep(v+ip)}
- e^{\vep(v+ip)}
e^{-2\pi i \tau_2 \frac{1}{K} (p-iv)}
\right\},
\nn
&&
\label{s1 int} 
\end{eqnarray}
where we set $\vep \equiv 2\pi \frac{\tau_2}{K} \ep \, (>0)$.
Using
\begin{equation}
2K\ell+v = K (kw+n) \left(\equiv Nw+ Kn \right), ~~~
2K\tell+v = K (kw-n) \left(\equiv Nw- Kn \right).
\end{equation}
we obtain 
\begin{eqnarray}
&&e^{-\pi \tau_2 \left(\frac{n^2}{k} + k w^2\right) + 2\pi i \tau_1 nw }
= q^{\frac{(n+kw)^2}{4k}} \bar{q}^{\frac{(n-kw)^2}{4k}}
= q^{\frac{(2K\ell+v)^2}{4NK}} \bar{q}^{\frac{(2K\tell+v)^2}{4NK}}\nn
&&\hskip3cm = q^{\frac{1}{N}\left(K\ell^2 + \ell v\right)} 
\bar{q}^{\frac{1}{N}\left(K\tell^2 + \tell v\right)} e^{- \pi \tau_2 \frac{v^2}{NK}},
\end{eqnarray}
We also note that  
\begin{equation}
y^{w+\frac{n}{k}} \bar{y}^{w-\frac{n}{k}} 
= y^{\frac{Nw+Kn}{N}}\bar{y}^{\frac{Nw-Kn}{N}}
= y^{\frac{2K}{N}\left(\ell + \frac{v}{2K}\right)}
\bar{y}^{\frac{2K}{N}\left(\tell + \frac{v}{2K}\right)}.
\end{equation}

\nopagebreak

%%%%%%%%%%%%%%%%%%%%%%%%%%%%%%%%%%%%%%%%%%%%%%%%%%%%%%%%%%%%%%%%%%%%

Combining all the pieces, we obtain the following expression of $Z_{\reg}(\tau,z;\ep)$;
\begin{eqnarray}
&& \hspace{-1.5cm}
Z_{\reg}(\tau,z;\ep) 
= 
%e^{-2\pi \frac{\hc}{\tau_2}z_2^2} \,
%%%%%%%%%%%%%%%%%%%%%%%%%%%%%%%%%%%%%%%
e^{2\pi \frac{\hc}{\tau_2}z_1^2} \,
%%%%%%%%%%%%%%%%%%%%%%%%%%%%%%%%%%%%%%%
\left|\frac{\th_1(\tau,z)}{\eta(\tau)^3}\right|^2 \,
\sum_{\stackrel{v,\ell,\tell\in \bz}{v+K(\ell+\tell) \in N\bz}}\,
\frac{1}{2\pi i} \,
\int_{\br-i0} dp\, 
\frac{e^{-\pi \tau_2 \frac{p^2+v^2}{NK}}} {p-iv}
\, \left\{ 
e^{-\vep (v+ip)} - e^{\vep (v+ip)}\,e^{-2\pi i \tau_2 \frac{1}{K} (p-iv)}
\right\}
\nn
&& \hspace{5cm} \times \,
\frac{(yq^{\ell})^{1+\frac{v}{N}}}{1-yq^{\ell}} \,
\left[
\frac{(yq^{\tell})^{1+\frac{v}{N}}}{1-yq^{\tell}}
\right]^*\,
y^{\frac{2K}{N}\ell} q^{\frac{K}{N}\ell^2} \, 
\left[y^{\frac{2K}{N}\tell} q^{\frac{K}{N}\tell^2}\right]^*
\nn
%%%
&&= 
%e^{-2\pi \frac{\hc}{\tau_2}z_2^2} \,
%%%%%%%%%%%%%%%%%%%%%%%%%%%%%%%%%%%%%%%
e^{2\pi \frac{\hc}{\tau_2}z_1^2} \,
%%%%%%%%%%%%%%%%%%%%%%%%%%%%%%%%%%%%%%%
\left|\frac{\th_1(\tau,z)}{\eta(\tau)^3}\right|^2 \,
\sum_{\stackrel{v,\ell,\tell\in \bz}{v+K(\ell+\tell) \in N\bz}}\,
\frac{1}{2\pi i} \, \left[
\int_{\br-i0} dp \, e^{-\vep (v+ip)}\,
yq^{\ell} \, \left[ yq^{\tell} \right]^* 
- \int_{\br+i(N-0)} dp \, e^{\vep(v+ip)}
\right]
\nn
&& \hspace{4cm}
\times \,
\frac{e^{-\pi \tau_2 \frac{p^2+v^2}{NK}}} {p-iv}
\, \frac{(yq^{\ell})^{\frac{v}{N}}}{1-yq^{\ell}} \,
\left[
\frac{(yq^{\tell})^{\frac{v}{N}}}{1-yq^{\tell}}
\right]^*\,
y^{\frac{2K}{N}\ell} q^{\frac{K}{N}\ell^2} \, 
\left[y^{\frac{2K}{N}\tell} q^{\frac{K}{N}\tell^2}\right]^*\label{hZreg-a}
\end{eqnarray}
%\nn
Here we have absorbed  the factor
$$
yq^{\ell} \cdot \left[y q^{\tell}\right]^* \cdot e^{-2\pi i \tau_2 \frac{1}{K}(p-iv)}
$$
by 
%%%%%%%%%%%%%%%%%%%%%%%%%%%%%%%%%%%%%%%%%%%%%%%%%%%%%%%
%the shift of contour  
%$$\br-i0 ~ \rightarrow ~ \br+ i(N-0), $$
%and 
%%%%%%%%%%%%%%%%%%%%%%%%%%%%%%%%%%%%%%%%%%%%%%%%%%%%%%%
the change of 
variables $p=: p'-iN$, $v=:  v'- N$ with
\begin{eqnarray*}
&& \frac{p^2+v^2}{NK} = \frac{p^{'2}+v^{'2}}{NK} - \frac{2i}{K}(p'-iv'), \hspace{1cm}
p-iv = p'-iv', 
\\
&& \left(yq^{\ell}\right)^{\frac{v}{N}} \, 
\left[\left( y q^{\tell}\right)^{\frac{v}{N}}\right]^* 
=  \left(yq^{\ell}\right)^{\frac{v'}{N}-1} \,
\left[ \left( y q^{\tell}\right)^{\frac{v'}{N}-1}\right]^* .
\end{eqnarray*}
(\ref{hZreg-a})  is further rewritten as 
\begin{eqnarray}
&& \hspace{-1.5cm}
Z_{\reg}(\tau,z;\ep) 
= e^{2\pi \frac{\hc}{\tau_2}z_1^2} \,
\left|\frac{\th_1(\tau,z)}{\eta(\tau)^3}\right|^2 \,
\sum_{\stackrel{v,\ell,\tell\in \bz}{v+K(\ell+\tell) \in N\bz}}\,
\frac{1}{2\pi i} \, \left[
\int_{\br-i0} dp \, e^{-\vep (v+ip)}\,
yq^{\ell} \, \left[yq^{\tell}\right]^* 
- \int_{\br+i(N-0)} dp \, e^{\vep(v+ip)}
\right]
\nn
&& \hspace{1cm}
\times \,
\frac{e^{-\pi \tau_2 \frac{p^2}{NK}}} {p-iv}
\, \frac{1}{1-yq^{\ell}} \,
\left[
\frac{1}{1-yq^{\tell}}
\right]^*\,
y^{\frac{2K}{N}\left(\ell+\frac{v}{2K}\right)} 
q^{\frac{K}{N}\left(\ell+ \frac{v}{2K} \right)^2} \, 
\left[y^{\frac{2K}{N}\left(\tell+ \frac{v}{2K}\right)} 
q^{\frac{K}{N}\left(\tell+ \frac{v}{2K}\right)^2}\right]^*.
\label{hZreg}
\end{eqnarray}
%%%%%%%%%%%%%%%%%%%%%%%%%%%%%%%%%%%%%%%%%%%%%%%%%%%%%%%%%%%%%%%%%%%%%
We note that the power series expansion in $v,\ell,\tell$  \eqn{hZreg} converges as is expected. 
In fact, as is obvious from the last line of \eqn{hZreg}, a potential 
divergence could happen when taking the limit $\ell,\tell\, \rightarrow \, \pm \infty$, 
$v \, \rightarrow \, \mp \infty$ with keeping $\ell+\frac{v}{2K}$, $\tell+\frac{v}{2K}$
finite. However, it is easy to confirm that the factors 
$e^{-\vep(v+ip)}q^{\ell}/(1-yq^{\ell})$, $e^{\vep(v+ip)}/(1-yq^{\tell})$ suitably produce damping effects to make 
the power series convergent in both sides of $v\, \rightarrow \, \pm \infty$.

~

%%%%%%%%%%%%%%%%%%%%%%%%%%%%%%%%%%%%%%%%%%%%%%%%%%%%%%%%%%%%%%%%%%%%%%%%%%%
%%%%%%%%%%%%%%%%%%%%%%%%%%%%%%%%%%%%%%%%%%%%%%%%%%%%%%%%%%%%%%%%%%%%%%%%%%%

\section{Character Decomposition  of Partition Function}

Now, we would like to discuss the decomposition of partition function \eqn{hZreg} in terms of ${\cal N}=2$ characters, which will be the main part of our analysis in this paper.

First of all, by shifting the integration contour in the second term as 
$
\br +i(N-0) \, \rightarrow \, \br-i0 ,
$
we can decompose \eqn{hZreg} into the pole contributions 
$Z_{\dis}(\ep)$ and the `remainder part' $Z_{\rem}(\ep)$ 
as follows;
\begin{eqnarray}
&& Z_{\reg}(\tau,z;\ep) 
= Z_{\dis}(\tau,z;\ep) + 
Z_{\rem}(\tau,z;\ep),
\nn
%%%
&& Z_{\dis}(\tau,z;\ep)
%%%%%%%%%%%%%%%%%%%%%%%%%%%%%%%%%%%%%%%%%%%%%
\equiv e^{2\pi \frac{\hc}{\tau_2}z_1^2} \,
%%%%%%%%%%%%%%%%%%%%%%%%%%%%%%%%%%%%%%%%%%%%%
\left|\frac{\th_1(\tau,z)}{\eta(\tau)^3}\right|^2 \,
\sum_{v=0}^{N-1}\, 
\sum_{\stackrel{\ell,\tell\in \bz}{v+K(\ell+\tell) \in N\bz}}\,
\frac{(yq^{\ell})^{\frac{v}{N}}}{1-yq^{\ell}} \,
\left[
\frac{(yq^{\tell})^{\frac{v}{N}}}{1-yq^{\tell}}\right]^*
\,
y^{\frac{2K}{N}\ell} q^{\frac{K}{N}\ell^2} \, 
\left[y^{\frac{2K}{N}\tell} q^{\frac{K}{N}\tell^2}\right]^*,
\nn
&& 
\label{Z dis 1}\\
%%%
&& 
Z_{\rem}(\tau,z;\ep) \equiv  
%%%%%%%%%%%%%%%%%%%%%%%%%%%%%%%%%%%%%%%
e^{2\pi \frac{\hc}{\tau_2}z_1^2} \,
%%%%%%%%%%%%%%%%%%%%%%%%%%%%%%%%%%%%%%%
\left|\frac{\th_1(\tau,z)}{\eta(\tau)^3}\right|^2 \,
\sum_{\stackrel{v,\ell,\tell\in \bz}{v+K(\ell+\tell) \in N\bz}}\,
\frac{1}{2\pi i} \,
\int_{\br-i0} dp \, \frac{e^{-\pi \tau_2 \frac{p^2+v^2}{NK}}} {p-iv}
\nn
&& \hspace{2cm}
\times \left\{ yq^{\ell} \cdot \left[ yq^{\tell} \right]^* e^{-\vep(v+ip)}-
e^{\vep (v+ip)} \right\}
\frac{(yq^{\ell})^{\frac{v}{N}}}{1-yq^{\ell}} \,
\left[
\frac{(yq^{\tell})^{\frac{v}{N}}}{1-yq^{\tell}}
\right]\,
y^{\frac{2K}{N}\ell} q^{\frac{K}{N}\ell^2} \, 
\left[y^{\frac{2K}{N}\tell} q^{\frac{K}{N}\tell^2}\right]^*
\nn
\label{Z rem 1}
\end{eqnarray}
Note that the $\ep$-dependence disappears in the discrete part 
$Z_{\dis}$, because the pole occurs at  
$p=iv$, $v=0,1,\ldots, N-1$.

We next elaborate on each of these contributions.

~

%%%%%%%%%%%%%%%%%%%%%%%%%%%%%%%%%%%%%%%%%%%%%%%%%%%%%%%%%%%%%%%%%%%%%%%%

\subsection{Discrete Part}

The discrete part \eqn{Z dis 1} is rewritten in terms of the (extended) discrete characters \cite{ES-L,ES-BH,ES-C}. To make notations simple, we here adopt a slightly different notation for extended characters 
(see Appendix C);
%$(a\in \bz_N, ~ 0\leq v \leq N-1)$;
%%%%%%%%%%%%%%%%%%%%%%%%%%%%%%%%%%%%%%%%%%%%%%%%%%%%
% discrete extended character
%%%%%%%%%%%%%%%%%%%%%%%%%%%%%%%%%%%%%%%%%%%%%%%%%%%%
\begin{eqnarray}
\chid (v,a;\tau,z) \equiv \frac{i\th_1(\tau,z)}{\eta(\tau)^3}\,
\sum_{n\in\bz} \,\frac{\left(yq^{N n+a}\right)^{\frac{v}{N}}}
{1-y q^{Nn+a}} \, y^{2K \left(n+\frac{a}{N}\right)} 
q^{NK \left(n+\frac{a}{N}\right)^2}. 
%%~~~ (a\in \bz_N, ~ 0\leq v \leq N-1),
\label{chi d}
\end{eqnarray}
%%%%%%%%%%%%%%%%%%%%%%%%%%%%%%%%%%%%%%%%%%%%%%%
%%%%%%%%%%%%%%%%%%%%%%%%%%%%%%%%%%%%%%%%%%%%%%%
This is the sum over spectral flows of 
the discrete (BPS) representation generated by 
Ramond vacuum with the $U(1)$-charge $ Q =\frac{v}{N}-\frac{1}{2}$,
whose flow momenta are taken to be 
$n\in a + N\bz$, ($a\in \bz_N$).
%%%%%%%%%%%%%%%%%%%%%%%%%%%%%%%%%%%%%%%%%%%%%%
%%%%%%%%%%%%%%%%%%%%%%%%%%%%%%%%%%%%%%%%%%%%%%
Setting $\ell= N n_L + a_L$, $\tell = N n_R + a_R$
($n_L,n_R \in \bz$, $a_L, a_R \in \bz_N$) in \eqn{Z dis 1}, 
we obtain 
\begin{equation}
Z_{\dis}(\tau,z; \ep)= 
%%%%%%%%%%%%%%%%%%%%%%%%%%%%%%%%%%%%%%%
e^{2\pi \frac{\hc}{\tau_2}z_1^2} \, 
%%%%%%%%%%%%%%%%%%%%%%%%%%%%%%%%%%%%%%%
\sum_{v=0}^{N-1} \, 
\sum_{\stackrel{a_L,a_R\in \bz_N}{v+K(a_L+a_R) \in N\bz}}\, 
\chid(v,a_L;\tau,z) \chid(v,a_R;\tau,z)^*.
\label{Z dis 2}
\end{equation}
This gives essentially the same result as in \cite{ES-BH}.

%%%%%%%%%%%%%%%%%%%%%%%%%%%%%%%%%%%%%%%%%%%%%%%%%%%%%%%%%%%%%%%%%%%%%%%%%

~

\subsection{Remainder Part}

To evaluate the remainder part $Z_{\rem}(\tau,z;\ep)$, 
it is convenient to decompose \eqn{Z rem 1} into three pieces;
\begin{eqnarray}
%\hspace{-1cm}
Z_{\rem}(\tau,z;\ep)
&=& - 
%%%%%%%%%%%%%%%%%%%%%%%%%%%%%%%%%%%%%%
e^{2\pi \frac{\hc}{\tau_2}z_1^2} \,
%%%%%%%%%%%%%%%%%%%%%%%%%%%%%%%%%%%%%%
\left|\frac{\th_1(\tau,z)}{\eta(\tau)^3}\right|^2 \,
\sum_{\stackrel{v,\ell,\tell\in \bz}{v+K(\ell+\tell) \in N\bz}}\,
\frac{1}{2\pi i} \,
\int_{\br-i0} dp \,\,
\frac{e^{-\pi \tau_2 \frac{p^2+v^2}{NK}}} {p-iv}
\nn
&&\hspace{2cm} 
\times\,
y^{\frac{1}{N}\left(2K\ell+v\right)} q^{\frac{1}{N}\left(K \ell^2+ \ell v\right)} \, 
\left[
y^{\frac{1}{N}\left(2K \tell+v\right)} q^{\frac{1}{N}\left(K \tell^2 + \tell v \right)}
\right]^*
\nn
&& \hspace{2cm}
\times 
\left[
\frac{yq^{\ell} e^{-\vep(v+ip)}}{1-yq^{\ell}} 
+  \frac{e^{\vep(v+ip)}}{1-\bar{y}\bar{q}^{\tell}} 
+ \frac{yq^{\ell} \left\{e^{\vep(v+ip)} - e^{-\vep(v+ip)}\right\}}
{(1-yq^{\ell}) (1- \bar{y} \bar{q}^{\tell})}
\right] 
\nn
& = : & Z_{(1)} + Z_{(2)} + Z_{(3)}.
\label{Zrem decomp} 
\end{eqnarray}
Here $Z_{(1)}$, $Z_{(2)}$, $Z_{(3)}$ correspond to the choice of three terms within the square bracket $[~~~]$.  
We here made use of a formula ;
$$
\frac{X Y \al^{-1}-\al }{(1-X)(1-Y)} = 
- \left[ \frac{X \al^{-1}}{1-X} +\frac{\al}{1-Y} 
+ \frac{X(\al-\al^{-1}) }{(1-X)(1-Y)}
\right].
$$

~

%%%%%%%%%%%%%%%%%%%%%%%%%%%%%%%%%%%%%%%%%%%%%%%%%%%%%%%%%%

\noindent
{\bf (1) ~ $Z_{(1)}$ :}\\

Let us first consider the contribution $Z_{(1)}$. 
To carry out the $v$-summation,  
we decompose it as $v= v_0+ N r$, $(r\in \bz, ~ v_0=0,1,\ldots, N-1)$,
and make use of the identities;
\begin{eqnarray}
\frac{(yq^{\ell})^{\frac{v_0}{N} +r+1}}{1-yq^{\ell}}
&=& \frac{(yq^{\ell})^{\frac{v_0}{N}}}{1-yq^{\ell}} - \sum_{j=0}^{r} \, 
(yq^{\ell})^{\frac{v_0}{N}+j}, 
\hspace{1cm} (r\geq 0),\label{ 1}
\\
%%%
\frac{(yq^{\ell})^{\frac{v_0}{N} +r+1}}{1-yq^{\ell}}
&=& \frac{(yq^{\ell})^{\frac{v_0}{N}}}{1-yq^{\ell}} + \sum_{j=-1}^{r+1} \, 
(yq^{\ell})^{\frac{v_0}{N}+j}, 
\hspace{1cm} (r\leq -2).\nonumber
\end{eqnarray}
%%%

Corresponding to various terms in the above expansion, we can further decompose $Z_{(1)}$
as follows;
%%%
\begin{description}
\item[(i)] 
We consider the contribution to $Z_{(1)}$ which is multiplied by the term $\dsp \frac{(yq^{\ell})^{\frac{v_0}{N}}}{1-yq^{\ell}}$ in the expansion (\ref{ 1})
and denote it as 
$Z_{\msc{(1),(i)}}$.

%%%
\item[(ii)]
We consider  the contribution to $Z_{(1)}$ which is multiplied by the term $- (yq^{\ell})^{\frac{v_0}{N}+r}$ in (\ref{ 1}). This contribution 
 is considered to be dominant 
as compared with those multiplied by the terms $(yq^{\ell})^{\frac{v_0}{N}+j}$ 
and will be denoted as $Z_{\msc{(1), (ii)}}$.
Note that the `saturation' $|j|=|r|$ is possible only for $r\ge 0$.
This fact is important in the following analysis. 

%%%
\item[(iii)]
We collect all the remaining terms, 
that is, $(yq^{\ell})^{\frac{v_0}{N}+j}$ with $|j|< |r|$
 and denote it as  $Z_{\msc{(1),(iii)}}$.

\end{description}

We first evaluate $Z_{\msc{(1),(i)}}$; 
\begin{eqnarray}
\hspace{-1cm}
Z_{\msc{(1),(i)}} &=& 
%%%%%%%%%%%%%%%%%%%%%%%%%%%%%%%%%%%%%%%%
- e^{2\pi \frac{\hc}{\tau_2}z_1^2} \,
%%%%%%%%%%%%%%%%%%%%%%%%%%%%%%%%%%%%%%%%
\left|\frac{\th_1(\tau,z)}{\eta(\tau)^3}\right|^2 \,
\sum_{v_0=0}^{N-1}\, \sum_{r\in \bz}\, \sum_{\stackrel{a_L,a_R\in \bz_N}{v_0+K(a_L+a_R) \in N\bz}}
\, \sum_{n_L,n_R\in \bz}\,
 \frac{1}{2\pi i} \,
\int_{\br-i0} dp \, \frac{e^{-\pi \tau_2 \frac{p^2 + (v_0+Nr)^2}{NK}}} {p-i(v_0+Nr)} \,
\nn
&&
\hspace{2cm} \times \,\,  
\frac{(yq^{N n_L+a_L})^{\frac{v_0}{N}}}{1-yq^{Nn_L+a_L}}\,
y^{2K\left(n_L+\frac{a_L}{N}\right)}\,
q^{NK\left(n_L+\frac{a_L}{N} \right)^2} 
\nn
&& \hspace{2cm} \times \,
\left[
y^{2K\left(n_R + \frac{v_0+Nr+ 2K a_R}{2NK} \right)}\,
q^{NK \left(n_R + \frac{v_0+Nr+ 2K a_R}{2NK} \right)^2}
\,q^{- \frac{(v_0+Nr)^2}{4NK}} \right]^* \times e^{-\vep (v_0+Nr+ip)}
\nn
%%%
&=& 
%%%%%%%%%%%%%%%%%%%%%%%%%%%%%%%%%%%%%%%
- e^{2\pi \frac{\hc}{\tau_2}z_1^2} \,
%%%%%%%%%%%%%%%%%%%%%%%%%%%%%%%%%%%%%%%
\sum_{v_0=0}^{N-1}\, \sum_{\stackrel{a_L,a_R\in \bz_N}{v_0+K(a_L+a_R) \in N\bz}}
\, \chid (v_0, a_L;\tau,z) \nn
&& \hspace{2cm}
\times \left[ \frac{i \th_1(\tau,z)}{\eta(\tau)^3}
\cdot \frac{1}{2} \sum_{j\in \bz_{2K}}\, R^{(+)}_{v_0+Nj, NK} \,
\Th{v_0+Nj+2Ka_R}{NK}\left(\tau, \frac{2z}{N}\right)\right]^*  + \cO(\vep).
\label{Z 1 i}
\end{eqnarray}
In the last line  we introduced the function
 $R^{(+)}_{m,k}$ \eqn{Rmk}, 
given explicitly as 
\begin{eqnarray}
R^{(+)}_{m,NK}&\equiv& \frac{1}{i\pi} \, \sum_{\la \in m + 2NK \bz}\,  
\int_{\br-i0} dp \, \frac{e^{-\pi \tau_2 \frac{p^2 + \la^2}{NK}}} {p-i\la} \, 
q^{-\frac{\la^2}{4NK}}
\nn
&\equiv & 
\sum_{\la \in m+2NK\bz}\, \sgn(\la + 0) 
\erfc\left(\sqrt{\frac{\pi \tau_2}{NK}} \left|\la\right|\right)\, 
q^{- \frac{\la^2}{4NK}},
\end{eqnarray}
where $\erfc (x)$ is the error-function defined in \eqn{erfc}.
The  power series in \eqn{Z 1 i} converges even  at $\vep=0$, 
and one can simply take the limit $\vep\, \rightarrow \, 0$.
%%%%%%%%%%%%%%%%%%%%%%%%%%%%%%%%%%%%%%%%%%%%%%%%%%%%%%%%%%%%%%%%%%%%%%%%%%%%%
The emergence of non-holomorphic function $R^{(+)}_{*,*}$ \eqn{Rmk}
is crucial in our analysis. 
%%%%%%%%%%%%%%%%%%%%%%%%%%%%%%%%%%%%%%%%%%%%%%%%%%%%%%%%%%%%%%%%%%%%%%%%%%%%%%%

On the other hand, $Z_{\msc{(1),(ii)}}$ is evaluated as 
\begin{eqnarray}
\hspace{-1cm}
Z_{\msc{(1),(ii)}} &=& 
%%%%%%%%%%%%%%%%%%%%%%%%%%%%%%%%%%%%%%%
- e^{2\pi \frac{\hc}{\tau_2}z_1^2} \,
%%%%%%%%%%%%%%%%%%%%%%%%%%%%%%%%%%%%%%%
\left|\frac{\th_1(\tau,z)}{\eta(\tau)^3}\right|^2 \,
\sum_{v_0=0}^{N-1}\, \sum_{r=0}^{\infty}\, 
\sum_{\stackrel{a_L,a_R\in \bz_N}{v_0+K(a_L+a_R) \in N\bz}}
\, \sum_{n_L,n_R\in \bz}\,
 \frac{1}{2\pi i} \,
\int_{\br-i0} dp \, \frac{e^{-\pi \tau_2 \frac{p^2 + (v_0+Nr)^2}{NK}}} {p-i(v_0+Nr)} \,
\nn
&&
\hspace{2cm} \times \,  
(-1) (yq^{N n_L + a_L})^{\frac{v_0+Nr}{N}}\, 
y^{2K\left(n_L+\frac{a_L}{N}\right)}\,
q^{NK\left(n_L+\frac{a_L}{N} \right)^2} 
\nn
&& \hspace{2cm} \times \,
\left[y^{2K\left(n_R + \frac{v_0+Nr+ 2K a_R}{2NK} \right)}\,
q^{NK \left(n_R + \frac{v_0+Nr+ 2K a_R}{2NK} \right)^2}
\,q^{- \frac{(v_0+Nr)^2}{4NK}} \right]^* \, e^{-\vep (v_0+Nr+ip)}
\nn
%%%
&=& 
%%%%%%%%%%%%%%%%%%%%%%%%%%%%%%%%%%%%%%%
e^{2\pi \frac{\hc}{\tau_2}z_1^2} \,
%%%%%%%%%%%%%%%%%%%%%%%%%%%%%%%%%%%%%%%
\left|\frac{\th_1(\tau,z)}{\eta(\tau)^3}\right|^2 \,
\sum_{v_0=0}^{N-1}\, \sum_{r=0}^{\infty}\, 
\sum_{\stackrel{a_L,a_R\in \bz_N}{v_0+K(a_L+a_R) \in N\bz}}
\, \sum_{n_L,n_R\in \bz}\,
 \frac{1}{2\pi i} \,
\int_{\br-i0} dp \, \frac{e^{-\pi \tau_2 \frac{p^2 }{NK}}} {p-i(v_0+Nr)} \,
\nn
&&
\hspace{3cm} 
\times \,  \,
y^{2K\left(n_L + \frac{v_0+Nr+ 2K a_L}{2NK} \right)}\,
q^{NK \left(n_L + \frac{v_0+Nr+ 2K a_L}{2NK} \right)^2}
\nn
&& \hspace{3cm} \times 
\,
\left[ y^{2K\left(n_R + \frac{v_0+Nr+ 2K a_R}{2NK} \right)}\,
q^{NK \left(n_R + \frac{v_0+Nr+ 2K a_R}{2NK} \right)^2} \right]^*
\, e^{-\vep (v_0+Nr+ip)}.
\label{Z 1 ii 1}
\end{eqnarray}
The power series is convergent due to the damping factor 
$e^{-\vep (v_0+Nr+ip)}$ and an expected logarithmic divergence 
emerges in the limit $\vep\, \rightarrow \, 0$.

We can rewrite \eqn{Z 1 ii 1} in a simpler form
by using the (extended) continuous characters \eqn{chic}.
Introducing new quantum numbers $n_0\in \bz_N$, $w_0\in \bz_{2K}$
by the relation
$$
v_0+Nr+2Ka_L \equiv N w_0+ Kn_0  ~ (\mod  2NK), \hspace{1cm} 
v_0+Nr+2Ka_R\equiv N w_0- Kn_0   ~ (\mod  2NK), 
%\hspace{1cm}
%(w_0\in \bz_{2K}, ~ n_0\in \bz_{N}),
$$
which solves the constraint $v_0+ K(a_L+a_R) \in N \bz$, 
we obtain
\begin{eqnarray}
%\hspace{-1cm}
Z_{\msc{(1),(ii)}} &=&  
%%%%%%%%%%%%%%%%%%%%%%%%%%%%%%%%%%%%%%%%%%%
e^{2\pi \frac{\hc}{\tau_2}z_1^2} \,
%%%%%%%%%%%%%%%%%%%%%%%%%%%%%%%%%%%%%%%%%%%
\sum_{n_0\in \bz_{N}, \, w_0 \in \bz_{2K}}\,
 \int_{\br-i0} dp\, \rho_{(1)} (p,n_0,w_0 ;\vep) \, \nn
&& \hspace{2cm} \times  
\chic(p,Nw_0+Kn_0;\tau,z) \chic(p,Nw_0-Kn_0;\tau,z)^*, 
\label{Z 1 ii 2}
\end{eqnarray}
where $\rho_{(1)}$ denotes some spectral density. 
To evaluate it, it is convenient to introduce the symbol `$[m]_{2K}$' defined by 
$$
[m]_{2K} \equiv m ~(\mod 2K), ~~~ 0\leq [m]_{2K} \leq 2K-1,
$$ 
and set $m_L := N w_0 + K n_0$, $m_R := N w_0 - Kn_0$. 
Note that 
\begin{equation}
v_0+ Nr  = [m_L]_{2K} + 2Ks = [m_R]_{2K} + 2K \tilde{s}, ~~~ 
({}^{\exists} \, s, \tilde{s} \in \bz_{\geq 0}).
\end{equation}
%%%
With the helps of \eqn{reg formula},
we obtain\footnote
   {We shall here adopt the `left-right symmetric form' of spectral density just for 
    convention.}
\begin{eqnarray}
\rho_{(1)}(p,n_0,w_0 ; \vep) &=& 
\frac{1}{4 \pi i} \left[
\sum_{s=0}^{\infty} \frac{e^{-\vep ([m_L]_{2K} +2Ks + ip)}}{p- i\left([m_L]_{2K}+ 2Ks\right)}
+ \sum_{\tilde{s}=0}^{\infty} \frac{e^{-\vep ([m_R]_{2K} +2K\tilde{s} + ip)}}
{p- i\left([m_R]_{2K}+ 2K\tilde{s}\right)}
\right] \nn
%%%
&=& C(\vep) - \frac{1}{4\pi i } \frac{\partial}{\partial p} 
\log \left[
\Gamma\left(\frac{[m_L]_{2K}}{2K} + \frac{ip}{2K}\right)
\Gamma\left(\frac{[m_R]_{2K}}{2K} + \frac{ip}{2K}\right)
\right] + \cO(\vep),
\nn
&&
\label{rho 1}
\end{eqnarray}
where $C(\vep)$ denotes some positive  constant independent of $n_0$, $w_0$, 
which logarithmically diverges in the $\vep\, \rightarrow \, 0$ limit.

~

%%%%%%%%%%%%%%%%%%%%%%%%%%%%%%%%%%%%%%%%%%%%%%%%%%%%%%%%%%%%%%%%%%%%%%%%%%%%%%%%%%%

Finally, $Z_{\msc{(1),(iii)}}$ has a complicated form; 
\begin{eqnarray}
\hspace{-5mm}
 Z_{\msc{(1),(iii)}} &=& 
%%%%%%%%%%%%%%%%%%%%%%%%%%%%%%%%%%%%%%
e^{2\pi \frac{\hc}{\tau_2}z_1^2}\,
%%%%%%%%%%%%%%%%%%%%%%%%%%%%%%%%%%%%%%
  \sum_{v_0=0}^{N-1}\, 
\sum_{\stackrel{a_L,a_R\in \bz_N}{v_0+K(a_L+a_R) \in N\bz}}\, 
\left[
\sum_{r=1}^{\infty}\,\sum_{j=0}^{r-1}\, - \sum_{r=-2}^{-\infty}\,\sum_{j=-1}^{r+1}\, 
\right]\,
\frac{1}{2\pi i} \int_{\br-i0}dp\, \frac{1}{p-i(v_0+Nr)}\,
\nn
&& \hspace{1cm}
\times \chic \left(p(r,j), v_0+Nj+2K a_L; \tau,z\right)
\, \chic \left(p, v_0+Nj+2K a_R; \tau,z\right)^*,
\label{Z 1 iii}
\end{eqnarray}
where we set 
$$
p(r,j)^2 : = p^2 + (v_0+Nr)^2- (v_0+Nj)^2.
$$
Note that we always have $p(r,j)^2> p^2$ for $0\leq j \leq r-1$ ($r\geq
1$) or $-1 \geq j \geq r+1$ ($r\leq -2$).

A few remarks are in order;
%%%
\begin{itemize}
\item In the IR region $\tau_2 \sim +\infty$, $Z_{\msc{(1),(iii)}}$ is negligible in comparison with 
the ``dominant" continuous part $Z_{\msc{(1),(ii)}}$ for each fixed values of left and right $U(1)$ charge. In fact, by construction 
$p(r,j)^2>p^2$ and this leads to 
$
|Z_{\msc{(1),(ii)}}| \gg |Z_{\msc{(1),(iii)}}|,
$
around $\tau_2\,\sim \, + \infty$.  
\item 
$Z_{\msc{(1), (iii)}}$ is expanded into continuous characters $\chic(p,m)$ which has left-right   {\em asymmetric\/} momenta;  
$$
\sim \chic (p, m_L;\tau,z) \chic (p', m_R;\tau,z)^*, \hspace{1cm}
p\neq p' ~ \mbox{in general.}
$$ 
It is, however, easy to check that the level matching condition 
$h_L-h_R \in \bz$, 
and invariance under T-transformation are satisfied.

\item The $q$-expansion of $Z_{\msc{(1),(iii)}}$ converges {\em even at $\vep=0$} (without the damping factor).

\end{itemize}

~

%%%%%%%%%%%%%%%%%%%%%%%%%%%%%%%%%%%%%%%%%%%%%%%%%%%%%%%%%%%%%%%%%%%%%%%%%%%%%%

\noindent
{\bf (2) ~ $Z_{(2)}$ :}

The evaluation of $Z_{(2)}$ is almost parallel to that of $Z_{(1)}$. We again make use of 
the expansion similar to \eqn{ 1} and decompose $Z_{(2)}$ as follows;
%%%
\begin{description}
\item[(i)] 
We extract the terms multiplied by  
$\dsp \left[\frac{(yq^{\tell})^{\frac{v_0}{N}}}{1-yq^{\ell}}\right]^*$
and denote them as 
$Z_{\msc{(2), (i)}}$.

%%%
\item[(ii)]
We extract the terms multiplied by
 $ \left[ (yq^{\ell})^{\frac{v_0}{N}+r}\right]^*$
for $v= v_0 + Nr $ and $r\leq  -1$.  
This contribution is denoted as $Z_{\msc{(2), (ii)}}$.

%%%
\item[(iii)]
The remaining part is denoted as $Z_{\msc{(2), (iii)}}$.

\end{description}

$Z_{\msc{(2),(i)}}$, $Z_{\msc{(2),(ii)}}$ are calculated as follows;
%%%%%%%%%%%%%%%%%%%%%%%%%%%%%%%%%%%%%%%%%%%%%%%%%%%%%%%%%%%%%%%%%%%%%
\begin{eqnarray}
%\hspace{-1cm}
&& Z_{\msc{(2),(i)}} 
= 
%%%%%%%%%%%%%%%%%%%%%%%%%%%%%%%%%%%%%%%%
- e^{2\pi \frac{\hc}{\tau_2}z_1^2} \,
%%%%%%%%%%%%%%%%%%%%%%%%%%%%%%%%%%%%%%%%
\sum_{v_0=0}^{N-1}\, \sum_{\stackrel{a_L,a_R\in \bz_N}{v_0+K(a_L+a_R) \in N\bz}}
\, \chid (v_0, a_R;\tau,z)^* \nn
&& \hspace{3cm}
\times \frac{i \th_1(\tau,z)}{\eta(\tau)^3}
\cdot \frac{1}{2} \sum_{j\in \bz_{2K}}\, R^{(+)}_{v_0+Nj, NK} \,
\Th{v_0+Nj+2Ka_L}{NK}\left(\tau, \frac{2z}{N}\right)  + \cO(\vep),
\label{Z 2 i} \\
%\end{eqnarray}
%%%%%%%%%%%%%%%%%%%%%%%%%%%%%%%%%%%%%%%%%%%%%%%%%%%%%%%%%%%%%%%%%%%%%
%\begin{eqnarray}
%\hspace{-1cm}
&& Z_{\msc{(2),(ii)}} =  
%%%%%%%%%%%%%%%%%%%%%%%%%%%%%%%%%%%%%
e^{2\pi \frac{\hc}{\tau_2}z_1^2} \,
%%%%%%%%%%%%%%%%%%%%%%%%%%%%%%%%%%%%%
\sum_{n_0\in \bz_{N}, \, w_0 \in \bz_{2K}}\,
 \int_{\br-i0} dp\, \rho_{(2)} (p,n_0,w_0 ;\vep) \, \nn
&& \hspace{3cm} \times  
\chic(p,Nw_0+Kn_0;\tau,z)\chic(p,Nw_0-Kn_0;\tau,z)^*, 
\label{Z 2 ii} \\
%\end{eqnarray}
%%%%%%%%%%%%%%%%%%%%%%%%%%%%%%%%%%%%%%%%%%%%%%%%%%%%%%%%%%%%%%%%%%%%%%%
%\begin{eqnarray}
&& \rho_{(2)}(p,n_0,w_0 ; \vep) = 
- \frac{1}{4 \pi i} \left[
\sum_{s=0}^{\infty} \frac{e^{-\vep ([-m_L]_{2K} +2Ks - ip)}}{p+ i\left([-m_L]_{2K}+ 2Ks\right)}
+ \sum_{\tilde{s}=0}^{\infty} \frac{e^{-\vep ([-m_R]_{2K} +2K\tilde{s} - ip)}}
{p+ i\left([-m_R]_{2K}+ 2K\tilde{s}\right)}
\right] \nn
%%%
&& \hspace{2cm}
=  C'(\vep) + \frac{1}{4\pi i } \frac{\partial}{\partial p} 
\log \left[
\Gamma\left(\frac{[-m_L]_{2K}}{2K} - \frac{ip}{2K}\right)
\Gamma\left(\frac{[-m_R]_{2K}}{2K} - \frac{ip}{2K}\right)
\right] + \cO(\vep),
\nn
&&
\label{rho 2}
\end{eqnarray}
Again $C'(\vep)$ is a positive logarithmically divergent constant independent of $p$, $n_0$, $w_0$.

The subleading part $Z_{\msc{(2),(iii)}}$ has a similar form to \eqn{Z 1 iii},
and we omit it here.

~

%%%%%%%%%%%%%%%%%%%%%%%%%%%%%%%%%%%%%%%%%%%%%%%%%%%%%%%%%%%%%%%%%%%%%%%%%%%%

\noindent
{\bf (3) ~ $Z_{(3)}$} :

Finally, let us consider $Z_{(3)}$. 
We note that the power series including the factor
$$
\frac{yq^{\ell}}{(1-yq^{\ell})(1-\bar{y}\bar{q}^{\tell})}
$$
is converging. Then it is easy to see
$$
\lim_{\vep\,\rightarrow\, +0} Z_{(3)} =0.
$$

%%%%%%%%%%%%%%%%%%%%%%%%%%%%%%%%%%%%%%%%%%%%%%%%%%%%%%%%%%%%%%%%%%%%%%%%%%%%%%%

~

\subsection{Summary of Decomposition}

Now, let us collect all the pieces of information on partition function. 
%%%%%%%%%%%%%%%%%%%%%%%%%%%%%%%%%%%%%%%%%%%%%%%%%%%%%%%%%%%%%%%%%%%%%%%%%%%%%%%%%
%%%%%%%%%%%%%%%%%%%%%%%%%%%%%%%%%%%%%%%%%%%%%%%%%%%%%%%%%%%%%%%%%%%%%%%%%%%%%%%%%
A crucial fact is that contributions of discrete (BPS) representations to 
$Z_{\dis}$,
$Z_{\msc{(1),(i)}}$ and $Z_{\msc{(2),(i)}}$ 
are precisely combined into the form of 
the {\em ``modular completion"} (see Appendix C),
%\eqn{hchid}}, 
%%%%%%%%%%%%%%%%%%%%%%%%%%%%%%%%%%%%%%%%%%%%%%%%%%%%%%%%%%%%%%%%%%%%%%%%%%%%%%%%%
%%%%%%%%%%%%%%%%%%%%%%%%%%%%%%%%%%%%%%%%%%%%%%%%%%%%%%%%%%%%%%%%%%%%%%%%%%%%%%%%%
\begin{eqnarray}
\hchid (v,a;\tau,z) 
%& : =& \frac{1}{N} \sum_{b\in\bz_N}\,
% e^{-2\pi i \frac{v b}{N}} q^{\frac{K}{N} a^2} y^{\frac{2K}{N} a}\,
% \hcK^{(2NK)}\left(\tau, \frac{z+a\tau+b}{N}\right)\, 
%\frac{i\th_1(\tau,z)}{\eta(\tau)^3}
%\nn
%%%
&\equiv & \chid (v,a;\tau,z) - \frac{1}{2} \sum_{j\in \bz_{2K}}\,
R^{(+)}_{v+Nj, NK}(\tau) \Th{v+Nj+2Ka}{NK}\left(\tau, \frac{2z}{N}\right)\,
\frac{i\th_1(\tau,z)}{\eta(\tau)^3}
\nn
&&
\label{hchid 0}
\end{eqnarray}
and we can safely take the $\ep\,\rightarrow\, 0$ limit. 
In the modular completion (\ref{hchid 0}) the anomalous transformation property of the discrete character $\chid (v,a;\tau,z)$ is compensated by the transformation law of the auxiliary function $R^{(+)}_{v+Nj, NK}(\tau)$ and the combination transforms like a Jacobi form.

Thus, we have found that the true discrete part of the partition function is given by the bilinear form of modular completions;
\begin{equation}
Z_{\dis} (\tau, z) := 
%%%%%%%%%%%%%%%%%%%%%%%%%%%%%%%%%%%%%%%
e^{2\pi \frac{\hc}{\tau_2}z_1^2} \,
%%%%%%%%%%%%%%%%%%%%%%%%%%%%%%%%%%%%%%%
\sum_{v=0}^{N-1}\, \sum_{\stackrel{a_L,a_R\in \bz_N}{v+K(a_L+a_R) \in N\bz}}
\, \hchid (v,a_L;\tau,z)\hchid (v, a_R;\tau,z)^*.
\label{Z dis}
\end{equation}
%%%%%%%%%%%%%%%%%%%%%%%%%%%%%%%%%%%%%%%%%%%%%%%%%%%%%%%%%%%%%%
%%%%%%%%%%%%%%%%%%%%%%%%%%%%%%%%%%%%%%%%%%%%%%%%%%%%%%%%%%%%%%
This is indeed modular invariant as is directly confirmed by using  
the modular transformation formula \eqn{S hchid}.
%%%%%%%%%%%%%%%%%%%%%%%%%%%%%%%%%%%%%%%%%%%%%%%%%%%%%%%%%%%%%%%
%%%%%%%%%%%%%%%%%%%%%%%%%%%%%%%%%%%%%%%%%%%%%%%%%%%%%%%%%%%%%%%
Process of completion cures the modular property of the BPS characters, however,
we have to pay the price of the non-holomorphic dependence of the function $R^{(+)}_{v+Nj, NK}(\tau)$ in this construction.
To be precise we need quadratic terms of $R^{(+)}_{m,NK}$
in order to obtain the expression \eqn{Z dis}. However, they can be naturally regarded 
as a part of $Z_{\msc{subleading}}$
defined below.

%%%%%%%%%%%%%%%%%%%%%%%%%%%%%%%%%%%%%%%%%%%%%%%%%%%%%%%%%%%%%%%%%%%%%%%%%

The rest of partition function can be expanded only in terms of 
the continuous (non-BPS) characters $\chic (p,m)$. 
%%%
We shall first combine $Z_{\msc{(1),(ii)}}$ and $Z_{\msc{(2),(ii)}}$ 
\begin{eqnarray}
Z_{\con}(\tau,z;\ep) &:=& Z_{\msc{(1),(ii)}} + Z_{\msc{(2),(ii)}}
\nn
&\equiv & \hskip-7mm 
\sum_{n_0\in \bz_{N}, \, w_0\in \bz_{2K}}\int_0^{\infty} dp\, \rho(p,n_0,w_0;\ep) \, 
\nn
&& \hspace{0.5cm} \times\,
\chic (p, Nw_0+ K n_0;\tau,z)\chic (p, Nw_0- K n_0;\tau,z)^* ,
\label{Z con}
\end{eqnarray}
where the spectral density $\rho(p,n_0,w_0;\ep)$ is given by 
\begin{eqnarray}
\rho(p,n_0,w_0;\ep)&:=& \rho_{(1)}(p,n_0,w_0;\ep) +
\rho_{(1)}(- p,n_0,w_0;\ep) +  \rho_{(2)}(p,n_0,w_0;\ep)
+  \rho_{(2)}(- p,n_0,w_0;\ep) 
\nn
&=& 
{\cal C}(\ep)
+ \frac{1}{4\pi i} \frac{\partial}{\partial p} \log 
\frac{\prod_{\al,\beta =\pm 1} \Gamma\left(\frac{1}{2K} [\al Nw_0 + \beta K n_0]_{2K}
-\frac{ip}{2K}
\right) }
{
\prod_{\al,\beta =\pm 1} \Gamma\left(\frac{1}{2K} [\al Nw_0 + \beta K n_0]_{2K}
+\frac{ip}{2K}
\right)
}
+ \cO(\ep).
\label{spectral density}
\end{eqnarray} 
$\cC(\ep)$ is a positive logarithmically divergent constant independent of 
$p$, $n_0$, $w_0$. 
Note here that the density function function $\rho(p,n_0,w_0;\ep)$ has no singularity at $p=0$, and thus 
the integral $\dsp \int_0^{\infty} dp$ is well-defined\footnote
   {It may be worthwhile to note that it is {\em not} the case 
 for each of $\rho_{(1)}$, $\rho_{(2)}$. In fact, 
each of $\rho_{(1)}(p, 0,0)$ and $\rho_{(2)}(p,0,0)$ shows  a singularity 
when approaching to $p=0$. After taking the sum, the singularity at $p=0$ is canceled, 
and one may simply replace the integration contour $\br-i0$ with $\br$.}.

We also have the `subleading part' of partition function 
$Z_{\msc{subleading}}$, 
which consists of contributions from $Z_{\msc{(1),(iii)}}$, $Z_{\msc{(2), (iii)}}$ 
as well as 
the quadratic term of $R_{*,*}^{(+)}$ appearing in the modular completion 
$\hchid$ \eqn{hchid 0}.
This is expressible as a convergent series of the terms such as 
$$
\int dp_L \, \int dp_R \, \sum_{m_L, m_R}\,\sigma (p_L,p_R,m_L, m_R) \, 
\chic (p_L, m_L;\tau,z) \chic (p_R, m_R;\tau,z)^*, 
%\hspace{1cm}
%p\neq p' ~ \mbox{in general},
$$ 
with some density $\sigma(p_L,p_R, m_L,m_R)$ (including delta-functions
in general) not specified here. 
$Z_{\msc{subleading}}$ damps more rapidly than other parts of 
partition function $Z_{\dis}$, $Z_{\con}$, 
when taking the IR limit $\tau_2\, \rightarrow \, + \infty$. 
%%%%%%%%%%%%%%%%%%%%%%%%%%%%%%%%%%%%%%%%%%%%%%%%%%%%%%%%%%%%%%%%%%%%%
%%%%%%%%%%%%%%%%%%%%%%%%%%%%%%%%%%%%%%%%%%%%%%%%%%%%%%%%%%%%%%%%%%%%%

The asymmetry of radial momenta $p_L$, $p_R$ appearing in $Z_{\msc{subleading}}$
may be an interesting feature. 
We also note that such an asymmetry is not observed 
in $Z_{\msc{con}}$, especially, in the divergent term proportional to the volume factor 
$\cC(\ep)$. 
These facts suggest a non-compact and curved geometry, which is asymptotic to 
a flat space-time. In the context of string compactification,
the divergent term in $Z_{\msc{con}}$ corresponds to the strings freely propagating 
in the asymptotic region of space-time. On the other hand, the sectors of 
$Z_{\msc{dis}}$ and $Z_{\msc{subleading}}$ could be contributed from 
strings localized in the strongly curved region around the tip of cigar. 
The former corresponds to massless sectors, 
whereas the latter would be regarded as  `too heavy' string modes to 
propagate in the asymptotic region,
and the asymmetry of radial momenta mentioned above 
would originate from curvature of cigar geometry.  
Note that neither $p_L$ nor $p_R$ are good quantum numbers in such a curved background,
while conformal weights still make sense. 
It can be checked that the level-matching condition 
is satisfied and the T-invariance is maintained.

%The asymmetry of radial momenta $p_L$, $p_R$ 
%suggests a non-compact and curved geometry, which is asymptotic to 
%a flat space-time. Thus, the emergence of such a sector would be natural 
%in our cigar geometry. It can be checked that the level-matching condition is satisfied and the %T-invariance is maintained. 

%%%%%%%%%%%%%%%%%%%%%%%%%%%%%%%%%%%%%%%%%%%%%%%%%%%%%%%%%%%%%%%%%%%%%%
%%%%%%%%%%%%%%%%%%%%%%%%%%%%%%%%%%%%%%%%%%%%%%%%%%%%%%%%%%%%%%%%%%%%%%

Here we notice a general phenomenon: when one tries to describe the string theory on some non-compact  target manifold,  there occurs in general a clash between the holomorphy and modular property of the theory. We have to either give up holomorphy 
and keep modular invariance or weaken the condition of modular invariance  and keep holomorphy. If one starts from the path-integral formulation of the theory, modular invariance is automatically enforced and one ends up with a non-holomorphic dependence in the character decomposition.

%%%%%%%%%%%%%%%%%%%%%%%%%%%%%%%%%%%%%%%%%%%%%%%%%%%%%%%%%%%%%%%%%%%%%%%%%%

~

\subsection{Orbifolding}

In the last part of this section,
we discuss some variants of the modular invariant\footnote
  {Precisely speaking, one should understand the partition functions given here 
as the ones defined with the regularization scheme 
presented above.}. 
Among others, we focus on the $\bz_N$-orbifold of $SL(2;\br)/U(1)$ supercoset. 
Geometrically it amounts to reducing the size of asymptotic circle of the cigar  
to $1/N$, and we  obtain 
\begin{eqnarray}
\hspace{-1cm}
Z_{\msc{orb}}(\tau,z) &=& 
%%%%%%%%%%%%%%%%%%%%%%%%%%%%%%%%%%%%%%%%%%%%%%%%%%%%%%%%%%%%%%%%%%%%%%
e^{\frac{2\pi}{\tau_2}\left( \hc |z|^2 -\frac{k+4}{k} z_2^2\right)}\,
%%%%%%%%%%%%%%%%%%%%%%%%%%%%%%%%%%%%%%%%%%%%%%%%%%%%%%%%%%%%%%%%%%%%%%
\frac{k}{N} \sum_{a,b\in \bz_N}\,  \int_{\bc} \frac{d^2u}{\tau_2}\, 
 e^{ 4\pi \frac{u_2 z_2}{\tau_2}}\,
\left|\frac{\th_1\left(\tau, -u + \left(1+ \frac{2}{k}\right)z\right)}
{\th_1\left(\tau, -u + \frac{2}{k} z \right)} \right|^2 
\, e^{-\frac{\pi k}{\tau_2}\left| u + \frac{a\tau+b}{N} \right|^2}.
\nn
&&
\label{Z0 tR orb}
\end{eqnarray}
It is obvious that this is also modular invariant.

%%%%%%%%%%%%%%%%%%%%%%%%%%%%%%%%%%%%%%%%%%%%%%%%%%%%%%%%%%%%%%%%%%%%%%%%%%%%

More generally, one may consider the `$\bz_N$-twisted partition function'
($\al,\beta \in \bz_N$ denotes the parameters of twisting); 
\begin{eqnarray}
%\hspace{-5mm}
Z_{[\al,\beta]}(\tau,z) &=& 
%%%%%%%%%%%%%%%%%%%%%%%%%%%%%%%%%%%%%%%%%%%%%%%%%%%%%%%%%%%%%%%%%%%%%%
e^{\frac{2\pi}{\tau_2}\left( \hc |z|^2 -\frac{k+4}{k} z_2^2\right)}\,
%%%%%%%%%%%%%%%%%%%%%%%%%%%%%%%%%%%%%%%%%%%%%%%%%%%%%%%%%%%%%%%%%%%%%%
\frac{k}{N} \sum_{a,b\in \bz_N}\,  
e^{- 2\pi i \frac{1}{N}(\al b-\beta a)}
\,\int_{\bc} \frac{d^2u}{\tau_2}\, 
 e^{ 4\pi \frac{u_2 z_2}{\tau_2}}
\,
\nn
&& \hspace{3cm} \times
\left|\frac{\th_1\left(\tau, -u + \left(1+ \frac{2}{k}\right)z\right)}
{\th_1\left(\tau, -u + \frac{2}{k} z \right)} \right|^2 
\, e^{-\frac{\pi k}{\tau_2}\left| u + \frac{a\tau+b}{N} \right|^2}.
\label{Z0 tR al beta}
\end{eqnarray}
It is not modular invariant, but rather behaves modular covariantly;
\begin{equation}
Z_{[\al,\beta]}(\tau+1,z) = Z_{[\al,\al+\beta]}(\tau), 
\hspace{1cm} 
Z_{[\al,\beta]}\left(-\frac{1}{\tau}, \frac{z}{\tau}\right)=
Z_{[\beta,-\al]} (\tau,z).
\end{equation}
%%%
We also note that 
\begin{equation}
Z_{[0,0]} (\tau,z) = Z_{\msc{orb}}(\tau,z), \hspace{1cm}
\frac{1}{N}\sum_{\al,\beta\in \bz_N} \,  Z_{[\al,\beta]} (\tau,z)
= Z(\tau,z). 
\label{rel Z al beta}
\end{equation}

~

%%%%%%%%%%%%%%%%%%%%%%%%%%%%%%%%%%%%%%%%%%%%%%%%%%%%%%%%%%%%%%%%%%%%%%%%%%%%%%%%5

The partition function \eqn{Z0 tR al beta} can be evaluated 
in almost the same way, though the analysis gets a  bit more complicated. 
Relevant changes are summarized as follows;
%%%%
\begin{itemize}
\item The winding number $w$ is replaced with $\frac{w}{N}$, while the KK momentum $n$ is
      changed into   
$Nn+\al$. Thus \eqn{cond n l} is replaced with 
\begin{equation}
N n + \al = \ell-\tell,
\label{cond n l 2}
\end{equation} 
and \eqn{v def} becomes 
\begin{equation}
v = w- K(\ell+\tell).
\label{v def 2}
\end{equation}

%%%%
\item By the same reason, 
\begin{eqnarray*}
&& w+ \frac{n}{k} ~ \longrightarrow ~ \frac{w}{N} + \frac{Nn+\al}{k} = \frac{w+ K(Nn+\al)}{N} = 
\frac{2K}{N} \left(\ell+ \frac{v}{2K}\right),
\\
&& w- \frac{n}{k} ~ \longrightarrow ~ \frac{w}{N} - \frac{Nn+\al}{k} = \frac{w- K(Nn+\al)}{N} = 
\frac{2K}{N} \left(\tell+ \frac{v}{2K}\right).
\end{eqnarray*}
Thus the combinations $\frac{2K}{N} \left(\ell+ \frac{v}{2K}\right)$, 
$\frac{2K}{N} \left(\tell+ \frac{v}{2K}\right)$ are unchanged.

\end{itemize}
%%%%%

The final expression is found to be 
\begin{eqnarray}
&& \hspace{-0.5cm}
%Z^{(\stR)}(\tau,\bar{\tau};z,\bar{z}) = 
Z_{[\al,\beta]}(\tau,z) = 
%%%%%%%%%%%%%%%%%%%%%%%%%%%%%%%%%%%%%%%%%
e^{2\pi \frac{\hc}{\tau_2}z_1^2} \,
%%%%%%%%%%%%%%%%%%%%%%%%%%%%%%%%%%%%%%%%%
\left|\frac{\th_1(\tau,z)}{\eta(\tau)^3}\right|^2 \,
\sum_{\stackrel{v,\ell,\tell\in \bz}{\ell-\tell \in \al+ N\bz}}\,
\frac{1}{2\pi i} \, \left[
\int_{\br-i0} dp \, (yq^{\ell}) \, \overline{ (yq^{\tell}) } 
- \int_{\br+i(N-0)} dp \,
\right]
\nn
&& \hspace{1.5cm}
\times 
e^{2\pi i \frac{\beta}{N} \left\{v+ K(\ell+\tell) \right\}}
\,
\frac{e^{-\pi \tau_2 \frac{p^2+v^2}{NK}}} {p-iv}
\, \frac{(yq^{\ell})^{\frac{v}{N}}}{1-yq^{\ell}} \,
\left[
\frac{(yq^{\tell})^{\frac{v}{N}}}{1-yq^{\tell}}
\right]^*\,
y^{\frac{2K}{N}\ell} q^{\frac{K}{N}\ell^2} \, 
\left[y^{\frac{2K}{N}\tell} q^{\frac{K}{N}\tell^2}\right]^*.
\label{Z2 tR al beta}
\end{eqnarray}
Especially, the discrete part is determined as follows;
\begin{equation}
Z_{[\al,\beta]\, \dis} (\tau, z) =  
%%%%%%%%%%%%%%%%%%%%%%%%%%%%%%%%%%%%%%%%%%
e^{2\pi \frac{\hc}{\tau_2}z_1^2} \, 
%%%%%%%%%%%%%%%%%%%%%%%%%%%%%%%%%%%%%%%%%%%
\sum_{v=0}^{N-1}\, \sum_{a \in \bz_N}
\, e^{2\pi i \frac{\beta}{N} \left\{v+K (\al + 2a)  \right\}} \,
\hchid (v,a+\al;\tau,z) \hchid (v,  a ;\tau,z)^*.
\label{Z dis al beta}
\end{equation}
One can easily check that \eqn{Z dis} and \eqn{Z dis al beta}
are consistent with the relation \eqn{rel Z al beta}.

~

%%%%%%%%%%%%%%%%%%%%%%%%%%%%%%%%%%%%%%%%%%%%%%%%%%%%%%%%%%%%%%%%%%%%%%%%%%
%%%%%%%%%%%%%%%%%%%%%%%%%%%%%%%%%%%%%%%%%%%%%%%%%%%%%%%%%%%%%%%%%%%%%%%%%%

\section{Elliptic Genus}

\subsection{Analysis Based on the Character Decomposition}

The elliptic genus \cite{Witten-E} is given by formally setting $\bar{z}=0$ in the
partition function of $\tR$-sector $Z$, 
while leaving $z$ at a generic value.
Since we have already obtained the character decomposition 
of the partition function, it is straightforward to calculate the elliptic genus. 
It is obvious that only BPS representations $Z_{\dis}$ 
contributes in the right-moving ($\bar{z}$-dependent) sector.
Using the result of \eqn{Z dis} 
and the formulas of Witten index \eqn{WI};
%%%%
\begin{equation}
\hchid (v,a;\tau,0) = - \delta^{(N)}_{a,0},
\end{equation}
%%%% 
we obtain the formula for the elliptic genus;
\begin{equation}
\cZ(\tau,z) = - \sum_{v=0}^{N-1} \,
\sum_{\stackrel{a\in \bz_N}{v+Ka \in N\bz}}\,
\hchid (v,a;\tau,z).
\label{EG1}
\end{equation}

Also, \eqn{Z dis al beta} yields
\begin{eqnarray}
&&\cZ_{[\al,\beta]}(\tau,z) = - \sum_{v=0}^{N-1} \,
e^{2\pi i \frac{\beta}{N} (v+K \al)}\, 
\hchid (v,\al;\tau,z),
\label{EG al beta}\\
&&\hspace{1cm} \equiv  - 
e^{2\pi i \frac{K}{N}\al \beta} q^{\frac{K}{N}\al^2}
y^{\frac{2K}{N}\al}\, 
\hcK^{(2NK)}\left(\tau, \frac{z+\al \tau+\beta}{N}\right)\,
\frac{i\th_1(\tau,z)}{\eta(\tau)^3}.
\label{EG al beta 2}
\end{eqnarray}
Here we have introduced the notation of Appell function $\hcK$, see Appendix C.

It is obvious that the following relations hold;
\begin{equation}
\cZ_{[0,0]} (\tau,z) = \cZ_{\msc{orb}}(\tau,z), \hspace{1cm}
\frac{1}{N}\sum_{\al,\beta\in \bz_N} \,  \cZ_{[\al,\beta]} (\tau,z)
= \cZ(\tau,z), 
\label{rel EG al beta}
\end{equation}
corresponding to \eqn{rel Z al beta}, where $\cZ_{\msc{orb}}(\tau,z)$ 
is the elliptic genus of $\bz_N$-orbifold \eqn{Z0 tR orb}.
Especially, 
\begin{equation}
\cZ_{\msc{orb}}(\tau,z) 
%\equiv \cZ_{[0,0]}(\tau,z) 
= -  \sum_{v=0}^{N-1} \, \hchid (v,0;\tau,z) \equiv 
- \hcK^{(2NK)}\left(\tau, \frac{z}{N}\right)\, \frac{i\th_1(\tau,z)}{\eta(\tau)^3},
\label{EG orb}
\end{equation}
which is the one proposed in \cite{Troost}.
%%%%
We also note
\begin{equation}
\cZ(\tau,z) = - \frac{1}{N} \sum_{\al,\beta \in \bz_{N}}\, 
e^{2\pi i \frac{K}{N}\al \beta} q^{\frac{K}{N}\al^2}
y^{\frac{2K}{N}\al}\, 
\hcK^{(2NK)}\left(\tau, \frac{z+\al \tau+\beta}{N}\right)\,
\frac{i\th_1(\tau,z)}{\eta(\tau)^3}.
\label{EG 2}
\end{equation}
%%%
Thanks to the good modular property of $\hcK^{(2k)}(\tau,z)$ \eqn{modular hcK},
the expressions \eqn{EG al beta 2}, \eqn{EG orb} and \eqn{EG 2} 
all behave modular covariantly. 
For instance, $\cZ_{[\al,\beta]}(\tau,z)$ \eqn{EG al beta 2} satisfies the following modular 
transformation formulas;
\begin{equation}
\cZ_{[\al,\beta]}(\tau+1,z) = \cZ_{[\al,\al+\beta]}(\tau,z),
\hspace{1cm}
\cZ_{[\al,\beta]}\left(- \frac{1}{\tau}, \frac{z}{\tau}\right) 
= e^{i\pi \frac{\hc}{\tau}z^2}\, \cZ_{[\beta,-\al]}(\tau,z).
\label{modular EG al beta}
\end{equation}

~

%%%%%%%%%%%%%%%%%%%%%%%%%%%%%%%%%%%%%%%%%%%%%%%%%%%%%%%%%%%%%%%%%%%%%%%%%%%%%%

\subsection{Path-integral Representation of Elliptic Genus}

It would be interesting to clarify the `path-integral representation' of elliptic genus
in a way analogous to \cite{Troost}. 
We begin with generalizing the partition function \eqn{Z0 tR} so as to include
two {\em independent\/} angular variables $z_L$, $z_R$ that couple with the 
left/right $U(1)$-charges. 
To this aim, it is enough to formally replace $z$, $\bar{z}$ with $z_L$, $\overline{z_R}$
in the expression \eqn{Z0 tR}. Note also that 
$z_2 \equiv \frac{z-\bar{z}}{2i}$ should be replaced with 
$\frac{z_L-\overline{z_R}}{2i}$. 
%%%
We thus obtain the partition function%\footnote
%  {Here it is crucial  that we have no $z$-dependence in the  
%factor $e^{-\frac{\pi k}{\tau_2}|u|^2}$ at the beginning. 
%Otherwise, it seems more complicated to obtain a modular invariant   
%partition function with two angle variables. 
%};
\begin{eqnarray}
%\hspace{-5mm}
Z(\tau,z_L,z_R) &=&
%%%%%%%%%%%%%%%%%%%%%%%%%%%%%%%%%%%%%%%%%%%%%%%%%%%%%%%%%%%%%%%%%%%%%%%%%%%%%%%%%%%%%%%%%
ke^{\frac{2\pi}{\tau_2}\left\{ \hc z_L \bar{z}_R
-\frac{k+4}{k} \left(\frac{z_L-\overline{z_R}}{2i}\right)^2 \right\}}\,
%%%%%%%%%%%%%%%%%%%%%%%%%%%%%%%%%%%%%%%%%%%%%%%%%%%%%%%%%%%%%%%%%%%%%%%%%%%%%%%%%%%%%%%%%
\int_{\bc} \frac{d^2u}{\tau_2}\, 
e^{ 4\pi \frac{u_2}{\tau_2} \left(\frac{z_L-\overline{z_R}}{2i} \right)}\,
\nn
%%%
&& \hspace{5mm}
\times 
\frac{\th_1\left(\tau, -u + \left(1+ \frac{2}{k}\right)z_L\right)}
{\th_1\left(\tau, -u + \frac{2}{k} z_L \right)} \,
\overline{
\frac{\th_1\left(\tau, -u + \left(1+ \frac{2}{k}\right)z_R \right)}
{\th_1\left(\tau, -u + \frac{2}{k} z_R \right)} 
}
\, e^{-\frac{\pi k}{\tau_2}\left| u \right|^2}.
%\nn
%&&
\label{Z0 cpx tR}
\end{eqnarray}
This partition function is complex, but still modular invariant, as can be directly 
confirmed.

Then, the desired elliptic genus should be given as  
\begin{equation}
\cZ(\tau,z) = 
\lim_{z_R\,\rightarrow\, 0}\, 
%%%%%%%%%%%%%%%%%%%%%%%%%%%%%%%%%%%%%%%%%%%%%%%%%%%%%%%%%%%%%%%%%%%%%%%%
e^{- 2\pi \frac{\hc}{\tau_2} \left(\frac{z+\overline{z_R}}{2}\right)^2 }\,
%%%%%%%%%%%%%%%%%%%%%%%%%%%%%%%%%%%%%%%%%%%%%%%%%%%%%%%%%%%%%%%%%%%%%%%%
 Z(\tau, z_L=z , z_R).
\label{Z0 cpx EG}
\end{equation}
(Recall that the partition function $Z(\tau,z)$ includes 
the anomaly factor $e^{2\pi \frac{\hc}{\tau_2}z_1^2}$, which is absent 
in the definition of elliptic genus.)
When setting $z_R=0$, the right-moving $\th_1$-factors drop off, as is expected from supersymmetry.
We thus obtain
\begin{eqnarray}
\hspace{-5mm}
\cZ(\tau,z) &=& k e^{\pi \frac{z^2}{k\tau_2}}\, \int_{\bc} \frac{d^2u}{\tau_2}\, 
\frac{\th_1\left(\tau, -u + \left(1+ \frac{2}{k}\right)z\right)}
{\th_1\left(\tau, -u + \frac{2}{k} z \right)} \, 
\, e^{-2\pi i z \frac{u_2}{\tau_2}}\,
e^{-\frac{\pi k}{\tau_2}\left| u \right|^2}.
\label{EG path int}
\end{eqnarray}
%%%
This possesses the expected modular properties;
\begin{equation}
\cZ(\tau+1,z) = \cZ(\tau,z), \hspace{1cm} 
\cZ\left(-\frac{1}{\tau}, \frac{z}{\tau}\right) = e^{i\pi \frac{\hc}{\tau}z^2}\,\cZ(\tau,z).
\label{EG modular}
\end{equation}
%This is confirmed most easily by rewriting \eqn{EG path int} with the shift of the integration variable
%$u\,\rightarrow \, u+ \frac{1}{k}z$ as 
% \begin{eqnarray}
%\hspace{-5mm}
%\cZ(\tau,z) &=& k e^{\pi \frac{|z|^2}{k\tau_2}}\, \int_{\bc} \frac{d^2u}{\tau_2}\, 
%\frac{\th_1\left(\tau, -u + \left(1+ \frac{1}{k}\right)z\right)}
%{\th_1\left(\tau, -u + \frac{1}{k} z \right)} \, 
%\, e^{-2\pi i z \frac{u_2}{\tau_2}}\,
%e^{-\frac{\pi k}{\tau_2}\left| u + \frac{1}{k}z \right|^2}.
%\end{eqnarray}

%%%%%%%%%%%%%%%%%%%%%%%%%%%%%%%%%%%%%%%%%%%%%%%%%%%%%%%%%%%%%%%%%%%%%%%%%

If we started with the twisted partition function 
$Z_{[\al,\beta]}$ \eqn{Z0 tR al beta}, we would similarly 
obtain
\begin{eqnarray}
%\hspace{-5mm}
\cZ_{[\al,\beta]}(\tau,z) 
&=& 
\frac{1}{K} e^{\pi \frac{z^2}{k\tau_2}}\, \sum_{a,b \in\bz_N} \,
e^{-2\pi i \frac{1}{N}(\al b-\beta a)}\,  
\int_{\bc} \frac{d^2 u}{\tau_2}\, 
\frac{\th_1\left(\tau, -u+ \left(1+\frac{2}{k}\right)z\right)}
{\th_1\left(\tau, -u+ \frac{2}{k} z \right)} 
\nn
&& \hspace{3cm}
\times 
e^{- 2\pi i z \frac{u_2}{\tau_2}}
\, e^{-\frac{\pi k}{\tau_2}  \left| u + \frac{a\tau+b}{N} \right|^2}
\hspace{1cm} \left(\al,\beta \in \bz_N \right).
\label{EG al beta path int}
\end{eqnarray}
%where the twist parameters $\al$, $\beta$ take values in $\bz_N$. 
%
This expression shows that
$\cZ_{[\al,\beta]}(\tau,z) $ has the expected modular properties \eqn{modular EG al beta}.
%It is again most easily proven by shifting 
%$u\,\rightarrow \, u+ \frac{1}{k}z$.

%%%%%%%%%%%%%%%%%%%%%%%%%%%%%%%%%%%%%%%%%%%%%%%%%%%%%%%%%%%%%%%%%%%%%%%%%%%%%%%%%%%%%%%%%%%%%%%%%

~

%\noindent
%\underline{\bf Direct evaluation of \eqn{EG al beta path int}}

\subsubsection{Direct Evaluation of \eqn{EG al beta path int}}

Let us try to rederive the formula \eqn{EG al beta 2} from 
the path-integral representation \eqn{EG al beta path int}.
This is parallel to the analysis presented in the previous section.
However, contrary to the case of partition function, {\em we need not introduce any
regularization.} This is because the integrand of \eqn{EG al beta path int}
possesses at most simple poles, and thus the $u$-integral already converges.

%%%%%%%%%%%%%%%%%%%%%%%%%%%%%%%%%%%%%%%%%%%%%%%%%%%%%%%%%%%%%%%%%%%%%
Set $u =: \s_1\tau+\s_2$. Since the integrand includes Gaussian factors 
which decrease rapidly at infinity, we can safely shift the contours of 
$\s_1$, $\s_2$-integrals as 
$$
\br~\longrightarrow ~ \br+i\xi_1, \hspace{1cm}
\br~\longrightarrow ~ \br+ i\xi_2, 
$$
with arbitrary $\xi_1, \xi_2 \in \br$, without  changing the value of the integral.
Namely, we can rewrite \eqn{EG al beta path int} as
\begin{eqnarray}
\cZ_{[\al,\beta]}(\tau,z) 
&=& 
\frac{1}{K} e^{\pi \frac{z^2}{k\tau_2}}\, \sum_{a,b \in\bz_N} \,
e^{-2\pi i \frac{1}{N}(\al b-\beta a)}\,  
\int_{\br+i\xi_1} d\s_1\, \int_{\br+i\xi_2} d\s_2\, 
\frac{\th_1\left(\tau, -\s_1\tau-\s_2+ \left(1+\frac{2}{k}\right)z\right)}
{\th_1\left(\tau, -\s_1\tau-\s_2 + \frac{2}{k} z \right)} 
\nn
&& \hspace{3cm}
\times 
e^{- 2\pi i z \sms_1}
\, e^{-\frac{\pi k}{\tau_2}  
\left[\left\{
\left(\sms_1+\frac{a}{N}\right)\tau_1 + \sms_2+\frac{b}{N}
\right\}^2 
+ \left(\sms_1+\frac{a}{N} \right)^2 \tau_2^2 \right]}.
\label{EG al beta path int 2}
\end{eqnarray}
Moreover, we introduce the `winding numbers' $w,m\in \bz$ and real parameters 
$\zeta_i$ as 
\begin{equation}
\s_1 = (\zeta_1+i\xi_1) + s_1 +w, \hspace{1cm}
\s_2 = (\zeta_2+i\xi_2) + s_2 +m, \hspace{1cm} (0 < s_1, s_2 < 1)
\end{equation}
 We shall choose suitable parameters $\xi_i$, $\zeta_i$ to simplify   
the relevant integral.
A good choice is given by
\begin{equation}
\zeta_1+i\xi_1 = -i \frac{z}{k\tau_2}, \hspace{1cm}
\zeta_2+i\xi_2= i \frac{\bar{\tau} z}{k\tau_2}.
\label{choice zeta xi}
\end{equation}
Relevant calculations are as follows;
\begin{eqnarray}
&&\frac{\th_1\left(\tau, -\s_1\tau-\s_2+ \left(1+\frac{2}{k}\right)z\right)}
{\th_1\left(\tau, -\s_1\tau-\s_2 + \frac{2}{k} z \right)} 
=
\frac{\th_1\left(\tau, -(s_1+w)\tau-(s_2+m)+ z\right)}
{\th_1\left(\tau, -(s_1+w)\tau-(s_2+m) \right)}
\nn
&&\hskip3cm = 
e^{2\pi i z w}\,
\frac{\th_1\left(\tau, -s_1\tau-s_2+ z\right)}
{\th_1\left(\tau, -s_1\tau-s_2 \right)},
\label{eval 1}\\
&& e^{-\frac{\pi k}{\tau_2}\left[
\left\{
\left(\sms_1+\frac{a}{N}\right)\tau_1 + \sms_2+\frac{b}{N}
\right\}^2 
+ \left(\sms_1+\frac{a}{N} \right)^2 \tau_2^2
\right]} 
\nn
&&
= e^{-\frac{\pi k}{\tau_2} \left[
\left\{
\left(s_1+w+\frac{a}{N}\right)\tau_1 + s_2+m+\frac{b}{N} + \frac{z}{k}
\right\}^2 
+ \left(s_1+w +\frac{a}{N} \right)^2 \tau_2^2
\right]} \cdot e^{\frac{\pi z^2}{k\tau_2}} \cdot e^{2\pi i z \left(s_1+w+\frac{a}{N}\right)},
\label{eval 2}\\
&& e^{-2\pi i z \sms_1} = e^{-2\pi i z (s_1+w)} \cdot e^{- 2\pi \frac{z^2}{k\tau_2}}.
\label{eval 3}
\end{eqnarray}
%%%%
Substituting \eqn{eval 1}, \eqn{eval 2}, and \eqn{eval 3} into \eqn{EG al beta path int 2},
we obtain 
\begin{eqnarray}
%\hspace{-5mm}
\cZ_{[\al,\beta]}(\tau,z) 
&=& 
\frac{1}{K}  
\sum_{w,m\in \bsz}\, \sum_{a,b\in \bz_N} \,
 e^{-2\pi i \frac{1}{N}(\al b -\beta a)}\,
\int_0^1 ds_1 \int_0^1 ds_2\, 
\frac{\th_1\left(\tau, -s_1\tau-s_2+ z\right)}
{\th_1\left(\tau, -s_1\tau-s_2\right)} 
\nn
&& \hspace{2cm}
\times 
y^{w+\frac{a}{N}}\, 
e^{-\frac{\pi k}{\tau_2} \left[\left\{ \left( s_1+ w+ \frac{a}{N} \right) \tau_1 
+ s_2+ m+  \frac{b}{N}  + \frac{z}{k} \right\}^2 
+ \left(s_1 + w+ \frac{a}{N} \right)^2 \tau_2^2 \right]}
\nn
&=&
\frac{1}{K}  
\sum_{w,m\in \bsz}\,
 e^{-2\pi i \frac{1}{N}(\al m -\beta w)}\,
\int_0^1 ds_1 \int_0^1 ds_2\, 
\frac{\th_1\left(\tau, -s_1\tau-s_2+ z\right)}
{\th_1\left(\tau, -s_1\tau-s_2\right)} 
\nn
&& \hspace{2cm}
\times 
y^{\frac{w}{N}}\, 
e^{-\frac{\pi}{NK \tau_2} \left[\left\{ \left(N s_1+ w \right) \tau_1 
+ N s_2+ m + Kz  \right\}^2 
+ \left(Ns_1 + w\right)^2 \tau_2^2 \right]}.
\label{EG al beta path int 3}
\end{eqnarray}
In the 2nd line we rewrote  $Nw+a$, $Nm+b$ as $w$, $m$.

%%%%%%%%%%%%%%%%%%%%%%%%%%%%%%%%%%%%%%%%%%%%%%%%%%%%%%%%%%%%%%%%%%%%%%%

The Poisson resummation \eqn{PR formula} yields 
\begin{eqnarray}
\cZ_{[\al,\beta]}(\tau,z) &=& 
\sqrt{k \tau_2} \, 
\sum_{w,n \in \bsz}\, e^{2\pi i \frac{\beta w}{N}}\, 
\int_0^1 ds_1 \, \int_0^1 ds_2\, 
\frac{\th_1\left(\tau, -s_1\tau-s_2+ z\right)}
{\th_1\left(\tau, -s_1\tau-s_2\right)} 
\nn
&& \hspace{0.5cm}
\times  y^{\frac{w}{N}}\
\, e^{-\pi \tau_2  \left\{ NK \left(n+\frac{\al}{N}\right)^2 
+ \frac{1}{NK} \left(N s_1+ w \right)^2\right\}
+ 2\pi i \left(n + \frac{\al}{N} \right) \left\{\left(N s_1+w \right)\tau_1 + N s_2 
+ Kz \right\}}.
\label{EG al beta path int 4}
\end{eqnarray}
%%%%%%%%%%%%%%%%%%%%%%%%%%%%%%%%%%%%%%%%%%%%%%%%%%%%%%%%%%%%%%%%%%%%%%%%%%%
Due to the identity \eqn{th1/th1 formula 2}, we find 
\begin{eqnarray}
&& \hspace{-1cm}
\frac{\th_1\left(\tau, -s_1\tau-s_2+ z\right)}
{\th_1\left(\tau, -s_1\tau-s_2\right)}  = 
\frac{-i \th_1(\tau,z)}{\eta(\tau)^3} \,
\sum_{\ell \in \bz} \, \frac{yq^{\ell} }
{1-yq^{\ell} } \, 
%\nn
%&& \hspace{4cm}
%\times 
e^{-2\pi i (s_1\tau_1+s_2) \ell + 2\pi s_1 \tau_2 \ell}.
\label{exp th1/th1 EG}
\end{eqnarray}
%%%
The $s_2$-integral simply gives
\begin{equation}
\ell = Nn+\al. 
\label{cond n l EG}
\end{equation} 
%%%
Relevant terms of the $s_1$-integral are now calculated as 
\begin{eqnarray}
&& e^{-\pi \tau_2 \frac{N}{K} s_1^2 
-2\pi s_1 \left\{\frac{1}{K} w \tau_2- i(N n+\al) \tau_1 + i \ell \tau_1
- \ell \tau_2 
\right\}
} 
= e^{-\pi \tau_2 \frac{N}{K} s_1^2- 2\pi s_1 \tau_2 \frac{v}{K}},
\end{eqnarray}
where we set 
\begin{equation}
v:= w- K \ell = w- K(N n+ \al),
\label{v def EG}
\end{equation}
under the constraint \eqn{cond n l EG}. 
%%%%%%%%%%%%%%%%%%%%%%%%%%%%%%%%%%%%%%%%%%%%%%%%%%%%%%%%%%%%%%%%%%%%%%%%%%%%%%%%%%%%%%%
The $s_1$-integral is performed in the same way;
\begin{eqnarray}
\int_0^1ds_1\, e^{-\pi \tau_2 \frac{N}{K}s_1^2 -2\pi \tau_2 \frac{s_1}{K} v}
&=& \sqrt{\frac{\tau_2}{NK}} \,  \int_0^1 ds_1\, \int_{\br-i0} dp\,
e^{-\frac{\pi}{NK}\tau_2 p^2 - 2\pi i \tau_2 \frac{s_1}{K}(p-iv)}
\nonumber \\
&=& \sqrt{\frac{K}{N\tau_2}} \frac{1}{2\pi i} \,
\int_{\br-i0} dp\, 
\frac{e^{-\frac{\pi}{NK} \tau_2 p^2  }} {p-iv}
\, \left\{ 
1- e^{-2\pi i \tau_2 \frac{1}{K} (p-iv)}
\right\}.
\label{s1 int EG} 
\end{eqnarray}
%%%%%%%%%%%%%%%%%%%%%%%%%%%%%%%%%%%%%%%%%%%%%%%%%%%%%%%%%%%%%%%%%%%%%%%%%%%%%%%%%%%%%%%%%
We also note 
\begin{eqnarray}
e^{- \pi \tau_2 \left\{NK \left(n+\frac{\al}{N}\right)^2+\frac{w^2}{NK}\right\}}\,
e^{2\pi i w \left(n+\frac{\al}{N}\right)}
&=& q^{\frac{1}{4NK} \left\{w+K(Nn+\al)\right\}^2} \,
\left[q^{\frac{1}{4NK} \left\{w-K(Nn+\al)\right\}^2}\right]^*
\nn
&=& e^{-\pi \tau_2 \frac{v^2}{NK}}\, q^{NK\left(n+\frac{\al}{N}\right)^2
+v\left(n+\frac{\al}{N}\right)}.
\end{eqnarray}
In the 2nd line we used \eqn{cond n l EG} and \eqn{v def EG}.
On the other hand, the power of $y$ is given by
$$
y^{\frac{w}{N} + K\left(n+\frac{\al}{N}\right)} 
= y^{2K\left(n+\frac{\al}{N}\right) +\frac{v}{N}}.
$$

%%%%%%%%%%%%%%%%%%%%%%%%%%%%%%%%%%%%%%%%%%%%%%%%%%%%%%%%%%%%%%%%%%%%%%%%%%%%%%%%%%%%%

Collecting all these factors, we obtain 
\begin{eqnarray}
%\hspace{-0.5cm}
\cZ_{[\al,\beta]}(\tau,z)&=& 
- \frac{i \th_1(\tau,z)}{\eta(\tau)^3} \,
\sum_{n,v\in \bz}\,
\frac{1}{2\pi i} \,
\int_{\br-i0} dp\, 
\frac{e^{-\pi \tau_2 \frac{p^2+v^2}{NK}}} {p-iv}
\, \left\{ 
1- e^{-2\pi i \tau_2 \frac{1}{K} (p-iv)}
\right\}
\nn
&& \hspace{2cm} \times e^{2\pi i \frac{\beta}{N}(v+K\al)}\, 
\frac{(yq^{N n +\al})^{1+\frac{v}{N}}}{1-yq^{N n +\al}} \,
y^{2K\left(n+\frac{\al}{N}\right)} q^{NK \left(n+\frac{\al}{N}\right)^2} \, 
\nn
&=&  - \frac{i \th_1(\tau,z)}{\eta(\tau)^3} \,
\sum_{n,v\in \bz}\, 
\frac{1}{2\pi i} \, \left[
\int_{\br-i0} dp \, \left(yq^{N n +\al} \right)\,   
- \int_{\br+i(N-0)} dp \,
\right] \, \frac{e^{-\pi \tau_2 \frac{p^2+v^2}{NK}}} {p-iv} 
\nn
&& \hspace{2cm}
\times e^{2\pi i \frac{\beta}{N}(v+K\al)}\, 
\frac{(yq^{N n +\al})^{\frac{v}{N}}}{1-yq^{N n +\al}} \,
y^{2K\left(n+\frac{\al}{N}\right)} q^{NK \left(n+\frac{\al}{N}\right)^2} .
\label{EG al beta 1}
\end{eqnarray}
%%%%%%%%%%%%%%%%%%%%%%%%%%%%%%%%%%%%%%%%%%%%%%%%%%%%%%%%%%%%%%%%%%%%%%
Finally, by shifting the integration contour;
$$
\br+i(N-0)~\longrightarrow ~ \br-i0,
$$
in the 2nd integral, we obtain the decomposition;
\begin{eqnarray}
\cZ_{[\al,\beta]}(\tau,z) &= &\cZ_{[\al,\beta]\, \dis}(\tau,z) +  \cZ_{[\al,\beta]\, \rem} (\tau,z).
\end{eqnarray}
%%%
The discrete part (pole contribution) is easily calculated as 
\begin{eqnarray}
\cZ_{[\al,\beta]\, \dis}(\tau,z)
% &\equiv & [\mbox{pole contributions}]
%\nn
&=& - \frac{i \th_1(\tau,z)}{\eta(\tau)^3} \,
\sum_{v=0}^{N-1}\, \sum_{n\in\bz}\, 
e^{2\pi i \frac{\beta}{N}(v+K\al)}\, 
\frac{(yq^{N n +\al})^{\frac{v}{N}}}{1-yq^{N n +\al}} \,
y^{2K\left(n+\frac{\al}{N}\right)} q^{NK \left(n+\frac{\al}{N}\right)^2}
\nn
&=& - e^{2\pi i \frac{K}{N}\al \beta} q^{\frac{K}{N}\al^2}
y^{\frac{2K}{N}\al}\, 
\cK^{(2NK)}\left(\tau, \frac{z+\al \tau+\beta}{N}\right)\,
\frac{i\th_1(\tau,z)}{\eta(\tau)^3}.
\label{EG dis 1}
\end{eqnarray}
%%%%%%%%%%%
Here we used the identity \eqn{rel chid cK}. 
This part \eqn{EG dis 1} is obviously holomorphic. 

%%%%%%%%%%%%%%%%%%%%%%%%%%%%%%%%%%%%%%%%%%%%%%%%%%%%%%%%%%%%%%%%%%%

The remainder part is now computed as 
\begin{eqnarray}
%\hspace{-0.5cm}
\cZ_{[\al,\beta], \rem}(\tau,z)&=& 
 - \frac{i \th_1(\tau,z)}{\eta(\tau)^3} \,
\sum_{n,v\in \bz}\, 
\frac{1}{2\pi i} \, 
\int_{\br-i0} dp \, \left( yq^{N n +\al}   
- 1 \right)\,  \frac{e^{-\pi \tau_2 \frac{p^2+v^2}{NK}}} {p-iv} 
\nn
&& \hspace{1.5cm}
\times e^{2\pi i \frac{\beta}{N}(v+K\al)}\, 
\frac{(yq^{N n +\al})^{\frac{v}{N}}}{1-yq^{N n +\al}} \,
y^{2K\left(n+\frac{\al}{N}\right)} q^{NK \left(n+\frac{\al}{N}\right)^2} ,
\nn
&=& 
\frac{i \th_1(\tau,z)}{\eta(\tau)^3} \,
\sum_{n,v\in \bz}\, 
\frac{1}{2\pi i} \, \int_{\br-i0} dp \,
\frac{e^{-\pi \tau_2 \frac{p^2+v^2}{NK}}} {p-iv} 
\nn
&& \hspace{1cm}
\times
e^{2\pi i \frac{\beta}{N}(v+K\al)}\, 
y^{2K\left(n+\frac{v+ 2K\al}{2NK} \right)} q^{NK \left(n+\frac{v+ 2K\al}{2NK}\right)^2}
q^{- \frac{v^2}{4NK}},
\nn
&=& \frac{i \th_1(\tau,z)}{\eta(\tau)^3} \cdot 
\frac{1}{2} \sum_{v \in \bz_{2NK}} \, e^{2\pi i \frac{\beta}{N}(v+K\al)}\, R^{(+)}_{v,NK}
\, \Th{v+2K\al}{NK}\left(\tau,\frac{2z}{N}\right), 
\nn
&=& \frac{i \th_1(\tau,z)}{\eta(\tau)^3} \cdot 
\frac{1}{2} \sum_{v \in \bz_{2NK}} \, 
e^{2\pi i \frac{K}{N}\al \beta}\, q^{\frac{K}{N}\al^2} y^{\frac{2K}{N} \al}\,
R^{(+)}_{v,NK}
\, \Th{v}{NK}\left(\tau,\frac{2(z+\al\tau+\beta)}{N}\right) .
\nn
&&
\label{EG rem 1}
\end{eqnarray}
Here we used the definition of function $R^{(+)}_{*,*}$ \eqn{Rmk}.
Combining \eqn{EG dis 1} and \eqn{EG rem 1}, and using the formula of modular completion 
\eqn{hcK}, 
we finally obtain the expected result;
\begin{eqnarray}
&& \cZ_{[\al,\beta]} (\tau,z) = \cZ_{[\al,\beta]\, \dis} (\tau,z) 
+ \cZ_{[\al,\beta] \, \rem}(\tau,z) 
\nn
&& \hspace{1cm}
= - e^{2\pi i \frac{K}{N}\al \beta}\, q^{\frac{K}{N}\al^2} y^{\frac{2K}{N} \al}\,
\nn
&& \hspace{1cm}
\times 
\left[
\cK^{(2NK)}\left(\tau, \frac{z+\al \tau+\beta}{N}\right)
-\frac{1}{2} \sum_{v\in \bz_{2NK}}\, R^{(+)}_{v,NK}
\, \Th{v}{NK}\left(\tau,\frac{2(z+\al\tau+\beta)}{N}\right) 
\right] \, \frac{i \th_1(\tau,z)}{\eta(\tau)^3}
\nn
&& \hspace{1cm}
=   - e^{2\pi i \frac{K}{N}\al \beta}\, q^{\frac{K}{N}\al^2} y^{\frac{2K}{N} \al}\,
\hcK^{(2NK)}\left(\tau, \frac{z+\al \tau+\beta}{N}\right) \, \frac{i \th_1(\tau,z)}{\eta(\tau)^3}.
\end{eqnarray}

~

A few small remarks are in order;

~

\begin{itemize}
%
%\noindent
%{\bf 1.} 
\item
The decomposition into $\cZ_{[\al,\beta] \dis}(\tau,z)$ and 
$\cZ_{[\al,\beta] \rem}(\tau,z)$ depends on the choice 
of integration contour of the momentum $p$. 
However, $\cZ_{[\al,\beta]}(\tau,z)$ itself is of course free from 
such an ambiguity. For example, suppose we instead takes the contour 
$\br+i0$ in \eqn{s1 int EG},  we reach a different decomposition;
\begin{eqnarray}
&& 
\hspace{-1cm}
\cZ'_{[\al,\beta] \dis}(\tau,z) = 
 e^{2\pi i \frac{K}{N}\al \beta} q^{\frac{K}{N}\al^2}
y^{\frac{2K}{N}\al}\, 
\cK^{(2NK)}\left(\tau, - \frac{z+\al \tau+\beta}{N}\right)\,
\frac{i\th_1(\tau,z)}{\eta(\tau)^3},
\label{EG dis 2}
\\
%%%%
&& \hspace{-1cm}
\cZ'_{[\al,\beta] \rem}(\tau,z) = 
 \frac{i \th_1(\tau,z)}{\eta(\tau)^3} \cdot 
\frac{1}{2} \sum_{v \in \bz_{2NK}} \, 
e^{2\pi i \frac{K}{N}\al \beta}\, q^{\frac{K}{N}\al^2} y^{\frac{2K}{N} \al}\,
R^{(-)}_{v,NK}
\, \Th{v}{NK}\left(\tau,\frac{2(z+\al\tau+\beta)}{N}\right) .
\nn
&&
\label{EG rem 2}
\end{eqnarray}
in place of \eqn{EG dis 1} and \eqn{EG rem 1}. 
(Here $R^{(-)}_{m,k}$ is again defined by \eqn{Rmk}.) 
It is obvious that 
$$
\cZ'_{[\al,\beta] \dis}(\tau,z) \neq \cZ_{[\al,\beta] \dis}(\tau,z), ~~~
 \cZ'_{[\al,\beta] \rem}(\tau,z) \neq \cZ_{[\al,\beta] \rem}(\tau,z),
$$
but, the sum of them is unchanged, as should be;
$$
\cZ'_{[\al,\beta] \dis}(\tau,z) + \cZ'_{[\al,\beta] \rem}(\tau,z) = \cZ_{[\al,\beta] \dis}(\tau,z)+ \cZ_{[\al,\beta] \rem}(\tau,z).
$$
See the formula \eqn{hcK} to check this equivalence directly.

%%%%%%%%%%%%%%%%%%%%%%%%%%%%%%%%%%%%%%%%%%%%%%%%%%%%%%%%%%%%%%%%%%%%%
%\noindent
%{\bf 2.} 

\item
$\cZ_{[\al,\beta]}(\tau,z)$ has the following simple parity property
with respect to $z$;
\begin{equation}
\cZ_{[\al,\beta]}(\tau,-z) = \cZ_{[-\al,-\beta]}(\tau,z),
\end{equation}
whereas the holomorphic part $\cZ_{[\al,\beta]\dis}(\tau,z)$ is not. 
Especially, $\cZ(\tau,z)$ and $\cZ_{\msc{orb}}(\tau,z) \equiv \cZ_{[0,0]}(\tau,z)$
are even functions, as is expected. 
See again the Appendix C.

\end{itemize}

~

%%%%%%%%%%%%%%%%%%%%%%%%%%%%%%%%%%%%%%%%%%%%%%%%%%%%%%%%%%%%%%%%%%%%%%%%%%%%%%%%%%%%%%%%%

%\noindent
%\underline{\bf Comment  to the relation with \cite{Troost}: }

\subsubsection{Relation with \cite{Troost}}

In the case of $\al=\beta=0$, we obtain 
\begin{equation}
\cZ_{[0,0]}(\tau,z) = - \hcK^{(2NK)}\left(\tau, \frac{z}{N}\right) 
\, \frac{i \th_1(\tau,z)}{\eta(\tau)^3}.
\label{EG 0 0}
\end{equation}
This is essentially the result given by Troost \cite{Troost}, as we already mentioned. 
At first glance, our path-integral 
representation \eqn{EG al beta path int} may look
different 
from the one of \cite{Troost}, at $\al=\beta=0$. The latter reads as\footnote
   {See eq. (18) in \cite{Troost}.
\cite{Troost} deals only with the case of $k\in \bz_{>0}$, however, 
it is straightforward to generalize the analysis to cases with $k = \frac{N}{K}$.} 
%%%%%%
\begin{eqnarray}
%\hspace{-5mm}
\cZ^{\msc{Troost}}(\tau,z) 
&\equiv & 
\frac{1}{K} \,
\sum_{w,m\in \bsz}\,
\int_0^1 ds_1\, \int_0^1  ds_2\, 
\frac{\th_1\left(\tau, -s_1\tau-s_2+ \left(1+\frac{1}{k}\right)z\right)}
{\th_1\left(\tau, -s_1\tau-s_2+ \frac{1}{k} z \right)} 
\nn
&& \hspace{3cm}
\times 
y^{\frac{w}{N}}
\, e^{-\frac{\pi k}{\tau_2} \left|\left(s_1+ \frac{w}{N} \right)\tau 
+ \left(s_2+ \frac{m}{N} \right)\right|^2}.
\label{EG Troost}
\end{eqnarray}

Now, we try to directly show the coincidence of 
$\cZ_{[0,0]}(\tau,z)$ and $\cZ^{\msc{Troost}}(\tau,z)$
 by using the contour deformation technique as above. 
In fact, rewriting the integration variable as; 
\begin{equation}
\s_1 = \widetilde{\s}_1 - i \frac{z}{k\tau_2}, \hspace{1cm} 
\s_2 = \widetilde{\s}_2 + i \frac{\tau_1 z}{k\tau_2},
\end{equation}
and choosing the contours with 
$\xi_1= - \frac{z_1}{k\tau_2}$, $\xi_2= \frac{\tau_1 z_1}{k\tau_2}$, 
so as to make new integration variables $\widetilde{\s}_1$, $\widetilde{\s}_2$ real, 
we obtain from \eqn{EG al beta path int 2};
\begin{eqnarray}
\cZ_{[0,0]}(\tau,z) 
&=&
\frac{1}{K} e^{\pi \frac{z^2}{k\tau_2}}\, \sum_{a,b \in\bz_N} \,  
\int_{\br+i\xi_1} d\s_1\, \int_{\br+i\xi_2} d\s_2\, 
\frac{\th_1\left(\tau, -\s_1\tau-\s_2+ \left(1+\frac{2}{k}\right)z\right)}
{\th_1\left(\tau, -\s_1\tau-\s_2 + \frac{2}{k} z \right)} \,
e^{- 2\pi i z \sms_1}
\,
\nn
&& \hspace{3cm}
\times \,
 e^{-\frac{\pi k}{\tau_2}  
\left[\left\{
\left(\sms_1+\frac{a}{N}\right)\tau_1 + \sms_2+\frac{b}{N}
\right\}^2 
+ \left(\sms_1+\frac{a}{N} \right)^2 \tau_2^2 \right]},
\nn
&=&
\frac{1}{K} e^{\pi \frac{z^2}{k\tau_2}}\, \sum_{a,b \in\bz_N} \,  
\int_{\br} d\widetilde{\s}_1\, \int_{\br} d\widetilde{\s}_2\, 
\frac{\th_1\left(\tau, -\widetilde{\s}_1\tau-\widetilde{\s}_2
+ \left(1+\frac{1}{k}\right)z\right)}
{\th_1\left(\tau, -\widetilde{\s}_1\tau-\widetilde{\s}_2 + \frac{1}{k} z \right)} 
\, e^{- 2\pi i z \widetilde{\sms}_1}
\,  e^{-2 \pi \frac{z^2}{k\tau_2}} 
\nn
&& \hspace{3cm}
\times \,
e^{-\frac{\pi k}{\tau_2}
\left[\left\{
\left(\widetilde{\sms}_1+\frac{a}{N}\right)\tau_1 + \widetilde{\sms}_2+\frac{b}{N}
\right\}^2 
+ \left(\widetilde{\sms}_1+\frac{a}{N} \right)^2 \tau_2^2 \right]} \, 
e^{\pi \frac{z^2}{k\tau_2}}\, e^{2\pi i z \left(\widetilde{\sms}_1 + \frac{a}{N}\right)},
\nn
%%%
&=& 
\frac{1}{K} \, \sum_{a,b \in\bz_N} \,  
\int_{\br} d\widetilde{\s}_1\, \int_{\br} d\widetilde{\s}_2\, 
\frac{\th_1\left(\tau, -\widetilde{\s}_1\tau-\widetilde{\s}_2
+ \left(1+\frac{1}{k}\right)z\right)}
{\th_1\left(\tau, -\widetilde{\s}_1\tau-\widetilde{\s}_2 + \frac{1}{k} z \right)}  
\, e^{2\pi i z \frac{a}{N}}
\nn
&& \hspace{3cm}
\times \,
e^{-\frac{\pi k}{\tau_2}
\left[\left\{
\left(\widetilde{\sms}_1+\frac{a}{N}\right)\tau_1 + \widetilde{\sms}_2+\frac{b}{N}
\right\}^2 
+ \left(\widetilde{\sms}_1+\frac{a}{N} \right)^2 \tau_2^2 \right]} ,
\nn
%%%%
&= &\cZ^{\msc{Troost}}(\tau,z).
\label{rel Z00 Troost}
\end{eqnarray}
Thus the agreement is shown. 
%\footnote{
%The proof of modular covariance given in \cite{Troost} 
%would be problematic. 
%Some computations of Poisson resummation 
%({\em e.g.} to show the equivalence of eq. (18) with eq. (23) in \cite{Troost})
%would not be justified except for the cases of $z\in \br$. 
%}. 

~

%%%%%%%%%%%%%%%%%%%%%%%%%%%%%%%%%%%%%%%%%%%%%%%%%%%%%%%%%%%%%
%%%%%%%%%%%%%%%%%%%%%%%%%%%%%%%%%%%%%%%%%%%%%%%%%%%%%%%%%%%%%%%

\section{Conclusions}

In this paper we have studied the $SL(2,\br)/U(1)$ SUSY gauged WZW model.  
%%%%%%%%%%%%%%%%%%%%%%%%%%%%%%%%%%%%%%%%%%%%%%%%%%%%%%%%%%%%%%%%%%%%%%%%%%%%%%%%%%%%
%%%%%%%%%%%%%%%%%%%%%%%%%%%%%%%%%%%%%%%%%%%%%%%%%%%%%%%%%%%%%%%%%%%%%%%%%%%%%%%%%%%%
After introducing a suitable IR regularization preserving good modular properties, 
%%%%%%%%%%%%%%%%%%%%%%%%%%%%%%%%%%%%%%%%%%%%%%%%%%%%%%%%%%%%%%%%%%%%%%%%%%%%%%%%%%%%
%%%%%%%%%%%%%%%%%%%%%%%%%%%%%%%%%%%%%%%%%%%%%%%%%%%%%%%%%%%%%%%%%%%%%%%%%%%%%%%%%%%%
we have found that the partition function \eqn{hZreg} 
is decomposed as 
\begin{equation}
Z(\tau,z; \ep)
= Z_{\dis}(\tau,z) + Z_{\con}(\tau,z;\ep) + 
Z_{\msc{subleading}}(\tau,z;\ep),
\end{equation}
%%%
and main observations are summarized as follows;
\begin{itemize}
\item $Z_{\dis}(\tau,z)$ \eqn{Z dis} as a sum of
a finite number of the modular completion $\hchid(\tau,z)$
\eqn{hchid 0} of discrete (BPS) characters and is by itself modular invariant.

\item The remaining part $Z_{\con}(\tau,z;\ep) + 
Z_{\msc{subleading}}(\tau,z;\ep)$ contain 
only continuous (non-BPS) characters $\chic(\tau,z)$. 
The leading contribution shows a logarithmic divergence 
originating from the non-compactness of target space 
and is written in a modular invariant form;
\begin{eqnarray}
\sim (- \log \ep) \,
\sum_{n_0\in \bz_{N}, \, w_0\in \bz_{2K}}\, 
\int_0^{\infty} dp\,  
%\nn
%&& \hspace{1.5cm} \times
\chic (p, Nw_0+ K n_0;\tau,z) \chic (p, Nw_0- K n_0;\tau,z)^*.
\nonumber
\end{eqnarray}
\end{itemize}

%%%
We should emphasize that partition function 
when formulated using path-integral must be modular invariant  and thus expanded in terms of Jacobi forms.
Since BPS characters do not have a good modular property, they must necessarily appear 
in modular completed forms in the partition function. Thus the partition function acquires non-holomorphic dependence.

In our previous attempt at describing elliptic genera for ALE spaces \cite{EST,EST2}, 
we have relaxed the condition of invariance under the full modular group and imposed invariance only under the $\Gamma(2)$ congruence subgroup. In this approach we did not encounter non-holomorphic dependence in the elliptic genera for ALE spaces.

We note that here is a basic clash between the holomorphy and modular invariance in string theory on non-compact background. Insistence on strict modular invariance would introduce non-holomorphy in the theory while 
the relaxed modular invariance may still be realized within holomorphic partition functions. 

These alternatives will correspond to the choice of different boundary conditions at infinity of non-compact manifolds. From the physical point of view, however, 
we perhaps prefer invariance under the full modular group at the price of  
 non-holomorphicity in the amplitudes. 
 We, however,  do not yet have a good physical feel on the effects of the non-holomorphic dependence in string theory amplitudes and detailed studies on non-holomorphic modular forms is just beginning to become initiated.
We hope that this work sheds light on novel perspective 
in studies of non-compact superconformal fields theories.

\section*{Acknowledgment}

Research of T.Eguchi is supported by Japan Ministry of Education,
Culture, Sports, Science and Technology under grant No.19GS0219.

%%%%%%%%%%%%%%%%%%%%%%%%%%%%%%%%%%%%%%%%%%%%%%%%%%%%%%%%%%%%%%%%%%%%%%%%%%%%%%%%%%%%%%%%%%%
%%%%%%%%%%%%%%%%%%%%%%%%%%%%%%%%%%%%%%%%%%%%%%%%%%%%%%%%%%%%%%%%%%%%%%%%%%%%%%%%%%%%%%%%%%%%
%%%%%%%%%%%%%%%%%%%%%%%%%%%%%%%%%%%%%%%%%%%%%%%%%%%%%%%%%%%%%%%%%%%%%%%%%%%%%%%%%%%%%%%%%%%

\newpage

\section*{Appendix A: ~ Conventions for Theta Functions}

\setcounter{equation}{0}
\def\theequation{A.\arabic{equation}}

We assume $\tau\equiv \tau_1+i\tau_2$, $\tau_2>0$ and 
 set $q:= e^{2\pi i \tau}$, $y:=e^{2\pi i z}$;
 \begin{equation}
 \begin{array}{l}
 \dsp \th_1(\tau,z)=i\sum_{n=-\infty}^{\infty}(-1)^n q^{(n-1/2)^2/2} y^{n-1/2}
  \equiv 2 \sin(\pi z)q^{1/8}\prod_{m=1}^{\infty}
    (1-q^m)(1-yq^m)(1-y^{-1}q^m), \\
 \dsp \th_2(\tau,z)=\sum_{n=-\infty}^{\infty} q^{(n-1/2)^2/2} y^{n-1/2}
  \equiv 2 \cos(\pi z)q^{1/8}\prod_{m=1}^{\infty}
    (1-q^m)(1+yq^m)(1+y^{-1}q^m), \\
 \dsp \th_3(\tau,z)=\sum_{n=-\infty}^{\infty} q^{n^2/2} y^{n}
  \equiv \prod_{m=1}^{\infty}
    (1-q^m)(1+yq^{m-1/2})(1+y^{-1}q^{m-1/2}), \\
%\hspace{8cm} \mbox{(Jacobi's triple product identity)} \\
 \dsp \th_4(\tau,z)=\sum_{n=-\infty}^{\infty}(-1)^n q^{n^2/2} y^{n}
  \equiv \prod_{m=1}^{\infty}
    (1-q^m)(1-yq^{m-1/2})(1-y^{-1}q^{m-1/2}) .
 \end{array}
\label{th}
 \end{equation}
 \begin{eqnarray}
 \Th{m}{k}(\tau,z)&=&\sum_{n=-\infty}^{\infty}
 q^{k(n+\frac{m}{2k})^2}y^{k(n+\frac{m}{2k})} .
%\\
% \tTh{m}{k}(\tau,z)&=&\sum_{n=-\infty}^{\infty} (-1)^n
% q^{k(n+\frac{m}{2k})^2}y^{k(n+\frac{m}{2k})}.
 \end{eqnarray}
% We use abbreviations; $\th_i (\tau) \equiv \th_i(\tau, 0)$
% ($\th_1(\tau)\equiv 0$), 
%$\Th{m}{k}(\tau) \equiv \Th{m}{k}(\tau,0)$.
% $\tTh{m}{k}(\tau) \equiv \tTh{m}{k}(\tau,0)$.
 We also set
 \begin{equation}
 \eta(\tau)=q^{1/24}\prod_{n=1}^{\infty}(1-q^n).
 \end{equation}
%
%The anti-symmetrized theta functions 
%are defined as 
%\begin{eqnarray}
% && \Th{m}{k}^{(-)}(\tau,z) = \frac{1}{2}\left(
%\Th{m}{k}(\tau,z) - \Th{m}{k}(\tau,-z) 
%\right)\equiv \frac{1}{2}\left(
%\Th{m}{k}(\tau,z) - \Th{-m}{k}(\tau,z) 
%\right)~, \nn
% && \tTh{m}{k}^{(-)}(\tau,z) = \frac{1}{2}\left(
%\tTh{m}{k}(\tau,z) - \tTh{m}{k}(\tau,-z) 
%\right)\equiv \frac{1}{2}\left(
%\tTh{m}{k}(\tau,z) - \tTh{-m}{k}(\tau,z) 
%\right)~,
%\end{eqnarray}
% 
The spectral flow properties of theta functions are summarized 
as follows;
\begin{eqnarray}
 && \th_1(\tau, z+m\tau+n) = (-1)^{m+n} 
q^{-\frac{m^2}{2}} y^{-m} \th_1(\tau,z) ~, \nn
&& \th_2(\tau, z+m\tau+n) = (-1)^{n} 
q^{-\frac{m^2}{2}} y^{-m} \th_2(\tau,z) ~, \nn
&& \th_3(\tau, z+m\tau+n) = 
q^{-\frac{m^2}{2}} y^{-m} \th_3(\tau,z) ~, \nn
&& \th_4(\tau, z+m\tau+n) = (-1)^{m} 
q^{-\frac{m^2}{2}} y^{-m} \th_4(\tau,z) ~, \nn
%%%
&& \Th{a}{k}(\tau, 2(z+m\tau+n)) = 
%e^{2\pi i n a} 
q^{-k m^2} y^{-2 k m} \Th{a}{k}(\tau,2z)~.
%\nn
%&& \tTh{a}{k}(\tau,2(z+m\tau+n)) = (-1)^{m}e^{2\pi i n a} 
%q^{-k m^2 } y^{-2k m} \tTh{a}{k}(\tau,2z)~,
\label{sflow theta}
\end{eqnarray}

~

%%%%%%%%%%%%%%%%%%%%%%%%%%%%%%%%%%%%%%%%%%%%%%%%%%%%%%%%%%%%
%%%%%%%%%%%%%%%%%%%%%%%%%%%%%%%%%%%%%%%%%%%%%%%%%%%%%%%%%%%%

\section*{Appendix B:~ Summary of $H^+_3$-Gauged WZW Model }

\setcounter{equation}{0}
\def\theequation{B.\arabic{equation}}

%%%%%%%%%%%%%%%%%%%%%%%%%%%%%%%%%%%%%%%%%%%%%%%%%%%%%%%%%%%%%%%%%%%%%%%%%5

In this appendix we summarize relevant formulas 
%for our calculations
about the gauged WZW model associated with $H^+_3 \cong SL(2,\bc)/SU(2)$.
We here denote the $H^+_3$-WZW action as $S_{\msc{WZW}}(g)$, defined by 
\begin{equation}
S_{\msc{WZW}} (g) \equiv -\frac{1}{8\pi} \int_{\Sigma} d^2v\,
\tr \left(\partial_{\al}g^{-1}\partial_{\al}g\right) +
\frac{i}{12\pi} \int_B \,\tr\left((g^{-1}dg)^3\right), ~~~ (g(v,\bar{v}) \in H^+_3).
\label{H3+ WZW action}
\end{equation}
This is formally defined as an analytic continuation of $- S^{SU(2)}_{\msc{WZW}}(g)$, 
and positive definite for $\any g(v,\bar{v}) \in H^+_3$.

%%%%%%%%%%%%%%%%%%%%%%%%%%%%%%%%%%%%%%%%%%%%%%%%%%%
%%%%%%%%%%%%%%%%%%%%%%%%%%%%%%%%%%%%%%%%%%%%%%%%%%%
%%%%%%%%%%%%%%%%%%%%%%%%%%%%%%%%%%%%%%%%%%%%%%%%%%%
The Polyakov-Wiegmann identity is written as 
\begin{equation}
S_{\msc{WZW}}(gh)= S_{\msc{WZW}}(g)+ S_{\msc{WZW}}(h)
+ \frac{1}{\pi} \int_{\Sigma} d^2v\, \tr \left(g^{-1}\partial_{\bar{v}} g \, 
\partial_v h h^{-1}\right).
\label{PW}
\end{equation}
which plays a fundamental role in the analysis of (gauged) WZW model. 
%%%
It is convenient to define the vector 
and axial-type gauged WZW actions in the forms of 
\begin{eqnarray}
&& S_{\msc{gWZW}}^{(V)} (g, h, h^{\dag}) \equiv S_{\msc{WZW}}(h g h^{\dag}) 
- S_{\msc{WZW}}(h h^{\dag}),
\label{gWZW V}
\\
&& S_{\msc{gWZW}}^{(A)} (g, h, h^{\dag}) \equiv S_{\msc{WZW}}(h g h^{\dag}) 
- S_{\msc{WZW}}(h h^{\dag\, -1}),
\label{gWZW A}
\end{eqnarray}
where we assume $h(v,\bar{v}) \in  \exp \left(\bc \sigma_2\right)$, 
and the gauge field (defined to be a hermitian 1-form) is identified as 
\begin{equation}
A_{\bar{v}}\frac{\sigma_2}{2} = \partial_{\bar{v}} h h^{-1}, 
\hspace{1cm}
A_v \frac{\sigma_2}{2} = \partial_v h^{\dag} h^{\dag\,-1}.    
\nonumber
\end{equation}
The chiral gauge transformation is defined as 
\begin{equation}
g~ \longmapsto ~ \Omega g \Omega^{\dag}, ~~~ h~ \longmapsto ~ h \Omega^{-1}, 
~~~ h^{\dag}~ \longmapsto ~ \Omega^{\dag\, -1} h^{\dag}, ~~~ (\any \Omega(v,\bar{v}) \in 
\exp \left( \bc \sigma_2 \right)),
\label{chiral gtrsf}
\end{equation}
and the vector-like (axial-like) action 
\eqn{gWZW V} (\eqn{gWZW A}) is anomaly free along the vector (axial) direction
$\Om (v,\bar{v}) \in  \exp \left( i \br \sigma_2 \right)$ 
($\Om (v,\bar{v}) \in  \exp \left(  \br \sigma_2 \right)$).
%%%%%%%%%%%%%%%%%%%%%%%%%%%%%%%%%%%%%%%%%%%%%%%%%%%%%
%%%%%%%%%%%%%%%%%%%%%%%%%%%%%%%%%%%%%%%%%%%%%%%%%%%%%
%%%%%%%%%%%%%%%%%%%%%%%%%%%%%%%%%%%%%%%%%%%%%%%%%%%%%

The next path-integral formula has been given in \cite{GawK,Gaw}, and is useful 
for our analysis (up to some normalization constant $C$;
\begin{eqnarray}
\hspace{-5mm}
\int \cD g\, \exp\left[ - \kappa S_{\msc{WZW}}(h[u] g h[u]^{\dag})\right]
\equiv  \tr\, \left(q^{L_0-\frac{c_g}{24}}
\bar{q}^{\tilde{L}_0-\frac{c_g}{24}}\,
 e^{2\pi i (u j^3_0 -\bar{u} \tilde{j}^3_0)} \right) 
%&& \hspace{2cm}
= C \frac{e^{-(\kappa -2)\pi \frac{u_2^2}{\tau_2}}}{\sqrt{\tau_2}
|\th_1(\tau,u)|^2}~, \label{Gaw formula 1} 
\end{eqnarray} 
where we used the notation $u_1 \equiv \Re (u) \equiv s_1\tau_1 +s_2$, 
$u_2 \equiv \Im (u) \equiv s_1 \tau_2$, and 
$h[u]$ is defined in \eqn{parameterization h}, \eqn{Phi u}, namely, 
$$
h[u] \equiv e^{i\Phi[u] \frac{\sigma_2}{2}}, \hspace{1cm} 
\Phi[u] (w,\bar{w})
%= \frac{i}{2\tau_2}\left\{(\bar{w}\tau-w\bar{\tau})s_1+(\bar{w}-w)s_2\right\}
\equiv \frac{1}{\tau_2} \Im (w \bar{u}).
$$
%%%
One can easily find 
$$ S_{\msc{WZW}}(h[u] h[u]^{\dag})= \frac{\pi u_2^2}{\tau_2},
\hspace{1cm} 
S_{\msc{WZW}}(h[u]h[u]^{\dag\,-1})= -\frac{\pi u_1^2}{\tau_2}
\equiv \frac{\pi u_2^2}{\tau_2} - \frac{\pi |u|^2}{\tau_2}
$$ 
by direct calculations, 
and thus we further obtain 
\begin{eqnarray}
&& \int \cD g\, \exp\left[-\kappa S^{(V)}(g,\, h[u],\, h[u]^{\dag})\right]
= C \frac{e^{2\pi \frac{u_2^2}{\tau_2}}}{\sqrt{\tau_2}
|\th_1(\tau,u)|^2}, \label{Gaw formula 2} \\ 
&& \int \cD g\, \exp\left[-\kappa S^{(A)}(g,\, h[u],\, h[u]^{\dag})\right] 
= C  \frac{e^{2\pi \frac{u_2^2}{\tau_2} 
- \pi \kappa \frac{|u|^2}{\tau_2}}}{\sqrt{\tau_2}
|\th_1(\tau,u)|^2}. \label{Gaw formula 3} 
\end{eqnarray}
%%%
Note that the expressions \eqn{Gaw formula 2} and \eqn{Gaw formula 3} are 
modular invariant, whereas \eqn{Gaw formula 1} is not.

~

%%%%%%%%%%%%%%%%%%%%%%%%%%%%%%%%%%%%%%%%%%%%%%%%%%%%%%%%%%%%

\section*{Appendix C:~ Extended Characters and Modular Completion}

\setcounter{equation}{0}
\def\theequation{C.\arabic{equation}}

%%%%%%%%%%%%%%%%%%%%%%%%%%%%%%%%%%%%%%%%%%%%%%%%%%%%%%%%%%%%%%%%%%%%%%%%%

We consider the case of $\hc \equiv 1+ \frac{2K}{N}$, and focus only on  the $\tR$-sector.

~

\noindent
{\bf Extended Continuous (non-BPS) Characters \cite{ES-L,ES-BH}:}

\begin{equation}
\chic (p,m;\tau,z) := q^{\frac{p^2}{4NK}} \Th{m}{NK}\left(\tau,\frac{2z}{N}\right)\,
\frac{i\th_1(\tau,z)}{\eta(\tau)^3}.
\label{chic}
\end{equation}
%%%%%%%%%%%%%%%%%%%%%%%%%%%%%%%%%%%%%%%%%%%%%%%%%%%%%%%%%%%%%%%%%%%%%%%%%%%%%%
%%%%%%%%%%%%%%%%%%%%%%%%%%%%%%%%%%%%%%%%%%%%%%%%%%%%%%%%%%%%%%%%%%%%%%%%%%%%%%
This corresponds to the spectral flow sum of the non-degenerate representation with
$h= \frac{p^2+m^2}{4NK} + \frac{\hc}{8}$, 
$Q = \frac{m}{N}\pm \frac{1}{2}$~($p\geq 0$, $m\in \bz_{2NK}$),
whose flow momenta are taken to be $n\in N \bz$.
%%%%%%%%%%%%%%%%%%%%%%%%%%%%%%%%%%%%%%%%%%%%%%%%%%%%%%%%%%%%%%%%%%%%%%%%%%%%
%%%%%%%%%%%%%%%%%%%%%%%%%%%%%%%%%%%%%%%%%%%%%%%%%%%%%%%%%%%%%%%%%%%%%%%%%%%%

~

%%%%%%%%%%%%%%%%%%%%%%%%%%%%%%%%%%%%%%%%%%%%%%%%%%%%%%%%%%%%

\noindent
{\bf Extended Discrete (BPS) Characters \cite{ES-L,ES-BH}:}

\begin{equation}
\chid (v,a;\tau,z) := \sum_{n\in\bz}\, \frac{(yq^{N n+ a})^{\frac{v}{N}}}
{1-yq^{Nn+a}} \, y^{2K\left(n+\frac{a}{N}\right)} q^{NK \left(n+\frac{a}{N}\right)^2}
\, \frac{i\th_1(\tau,z)}{\eta(\tau)^3}.
\label{chid}
\end{equation}
%%%%%%%%%%%%%%%%%%%%%%%%%%%%%%%%%%%%%%%%%%%%%%%%%%%%%%%%%%%%%%%%%%%%%%%
%%%%%%%%%%%%%%%%%%%%%%%%%%%%%%%%%%%%%%%%%%%%%%%%%%%%%%%%%%%%%%%%%%%%%%%
This again corresponds to the  sum of the 
Ramond vacuum representation with $h= \frac{\hc}{8}$, $Q= \frac{v}{N}-\frac{1}{2}$
~($v=0,1,\ldots , N-1$) over spectral flow with flow momentum $m$  taken to be mod.$N$, as 
$m= a +N\bz$ ~ ($a\in \bz_N$).
%%%%%%%%%%%%%%%%%%%%%%%%%%%%%%%%%%%%%%%%%%%%%%%%%%%%%%%%%%%%%%%%%%%%%%%
%%%%%%%%%%%%%%%%%%%%%%%%%%%%%%%%%%%%%%%%%%%%%%%%%%%%%%%%%%%%%%%%%%%%%%%
If one introduces the notation of Appell function or Lerch sum
\cite{Pol,STT,Zwegers}, 
\begin{equation}
 \cK^{(2k)}(\tau,z) 
:= \sum_{n\in \bsz} \frac{q^{kn^2} y^{2kn}}
{1-yq^n}
\label{Appell}
\end{equation}
one can write as
\begin{eqnarray}
&& \chid (v,a;\tau,z) = \frac{1}{N} \sum_{b\in\bz_N}\,
 e^{-2\pi i \frac{v b}{N}} q^{\frac{K}{N} a^2} y^{\frac{2K}{N} a}\,
 \cK^{(2NK)}\left(\tau, \frac{z+a\tau+b}{N}\right)\, 
\frac{i\th_1(\tau,z)}{\eta(\tau)^3},
\nn
%%%%
&& q^{\frac{K}{N} a^2} y^{\frac{2K}{N} a}\,
 \cK^{(2NK)}\left(\tau, \frac{z+a\tau+b}{N}\right)\, 
\frac{i\th_1(\tau,z)}{\eta(\tau)^3} 
= \sum_{v=0}^{N-1} \,  e^{2\pi i \frac{v b}{N}} \, \chid (v,a;\tau,z).
\label{rel chid cK} 
\end{eqnarray}

The anomalous modular transformation formula of $\chid (v,a)$ and $\cK^{(2k)}$ 
can be expressed as \cite{ES-L,ES-BH,Zwegers,STT};
%%%
\begin{eqnarray}
&& 
\hspace{-1cm}
\chid \left(v,a ; - \frac{1}{\tau}, \frac{z}{\tau}\right)
= e^{i\pi \frac{\hc}{\tau}z^2}\,\left[
 \sum_{v=0}^{N-1} \,\sum_{a\in \bz_N}\,
\frac{1}{N} e^{2\pi i \frac{vv' - (v+2Ka)(v'+2Ka')}{2NK}}
\, \chid  (v',a';\tau,z) \right.
\nn
&& \hspace{1.5cm}
\left. -\frac{i}{2NK} \sum_{m' \in \bz_{2NK}} \, e^{-2\pi i \frac{(v+2Ka) m'}{2NK}}\,
\int_{\br+i0} dp'\, \frac{e^{-2\pi \frac{vp'}{2NK}}}{1-e^{-2\pi \frac{p'+im'}{2K}}}
\, \chic (p',m';\tau,z)
\right],
\label{S chid}
\end{eqnarray}
%%%
\begin{equation}
\cK^{(2 k)}\left(-\frac{1}{\tau}, \frac{z}{\tau}\right)
= \tau e^{i\pi  \frac{ 2k z^2}{\tau}}\,
\left[ \cK^{(2k)}(\tau,z) - \frac{i}{\sqrt{2 k}}\, \sum_{m\in \bz_{2k}}\,
\int_{\br+i0} dp' \, \frac{q^{\frac{1}{2}p^{'2}}}
{1-e^{-2\pi \left(\frac{p'}{\sqrt{2k}}+i\frac{m}{2k}\right)}}\,
\Th{m}{k}(\tau,2z)
\right]
\label{S cK}
\end{equation}
Integral over $p'$ in the above formulas  
is called the Mordell's integral \cite{Mordell,Watson}.

%%%%%%%%%%%%%%%%%%%%%%%%%%%%%%%%%%%%%%%%%%%%%%%%%%%%%%%%%%%%%%%%%%%

~

\noindent
{\bf Modular Completion :}

Following \cite{Zwegers}\footnote
  {See Chapter 3 (especially, `Definition 3.4' and related propositions) 
in \cite{Zwegers}.},
we define the modular completion of Appell function $\cK^{(2k)}$ 
by the following combination;
\begin{eqnarray}
\hcK^{(2k)} (\tau, z) &:=& \cK^{(2k)}(\tau,z) - \frac{1}{2} \sum_{m\in \bz_{2k}}\,
R^{(+)}_{m,k}(\tau) \, \Th{m}{k}(\tau,2z)
\nn
&\equiv & - \cK^{(2k)}(\tau,-z) - \frac{1}{2} \sum_{m\in \bz_{2k}}\,
R^{(-)}_{m,k}(\tau) \, \Th{m}{k}(\tau,2z).
\label{hcK}
\end{eqnarray}
Here we have 
\begin{equation}
R^{(\pm)}_{m,k}(\tau) := \sum_{\la \in m+2k\bz}\, \sgn(\la \pm 0) 
\erfc\left(\sqrt{\frac{\pi \tau_2}{k}} \left|\la\right|\right)\, q^{- \frac{\la^2}{4k}},
\label{Rmk}
\end{equation}
and $\erfc(*)$ denotes the error-function \eqn{erfc}. In (\ref{Rmk}) it supplies a strong enough damping factor to make the power series convergent.
Note that $\hcK^{(k)}(\tau,z)$ is holomorphic with respect to $z$, but {\em  not}
with respect to $\tau$, since $R^{(\pm)}_{m,k}$ depends on $\tau_2$.  
%%%%%%%%%%%%%%%%%%%%%%%%%
%We note 
%\begin{eqnarray}
%&& R^{(-)}_{m,k}(\tau) = - R^{(+)}_{-m,k}(\tau),\,\,
% R^{(+)}_{m,k} (\tau) - R^{(-)}_{m,k} (\tau) = 2 \delta^{(2k)}_{m,0}.
%\label{R identity}
%\end{eqnarray}
%%%%%%%%%%%%%%%%%%%%%%%%%%%%%%%%%%%%%%
%It is worthwhile to note that $R^{(\pm)}_{m,k}(\tau)$ satisfies an
%analogue of Laplace equation\footnote
%   {A (, in general) modular form of weight $k$ 
%is called the `weak Maass form of weight $k$', if it 
%is an eigen-function of the `weight $k$ Laplacian' defined as 
%$$
%\Delta_k := - y^2 \left(\frac{\partial^2}{\partial x^2}+ \frac{\partial^2}{\partial y^2}
%\right) 
%+ iky \left(\frac{\partial}{\partial x}+i\frac{\partial}{\partial y} \right) 
%\left(\equiv -4 y^{2-k}\, \frac{\partial}{\partial \tau} \, y^k \, \frac{\partial}{\partial \bar{\tau}}\right), ~~~ 
%(\tau\equiv x+iy),
%$$ 
%and an weak Maass form is called `harmonic', if its eigen-value vanishes.  
%See {\em e.g.} \cite{BOno} and references therein for more precise arguments. } 
%\begin{equation}
%\frac{\partial}{\partial \tau} \, \sqrt{\tau_2} \, \frac{\partial}{\partial \bar{\tau}}\,
%R^{(\pm)}_{m,k}(\tau) =0.
%\end{equation}
%%%%
%($\hcK^{(2k)}(\tau,z)$ itself is not a weak Maass form, however.)
%%%%%%%%%%%%%%%%%%%%%%%%%%%%%%%%%%%%%%%%%%%%%%%%%%%%%%%%%%%%%%%%%%%%%%%%%%%%%%
$R^{(+)}_{m,k}(\tau)$ is constructed in such a way that it generates the Mordell's integral under S-transformation \cite{Zwegers} ($0<t<1$);
\begin{eqnarray}
R^{(+)}_{m,k}(\tau)+{i\over \sqrt{-i\tau}}{1\over \sqrt{2k}}\sum_{\ell \in \bz_{2k}}
\, e^{-{i \pi m\ell\over k}}R^{(+)}_{\ell,k}\left(- {1\over \tau}\right)
=2ie^{-{i \pi  m^2\tau\over 2k}}\int_{\br-it} dp\, 
{e^{2\pi i k\tau p^2-2\pi m \tau p}\over 1-e^{2\pi p}}.
\end{eqnarray}
Mordell integrals then cancel in the combination $\hcK^{(2k)}(\tau,z)$ and the completed Appell function has a good transformation law
 \cite{Zwegers};
%%%%%%%%%%%%%%%%%%%%%%%%%
\begin{equation}
\hcK^{(2k)}\left(-\frac{1}{\tau}, \frac{z}{\tau}\right)
= \tau e^{i\pi \frac{2k z^2}{\tau}}\, \hcK^{(2k)}(\tau,z).
\label{S hcK}
\end{equation}
%whereas $\cK^{(2k)}(\tau,z)$ does not. (See \eqn{S cK}.)
%%%
To be precise, $\hcK^{(2k)}(\tau,z)$ is a {\em real-analytic\/} Jacobi form \cite{EZ}
of weight 1 and index $k$;
\begin{eqnarray}
&& \hspace{-5mm}
\hcK^{(2k)}\left(
\frac{a\tau+b}{c\tau+d}, \frac{z}{c\tau+d}
\right) = (c\tau+d)\,  e^{2\pi ik \frac{c z^2}{c\tau+d}} \, \hcK^{(2k)}(\tau,z)~, ~~~
%%%%
\hskip-3mm \left(
\begin{array}{cc}
 a & b \\
 c & d
\end{array}
\right) \in 
SL(2;\bz) 
\label{modular hcK}
\\
%\end{eqnarray}
%If $\cG$ is the full modular group $SL(2;\bz)$, this is equivalent with 
% \begin{eqnarray}
%  && Z(\tau+1,z) = Z(\tau,z) ~, \nn
%  && Z\left(-\frac{1}{\tau}, \frac{z}{\tau}\right)
%  = e^{i\pi \frac{\hc z^2}{\tau}} Z(\tau,z)~.
%\label{modular Z 2}
% \end{eqnarray}
%%%
%\item[(ii) Double Quasi-periodicity (`Spectral Flow Invariance')]
%\begin{eqnarray}
 && \hspace{-5mm}
\hcK^{(2k)} (\tau,z+\ell\tau+m) =  q^{-k \ell^2} y^{-2k \ell} 
\hcK^{(2k)} (\tau,z)~, \hskip1cm 
\ell, m \in \bz.
%%&& \hspace{3cm} \mbox{(when $N \hc$ is a positive integer )}
\label{s flow hcK}
\end{eqnarray}

%We also point out that $\hcK^{(2k)}(\tau,z)$ is an odd function;
%\begin{equation}
%\hcK^{(2k)}(\tau,-z) = - \hcK^{(2k)}(\tau,z),
%\label{hcK odd fn}
%\end{equation}
%while $\cK^{(2k)}$ is not. 
%\eqn{hcK odd fn} is easily confirmed by the identity 
%\begin{equation}
%\cK^{(2k)}(\tau,z) + \cK^{(2k)}(\tau,-z) = \Th{0}{k}(\tau,2z),
%\end{equation}
%and the relation \eqn{R identity}.

%%%%%%%%%%%%%%%%%%%%%%%%%%%%%%%%%%%%%%%%%%%%%%%%%%%%%%%%%%%%%%%%%%%%

The modular completion of the discrete character \eqn{chid}
is defined as 
\begin{eqnarray}
\hchid (v,a;\tau,z) %%%
&:=  &  \chid (v,a;\tau,z) - \frac{1}{2} \sum_{j\in \bz_{2K}}\,
R^{(+)}_{v+Nj, NK}(\tau) \Th{v+Nj+2Ka}{NK}\left(\tau, \frac{2z}{N}\right)\,
\frac{i\th_1(\tau,z)}{\eta(\tau)^3},
\nn
&\equiv & \frac{1}{N} \sum_{b\in\bz_N}\,
 e^{-2\pi i \frac{v b}{N}} q^{\frac{K}{N} a^2} y^{\frac{2K}{N} a}\,
 \hcK^{(2NK)}\left(\tau, \frac{z+a\tau+b}{N}\right)\, 
\frac{i\th_1(\tau,z)}{\eta(\tau)^3}.
\label{hchid}
\end{eqnarray}
%%%
The modular completion $\hchid$ 
%\eqn{hchid} 
is again non-holomorphic with respect to $\tau$, 
however, has a good modular property;
\begin{eqnarray}
&& \hspace{-1cm}
\hchid \left(v,a ; - \frac{1}{\tau}, \frac{z}{\tau}\right)
= e^{i\pi \frac{\hc}{\tau}z^2}\, \sum_{v'=0}^{N-1} \,\sum_{a'\in \bz_N}\,
\frac{1}{N} e^{2\pi i \frac{vv' - (v+2Ka)(v'+2Ka')}{2NK}}
\, \hchid (v',a';\tau,z).
\label{S hchid}
\end{eqnarray}
This is easily proven by using the modular transformation formula of $\hcK^{(2k)}$
\eqn{S hcK} and the second line of \eqn{hchid}.

~

\noindent
{\bf Witten Index :}
\begin{equation}
\lim_{z\,\rightarrow\, 0}\, \chid (v,a;\tau,z) = 
\lim_{z\,\rightarrow\, 0}\, \hchid (v,a;\tau,z) = - \delta_{a,0}^{(N)}.
\label{WI}
\end{equation}

~

%%%%%%%%%%%%%%%%%%%%%%%%%%%%%%%%%%%%%%%%%%%%%%%%%%%%%%%%%%%

\section*{Appendix D: ~ Useful Formulas}

\setcounter{equation}{0}
\def\theequation{D.\arabic{equation}}

\noindent
{\bf Poisson Resummation Formula : }
%%%
\begin{eqnarray}
&& \hspace{-5mm}
\sum_{n\in\bsz}\exp\left(-\pi \al (n+a)^2+2\pi i b (n+a)\right)
=\frac{1}{\sqrt{\al}}\sum_{m\in\bsz}\exp
\left(-\frac{\pi(m-b)^2}{\al}+2\pi i m a\right), 
%\hspace{1cm}
%(\Re \, \al>0)
\nn
&& \hspace{12cm} 
(\Re \, \al>0)
\label{PR formula}
%\\
%&& \sum_{n\in \bsz}\delta(x+nL)=\frac{1}{L}\sum_{m\in \bsz}
%e^{\frac{2\pi i m x}{L}}~.
%\end{eqnarray}
%%%%%%%%%%%%%%%%%
%\begin{eqnarray}
% && \hspace{-1cm}
%\sum_{m,w\in \bsz}
%e^{-\frac{\pi \al}{\tau_2}\left|(w+s_1)\tau + (m+s_2)\right|^2}
%  e^{2\pi i (m t_1-w t_2)}
%= \frac{1}{\al} \sum_{n,l\in \bsz}
%e^{-\frac{\pi}{\al \tau_2}\left|(n+t_1)\tau + (l+t_2)\right|^2}
%  e^{2\pi i \left\lb (l+t_2) s_1- (n+t_1) s_2\right\rb}~, 
%\hspace{10cm} 
%( \Re \, \al>0, ~~~
%s_i, t_i \in \br).
%\label{dualizing}
\end{eqnarray}
%The identity \eqn{dualizing} is proven by using twice the Poisson resummation formula
%\eqn{PR formula}. 

~

\noindent
{\bf Error-functions : }

We define the error-functions as 
\begin{eqnarray}
\erf(x)&:=& \frac{2}{\sqrt{\pi}} \int_0^x e^{-t^2} \, dt
\label{erf}
\\
\erfc(x)&:=& \frac{2}{\sqrt{\pi}} \int_x^{\infty} e^{-t^2} \, dt
\equiv 1- \erf(x)
\label{erfc}
\end{eqnarray}
Here we normalize $\erf(x)$ as $\erf(\infty)=1$.

%%%
A relevant integration formula is as follows;
%\footnote
%  {See, {\it e.g.} Iwanami Koshikisyu I, p232.};
\begin{eqnarray}
\int_0^{\infty} \frac{e^{-a^2 x^2}}{x^2+b^2} \, dx 
&=& \frac{\sqrt{\pi}}{b} e^{a^2 b^2} \, \int_{ab}^{\infty} e^{-t^2}\, dt
\nonumber\\
%&\equiv & \frac{\pi}{2b} e^{a^2 b^2} \,\left\{1 - \erf(ab)\right\}, ~~~ (a,b >0),
&\equiv & \frac{\pi}{2b} e^{a^2 b^2} \, \erfc(ab), ~~~ (a,b >0).
\label{formula erfc}
\end{eqnarray}
From this formula we readily obtain
\begin{equation}
\int_{\br\mp i0} dp\, \frac{e^{-\al p^2}}{p-i\la} = i\pi e^{\al \la^2} \sgn(\la\pm 0) 
\erfc(\sqrt{\al}|\la|), 
\hspace{1cm}(\la \in \br, ~ \al >0),
\label{formula erfc 2}
\end{equation}
which is useful for our calculation.
Especially, the function
%`indefinite theta function' 
$R^{(\pm)}_{m,k}(\tau)$ \eqn{Rmk}
is expressible as 
%%%%%%%%%%%%%%%%%%%%%%%%%%%%%%%%%%%%%%%%%%%%%%%
\begin{equation}
R^{(\pm)}_{m,k}(\tau) = \frac{1}{i\pi}\, \sum_{\la \in m+2k\bz}\,
\int_{\br\mp i0} dp \, \frac{e^{-\pi \tau_2 \frac{p^2+\la^2}{k}} }{p-i\la}\,
q^{- \frac{\la^2}{4k}}.
\label{id Rmk} 
\end{equation}

%%%

~

%%%%%%%%%%%%%%%%%%%%%%%%%%%%%%%%%%%%%%%%%%%%%%%%%%%%%%%%%%%%%%%%%
%%%%%%%%%%%%%%%%%%%%%%%%%%%%%%%%%%%%%%%%%%%%%%%%%%%%%%%%%%%%%%%%%

\noindent
{\bf Useful Formulas for $\th_1(\tau,z)$ :}

Following expansion  is useful for our calculations;
($u\equiv s_1\tau+s_2$, $0< s_1 < 1$);
\begin{eqnarray}
\frac{\th_1(\tau,u+z)}{\th_1(\tau,u)} &=&
\frac{-i \th_1(\tau,z)}{\eta(\tau)^3} \, \sum_{n\in\bsz}\, 
\frac{e^{2\pi i n u}}{1-yq^n}, \hspace{1cm}\left( y\equiv e^{2\pi i z}\right)
\label{th1/th1 formula 1}
\\
%%%
\frac{\th_1(\tau,-u+z)}{\th_1(\tau,-u)} &=&
\frac{-i \th_1(\tau,z)}{\eta(\tau)^3} \, \sum_{n\in\bsz}\, 
\frac{y q^n  e^{- 2\pi i n u}}{1-yq^n}.
\label{th1/th1 formula 2}
\end{eqnarray}
Note that the convergence of power series in R.H.S of \eqn{th1/th1 formula 1} 
and \eqn{th1/th1 formula 2} requires  the condition $0< s_1 <1$. 
These identities are essentially the same  
as the branching relation of $SL(2;\br)/U(1)$ supercoset 
theory analysed {\em e.g.} in \cite{ES-BH} (see also \cite{DPL}).

A proof of the first identity \eqn{th1/th1 formula 1} is given 
by comparing the poles and their residues of both sides. 
In fact, the L.H.S has simples poles at $u=r\tau+s$, ($r,s\in \bz$), 
whose residues are found to be 
\begin{equation}
\mbox{Res}_{u=r\tau+s}[\mbox{L.H.S}] = \frac{y^{-r}}{2\pi} \frac{ \th_1(\tau,z)}{\eta(\tau)^3}.
\label{res formula}
\end{equation}
On the other hand, the R.H.S of \eqn{th1/th1 formula 1} is analytically 
continued into a meromorphic function of $u\in \bc$
through a double power series expansion ($\xi\equiv e^{2\pi i u}$);
\begin{eqnarray}
\mbox{R.H.S} &=& \frac{-i\th_1(\tau,z)}{\eta(\tau)^3}\,
\left[
\sum_{n=0}^{\infty}\sum_{m=0}^{\infty} \, \xi^n y^m q^{nm}
-  \sum_{n=1}^{\infty}\sum_{m=1}^{\infty} \, \xi^{-n} y^{-m} q^{nm}
\right]
\nn
&=& \frac{-i\th_1(\tau,z)}{\eta(\tau)^3}\,
\sum_{m \in \bz} \frac{y^m}{1-\xi q^m}.
\label{id RHS}
\end{eqnarray}
(The power series in the second line is convergent for $z=t_1\tau+t_2$, $0<t_1<1$.)
It is easy to check that \eqn{id RHS} also possesses  simple poles 
at $u=r\tau+s$ ($r,s\in \bz$) with the residues equal to \eqn{res formula},
which completes the proof. 

\eqn{th1/th1 formula 2} is readily derived from  \eqn{th1/th1 formula 1}.

~

%%%%%%%%%%%%%%%%%%%%%%%%%%%%%%%%%%%%%%%%%%%%%%%%%%%%%%%%%%%%%%%%%%%%%

\newpage

\noindent
{\bf `Regularization formula' :}

\begin{eqnarray}
\sum_{n=0}^{\infty} \frac{e^{-\vep(n+z)}}{n+z} &=& \int_{\vep}^{\infty}\frac{e^{-t}}{t} \,dt  
- \frac{d}{d z} \log \Gamma(z) + \cO(\vep)
\nn
&=& - \log (\vep) + C 
- \frac{d}{d z} \log \Gamma(z) + \cO(\vep)
\hspace{1cm} (\vep >0, ~ \Re\, z>0),
\label{reg formula}
\end{eqnarray}
where $C$ expresses some $\cO(\vep^0)$-constant independent of $z$.

This formula is derived from the identity of the digamma function;
$$
\psi(z) \equiv \frac{d}{dz} \log \Gamma(z) = \int_0^{\infty} \, 
\left(\frac{e^{-t}}{t} - \frac{e^{-tz}}{1-e^{-t}} \right)\, dt, 
\hspace{1cm} (\Re\, z>0).
$$

~

%\section*{Appendix D: ~ Jacobi Forms}

%\setcounter{equation}{0}
%\def\theequation{D.\arabic{equation}}

%Let $\hat{c}$ be an positive integer. 
%The (weak) Jacobi form $Z(\tau,z)$ 
%of weight 0 and $\hat{c}/2$ with the congluence 
%group $\cG (\subset SL(2;\bz))$
%is characterized 
%by the following properties ({\em e.g.} \cite{KYY});
%\begin{description}
% \item[(i) Modular covariance : ]
%\begin{eqnarray}
%&& Z\left(
%\frac{a\tau+b}{c\tau+d}, \frac{z}{c\tau+d}
%\right) = e^{i\pi \hc \frac{c z^2}{c\tau+d}} \, Z(\tau,z)~, ~~~
%%%%
%\left(
%\begin{array}{cc}
% a & b \\
% c & d
%\end{array}
%\right) \in \cG ~.
%%SL(2;\bz) ~,
%\label{modular Z}
%\end{eqnarray}
%If $\cG$ is the full modular group $SL(2;\bz)$, this is equivalent with 
% \begin{eqnarray}
%  && Z(\tau+1,z) = Z(\tau,z) ~, \nn
%  && Z\left(-\frac{1}{\tau}, \frac{z}{\tau}\right)
%  = e^{i\pi \frac{\hc z^2}{\tau}} Z(\tau,z)~.
%\label{modular Z 2}
% \end{eqnarray}
%%%
%\item[(ii) Double Quasi-periodicity (`Spectral Flow Invariance')]
%\begin{eqnarray}
% &&q^{\frac{\hc}{2}a^2} y^{\hc a} 
%Z(\tau,z+a\tau+b) = (-1)^{\hc(a+b)} Z(\tau,z)~, ~~~
%{}^{\forall} a, b \in \bz ~ , \nn
%%&& \hspace{3cm} \mbox{(when $N \hc$ is a positive integer )}
%\label{s flow Z}
%\end{eqnarray}

%\end{description} 

%%%%%%%%%%%%%%%%%%%%%%%%%%%%%%%%%%%%%%%%%%%%%%%%%%%%%%%%%%%%%%%%%%%%%
%%%%%%%%%%%%%%%%%%%%%%%%%%%%%%%%%%%%%%%%%%%%%%%%%%%%%%%%%%%%%%%%%%%%%
%%%%%%%%%%%%%%%%%%%%%%%%%%%%%%%%%%%%%%%%%%%%%%%%%%%%%%%%%%%%%%%%%%%%%

\end{document}